\documentclass[%
 aip,
 amsmath,amssymb,
 reprint,%
]{revtex4-1}

\usepackage{graphicx}
\usepackage{dcolumn}
\usepackage{bm}
\usepackage{xcolor}

\usepackage[utf8]{inputenc}
\usepackage[T1]{fontenc}
\usepackage{mathptmx}
\usepackage{etoolbox}

\usepackage{ulem}

\usepackage{color}
\usepackage{subfigure}

\newtheorem{theorem}{Theorem}

\newcommand{\bfr}{{\mathbf{r}}}
\newcommand{\bfrprime}{{\mathbf{r'}}}

\def\R{\mathbb{R}}
\def\N{\mathbb{N}}

\def\Z{\mathbb{Z}}

\newcommand{\rhotilde}{\tilde{\rho}}
\newcommand{\mi} {{\! - \! }}

\newcommand{\bfx}{{\bf x}}
\newcommand{\bfa}{{\bf a}}
\newcommand{\bfy}{{\bf y}}

\newcommand{\bfrtilde}{{ \tilde{{\bf r}}}}
\newcommand{\s}{{\bf s}}
\newcommand{\bfR}{{\mathbf R}}
\newcommand{\rhobar}{{\overline{\rho}}}

\newcommand{\rmv}{{\rm v}}

\makeatletter
\def\@email#1#2{%
 \endgroup
 \patchcmd{\titleblock@produce}
  {\frontmatter@RRAPformat}
  {\frontmatter@RRAPformat{\produce@RRAP{*#1\href{mailto:#2}{#2}}}\frontmatter@RRAPformat}
  {}{}
}%
\makeatother
\begin{document}

\preprint{AIP/123-QED}

\title[Copula methods in density functional theory]{Copula methods for modeling pair densities in density functional theory}
\author{Genevi\`eve Dusson}
 \email{genevieve.dusson@math.cnrs.fr}
 \affiliation{Université Marie et Louis Pasteur, CNRS, LmB (UMR 6623), F-25000 Besançon, France}
\author{Claudia Kl\"uppelberg}
\email{cklu@ma.tum.de}
\affiliation{%
Department of Mathematics,
School of Computation, Information and Technology,
Technical University of Munich,
Boltzmanstrasse 3,
85748 Garching, Germany
}%
\author{Gero Friesecke}%
\email{gf@ma.tum.de}
\affiliation{%
Department of Mathematics,
School of Computation, Information and Technology,
Technical University of Munich, 
Boltzmanstrasse 3,
85748 Garching, Germany
}%

\date{\today}

\begin{abstract}
We propose a new approach towards approximating the density-to-pair-density map based on copula theory from statistics. We extend the copula theory to multi-dimensional marginals, 
and deduce that one can describe any (exact or approximate) pair density by the single-particle density and a copula. 
We present analytical formulas for the exact copula in scaling limits, numerically compute the copula for dissociating systems with two to four particles in one dimension, and propose accurate approximations of the copula between equilibrium and dissociation for two-particle systems.

\end{abstract}

\maketitle

\section{Introduction}

The applicability of Kohn--Sham (KS) density functional theory (DFT) to large many-electron systems relies on the use of exchange-correlation functionals that model the intricate many-electron interaction energy
by explicit and computationally cheap expressions in terms of the one-body density (or one-body KS orbitals). Despite their remarkable successes, the  standard DFT models like local density approximation (LDA) or generalized gradient approximations (GGA)\cite{Kohn1965-xz,Becke1988-vz,Perdew1996-lu} exhibit significant failures when strong correlations are important, as in the breaking of chemical bonds\cite{Cohen2012-xk}. Therefore, finding a computationally cheap single-particle formalism that remains accurate in strongly correlated regimes remains an important challenge. 

Here we propose a novel approach to this challenge,  by 
\begin{enumerate}
    \item taking as starting point the exact density-to-pair-density map $\rho\mapsto\rho_2$ rather than the exact density-to-interaction-energy map $\rho\mapsto V_{ee}$
    \item studying and approximating this map via \textit{copula methods} from statistics, which separates marginals (here: the  density $\rho$) from the dependence structure (the pair density $\rho_2$). 
\end{enumerate}

The viewpoint 1.~was first advocated in a celebrated paper by Gunnarsson and Lundqvist\cite{Gunnarsson1976-tu}, on the grounds that the pair density, a function on two-electron configuration space, encodes a wealth of physically interesting information about a many-electron system which is averaged out in the interaction energy 
\begin{equation} \label{VeeViaRho2}
V_{ee}=\int |\bfr-\bfr'|^{-1}\rho_2(\bfr,\bfr')d\bfr \, d\bfr', 
\end{equation}
a single number. 
In particular, the pair density allows to interpret interaction energy errors of approximate models by localizing the errors in two-body configuration space, e.g. by exhibiting the associated exchange-correlation holes. 

The approach 2.~appears to be novel in the context of density functional theory, but is standard in  multivariate statistics, for instance for modeling the dependence between the expected yield of different assets in quantitative finance\cite{Kluppelberg2009-ye,Genest2009-iy}. It has also found useful applications in energy, forestry, and environmental sciences\cite{Bhatti2019-gs}.
For a textbook account see  Czado\cite{czado_2019}. 

In fact, the theory of copulas does not apply directly to electronic structure, since it decomposes a multivariate distribution on $\R^N$ into its $N$ univariate marginals on $\R$ and a copula. Instead, for electrons moving in the physical space $\R^3$, one should factorize the configuration space $\R^{3N}$ into $N$ factors $\R^3$ (instead of $3N$ factors $\R$), and separate the $N$-electron distribution on $\R^{3N}$, or the pair density on $\R^6$, into a copula and its marginals on $\R^3$, which are given by the single-particle density $\rho$. Our first achievement in this paper is to generalize the theory of copulas to precisely such factorizations, using optimal transport theory\cite{Villani03, Fr2024}. Roughly, a generalized copula (introduced in detail in section \ref{sec:copula}) is a pair density $c$ with uniform single-particle density on a fixed reference domain; it gives rise to a unique density-to-pair-density map, by ``scaling'' the density to the uniform reference density and  applying the copula.

The usefulness of such generalized copulas depends on the following basic questions about pair densities: 
\begin{enumerate}
    \item[(1)] Do the generalized copulas representing exact ground state pair densities possess any universal or transferrable features? 
    \item[(2)] Can these features be efficiently approximated or learned, for example by neural networks? 
    \item[(3)] Do such approximated copulas provide an accurate electron-electron interaction energy?
\end{enumerate}
Our findings regarding these questions, while far from comprehensive, appear to us to be very encouraging. We 
\begin{itemize}
    \item prove that the exact generalized copula is \textit{universal} (i.e. density-independent) in the high and low density limits 
    \item prove for 1D systems that in the dissociation limit the generalized copula is given by a 
    \textit{universal analytic expression} in terms of the subsystem copulas 
    (we believe this to also hold in 3D, and prove it for certain dimers like N$_2$) 
    \item numerically compute the  exact copulas for 1D systems at different interatomic distances, and observe that these copulas exhibit fascinating multiscale properties and act as a revealing "magnification lens" for the true electron-electron dependence and for the errors of simple approximations like mean field, LDA, or strictly correlated electrons (SCE)
    \item numerically confirm our asymptotic results in the dissociation limit and moreover reveal a smooth path in copula space between equilibrium and dissociated states
    \item present simplified models for this path 
    using just one fitting parameter, finding  that a small one-parameter neural network performs best (far outperforming the LDA) and gives excellent results.
\end{itemize}
These preliminary findings suggest that developing machine-learned generalized copula models, trained on suitable reference pair densities with   asymptotic results built in as exact constraints, might be an interesting route towards novel DFT models which remain accurate in strongly correlated regimes.

The outline of the article is as follows. In Section~\ref{sec:sec2_context} we provide some notation and context. 
We then generalize, in Section~\ref{sec:copula}, the theory of copulas from one-dimensional to multi-dimensional marginals via optimal transport theory (as needed for making it applicable to DFT and describing the pair density by the single-particle density and a copula).  
In Section~\ref{sec:DFT_copulas} we explicitly compute the copulas for certain basic DFT models, namely 
independent electrons, strictly correlated electrons, and the (exchange-only) local density approximation.
In Section~\ref{sec:dissociation_theory} we present theoretical results on the structure of the copula at dissociation.
Section~\ref{sec:representability}  discusses representability constraints on the copula.
In Section~\ref{sec:cop_struct} we present our numerial results. Finally,  Section~\ref{sec:fitting} is devoted to several one-parameter models for fitting the copula.

\section{Notation and context}
\label{sec:sec2_context}

In electronic structure calculations, we are interested in calculating the wavefunction of a system, 
minimizing the (nonrelativistic) electronic energy of $N$ electrons in an external potential, 
\begin{align}
    E[\Psi] =&  
   \sum_{\s\in \Z_2^N}
   \int_{\Omega^N}
   \Big[
       |\nabla \Psi(\bfr,\s)|^2 
       + \sum_{i=1}^N V_{\rm ext} (x_i) |\Psi(\bfr,\s)|^2 \nonumber \\
       &  + \sum_{1\le i<j \le N} \rmv_{\rm ee}(|\bfr_i-\bfr_j|) |\Psi(\bfr,\s)|^2 
   \Big] d\bfr, \label{eq:energy}
\end{align}
where $\Psi$ is antisymmetric and of norm 1. The electrons are moving in a region $\Omega$ in $\R^d$, with our overall interest and some of our theoretical results covering the physical dimension $d=3$ and our numerical work focusing, for simplicity, on $d=1$. The kinetic, external, and interaction parts of the energy will, as usual, be denoted $T[\Psi]$, $V_{ext}[\Psi]$, and $V_{ee}[\Psi]$. 

The electronic density of an $N$-electron wavefunction is
\begin{equation}
    \label{eq:rho}
       \rho^\Psi(\bfr_1) = N \sum_{\s\in \Z_2^N}
   \int_{_{\Omega^{N-1}}}
   |\Psi(\bfr_1,\s_1,\ldots,
   \bfr_N,\s_N)|^2
   d\bfr_2 \ldots d\bfr_N,
\end{equation}
and the pair density is
    \begin{equation}
    \label{eq:rho2}
      \rho_2^\Psi(\bfr_1,\bfr_2) =  {N \choose 2} \!\! \sum_{\s\in \Z_2^N} \int_{_{\Omega^{N-2}}} \!\!\!\!\!\!\!\!\!\!\!\!
   |\Psi(\bfr_1,\s_1,\ldots,
   \bfr_N,\s_N)|^2
   d\bfr_3 \ldots d\bfr_N.
\end{equation}
With these definitions,
the electron-electron interaction energy 
is a simple explicit functional of the pair density,
\begin{equation}
    \label{eq:vee_tot}
    V_{\rm ee}[\Psi] = \int_{\Omega^2} \rmv_{ee}(|\bfr-\bfrprime|) \rho_2^\Psi(\bfr,\bfrprime) d\bfr d\bfrprime.
\end{equation}
This formula shows that providing an accurate pair density is enough to precisely compute the electron-electron interaction energy. This is one original motivation for this work which aims at providing novel insights into the pair density and its  approximation.

Minimizers of \eqref{eq:energy} satisfy the electronic Schr\"odinger equation
\begin{equation}
    \label{eq:HpsiEpsi}
    H \Psi = E\Psi,
\end{equation}
with the Hamiltonian $H$ given by
\[
    H = - \tfrac{1}{2} \sum_{i=1}^N \Delta_{\bfr_i} + \sum_{i=1}^N V_{\rm ext}(\bfr_i) + \sum_{1\le i<j \le N} {\rm v}_{\rm ee}(|\bfr_i-\bfr_j|).
\]

\section{Generalized copula}
\label{sec:copula}

An important principle in multivariate statistics, which to our knowledge has not previously been applied to electronic structure, is to separate the marginals (in electronic structure, the single-particle density) from the dependence structure (the $N$-point density). 

Such a separation is achieved by Sklar's theorem (see e.g. Czado\cite{czado_2019}), which expresses multivariate probability densities on $\R^{N}$ in terms of their $N$ univariate marginals on $\R$ and a function $c$ on the unit cube $[0,1]^N$ called a \textit{copula density} (see eq.~\eqref{p'} below). 
This concept has been extensively used in multivariate statistics and its applications~\cite{Genest2009-iy,Kluppelberg2009-ye,Bhatti2019-gs}. 

The copula concept cannot be applied directly to electronic structure. This is because, for electrons moving in the physical space $\R^3$ -- or more generally in $\R^d$ with any $d\!>\!1$ -- it is natural to separate the $N$-particle density on $\R^{dN}$ into $N$ single-particle marginals on $\R^d$ (instead of $d\cdot N$ single-particle marginals on $\R$) and a copula. To facilitate such a decomposition we propose the following natural --and to our knowledge novel-- generalization of a copula, based on optimal transport theory\cite{Villani03, Fr2024}. 

1. Pick a reference region $D\subseteq\R^d$ and a positive reference probability density $f_0$ on $D$ (that is, a function $f_0 \, : \, D\to\R$ with $f>0$ in $D$ and $\int_D f_0 = 1$). Prototypical choices are: \\
(i) $D$ bounded, $f_0\equiv \tfrac{1}{{\rm vol}(D)}$ (uniform distribution on $D$) \\
(ii) $D=\R^d$, $f_0(\bfr)=\tfrac{1}{(2\pi)^{d/2}}e^{-|\bfr|^2/2}$ (normal distribution).
\\
In the copula theory, the reference region $D$ is the unit interval $[0,1]$ and $f_0$ is the uniform distribution on $[0,1]$. In our general theory, one can also use any non-uniform reference density; for instance a Gaussian corresponds to the ground state density of two non-interacting electrons in a harmonic potential.

2. For any probability density $f_i$ on $\R^d$, let $T_i \, : \, \R^d\to D$ be the Brenier map which transports $f_i$ to $f_0$. Here, a map $T$ transports $f_i$ to $f_0$ if   
\begin{equation} \label{pushfwd}
     f_i(\bfr_i) = f_0(T(\bfr_i))|\det \nabla T(\bfr_i)|,
\end{equation}
and the Brenier map is the one minimizing  
the transport cost
$$
   \int_{\R^d} |\bfr_i - T(\bfr_i)|^2 f_i(\bfr_i)d\bfr_i
$$
among all maps satisfying \eqref{pushfwd}. Such a minimizing map exists and is unique\cite{Brenier91, Villani03, Fr2024}. 
In copula theory (that is, when $d=1$ and $f_0$ is the uniform density on the unit interval), the Brenier map is given by $T_i=F_i$, where  $F_i$ is the cumulative distribution function (CDF)
of $f_i$,
\begin{equation} \label{eq:cdf}
   F_i(x_i) = \int_{-\infty}^{x_i} f_i(t) \, dt.
\end{equation}
This is a standard result of optimal transport theory\cite{Villani03, Fr2024}. In higher dimensions, there is no explicit formula for the Brenier map, but it shares the key property of the CDF of $p_i$  
of being monotone, which in general dimension $d$ means $(T(\bfr_i)-T(\bfr_i'))\cdot (\bfr_i-\bfr_i')\ge 0$ for all $\bfr_i$ and $\bfr_i'$. This implies in particular that for the Brenier map, the determinant in \eqref{pushfwd} is nonnegative. For further properties of the Brenier map see the optimal transport literature\cite{Villani03, Fr2024}. 

The following novel result 
provides the theoretical foundation of copula methods for electronic structure.

\begin{theorem}[Generalized Sklar's theorem] \label{T:genSklar} For any probability density $f$ on $\R^{dN}=\R^d \times ... \times \R^d$ with marginal densities $f_1,...,f_N$ on $\R^d$, there exists a probability density $c$ on the $N$th power $D^{N}$ of the reference region $D$, which we propose to call {\rm generalized copula density},  such that 
\begin{equation} \label{p'multidim}
    f(\bfr_1,...,\bfr_N) = 
\frac{c(T_1(\bfr_1),...,T_N(\bfr_N))}{f_0(T_1(\bfr_1)) \cdots f_0(T_N(\bfr_N))} f_1(\bfr_1)  \cdots f_N(\bfr_N),
\end{equation}
where $T_i$ is the Brenier map transporting $f_i$ to $f_0$. 
The generalized copula density is unique, and is given by 
\begin{equation} \label{c'multidim2}
  c(\bfr'_1,...,\bfr'_N) = 
  \frac{f(T_1^{-1}(\bfr'_1),...,T_N^{-1}(\bfr'_N))}{f_1(T_1^{-1}(\bfr'_1)) \cdots f_N(T_N^{-1}(\bfr'_N))} f_0(\bfr'_1)\cdots f_0(\bfr'_N).
\end{equation}
\end{theorem}
In copula theory (that is, in dimension $d=1$ and with the reference density $f_0$ given by the uniform density on $[0,1]$), the Brenier maps $T_i$ are given by $T_i(x_i)=F_i(x_i)$, and the above theorem reduces to Sklar's theorem\cite{czado_2019} 
which asserts the existence of a unique function $c$ on $[0,1]^N$ such that 
\begin{equation}
    f(x_1,\ldots,x_N)  = c\bigl(F_1(x_1),\ldots,F_N(x_N)\bigr)\,  f_1(x_1) \cdots f_N(x_N). \label{p'}
\end{equation}
The function $c$ in \eqref{p'} is called \textit{copula density}\cite{czado_2019}, and can be recovered from $f$ via
\begin{equation}
    c(x_1,\ldots,x_N)  = \frac{f\bigl(F_1^{-1}(x_1),\ldots,F_1^{-1}(x_N)\bigr)}{f_1\bigl(F_1^{-1}(x_1)\bigr) \cdots f_N\bigl(F_N^{-1}(x_N)\bigr)}. \label{c'}
\end{equation}
We remark that eq.~\eqref{p'} is usually formulated in terms of the cumulative distribution functions $C(x_1,\ldots,x_N) = \int_{0}^{x_1} \ldots \int_{0}^{x_N}  c(t_1,\ldots,t_N)\, dt_1 \ldots dt_N$ of $c$ and $F(x_1,...,x_N)=\int_{-\infty}^{x_1} \ldots \int_{-\infty}^{x_N}  f(t_1,\ldots,t_N)\, dt_1 \ldots dt_N$ of $f$, as 
$$
   F(x_1,\ldots,x_N) = C\bigl(F_1(x_1),\ldots,F_N(x_N)\bigr). 
$$
The function $C$ is called \textit{copula}. 
Applying the differential operator ${\partial^N}/{\partial x_1...\partial x_N}$ to both sides recovers the formulation \eqref{p'} in terms of densities. 
\\[2mm]
{\bf Proof of Theorem \ref{T:genSklar}}  Define $c$ as the push-forward or image-measure of $f$ under the map $(T_1,...,T_N)$, that is to say, $c=(T_1,...,T_N)_\sharp f$ or  
\begin{equation} \label{c'multidim}
  c(\bfr'_1,...,\bfr'_N) = 
  \frac{f(T_1^{-1}(\bfr'_1),...,T_N^{-1}(\bfr'_N))}{\det \nabla T_1(T_1^{-1}(\bfr'_1)) \cdots \det \nabla T_N(T_N^{-1}(\bfr'_N))}.
\end{equation}
Eliminating the determinants using \eqref{pushfwd} gives
\eqref{c'multidim2}. The representation \eqref{p'multidim} now follows by changing variables $\bfr_i'=T_i(\bfr_i)$, $d\bfr'_i=\det \nabla T_i(\bfr_i)d\bfr_i$, completing the proof.
\\[2mm]
The generalized copula density satisfies the marginal conditions 
\begin{equation} \label{margsmultidim}
 \int_{D^{N-1}}c(\bfr'_1,...,\bfr'_N)d\bfr'_1...d\bfr'_{i-1}d\bfr'_{i+1}...d\bfr'_N = f_0(\bfr'_i) \mbox{ for all }i.
\end{equation}
In particular, when the reference density $f_0$ is uniform, the marginals are uniform, in line with classical copula theory. To prove \eqref{margsmultidim}, use \eqref{c'multidim}, change variables $\bfr'_j=T_j(\bfr_j)$, $d\bfr'_j=\det \nabla T_j(\bfr_j)d\bfr_j$ ($j\neq i$), and use that $f$ has $i$th marginal $f_i$, giving that the left hand side of \eqref{margsmultidim} equals $f_i(T_i^{-1}(\bfr'_i))/\det \nabla T_i(T_i^{-1}(\bfr'_i))$, which by \eqref{pushfwd} equals $f_0(\bfr'_i)$. 

The generalized copula density $c$ in \eqref{p'multidim}--\eqref{c'multidim2} describes the interaction between the variables independently of the marginals, in the setting of densities. 

Basic intuition about copula densities is provided by the following classical examples from statistics (for $d=1$ and the reference density $f_0\equiv 1$ on $[0,1]$). For $N$ independent probability densities, $f(x_1,\ldots,x_N)=f_1(x_1)\cdot \ldots \cdot f_N(x_N)$, 
\begin{equation} \label{indep}
   C(x_1,\ldots,x_N) = x_1 \cdot \ldots \cdot x_N, \;\;\; c(x_1,\ldots,x_N) \equiv 1;
\end{equation}
and for two fully dependent probability measures, $f(x_1,x_2)=\delta(x_2-S(x_1))$ for some increasing map $S \, : \, \R\to\R$ (corresponding to fully and monotonically dependent random variables $X_2=S(X_1)$),  
\begin{equation} \label{dep}
   C(x_1,x_2) = \min\{x_1,x_2\}, \;\;\; c(x_1,x_2) = \delta(x_1-x_2).
\end{equation}
Thus independence corresponds to a constant copula density on the unit cube, whereas full and monotone dependence yields a delta function concentrated on the diagonal. Notice that the formulas for the copula in \eqref{indep}, \eqref{dep} are \textit{universal}, that is, independent of the marginals $f_i$ and the function $S$; in this sense Sklar's theorem succeeds in separating the dependence structure from the single-particle density.
As we shall see in section \ref{sec:DFT_copulas}, example \eqref{indep} is closely related to the Hartree functional, and \eqref{dep} is loosely related to the strongly correlated limit of DFT.

Of particular interest for electronic structure theory is the dependence structure of the pair density as a function of the positions $\bfr$, $\bfr'$ of an electron pair. Taking the reference density $f_0$ to be the uniform density in a bounded domain $D$ in $\R^d$, 
taking into account that the normalized pair density 
$$ 
  {N \choose 2}^{-1}\rho_2 
$$
is a probability density on $\R^d\times \R^d$ with equal marginals $N^{-1}\rho$ on $\R^d$, and applying the Generalized Sklar's Theorem gives the following two-particle generalized copula density:
\begin{equation} \label{defcopmulti}
   c(\bfr,\bfr') \, = \, \frac{2N}{N-1} \; \frac{\rho_2(T^{-1}(\bfr),T^{-1}(\bfr'))}{\rho(T^{-1}(\bfr)) \rho(T^{-1}(\bfr'))} \; \frac{1}{{\rm vol}(D)^2},
\end{equation}
where $T$ is the Brenier map transporting  $\rho/N$ to the uniform density $f_0\equiv {1}/{{\rm vol}(D)}$ on $D$. Conversely, the pair density in terms of the density and the generalized copula density is
$$
   \rho_2(\bfr,\bfr') = \frac{N-1}{2N}c\bigl(T(\bfr), T(\bfr')\bigr) \, \rho(\bfr)\,  \rho(\bfr') \cdot {\rm vol}(D)^2.
$$
Note that physical pair densities for $N$-particle systems must be $N$-representable, i.e.~they must arise from some $N$-particle quantum system via eq.~\eqref{eq:rho2}. An analogous $N$-representability condition holds for two-particle copula densities, and known explicit constraints on pair densities translate into explicit constraints on copula densities.  See Section~\ref{sec:representability}.

A main goal in the remainder of the paper is to understand the exact two-particle copula density \eqref{defcopmulti} for ground states of molecules, as well as the approximations to \eqref{defcopmulti} implied by  standard DFT approximations, in various situations of interest. For a basic impression of what exact copulas look like see the right column of Figure \ref{fig:111dissociation}.

\section{The copulas for some standard DFT models}
\label{sec:DFT_copulas}

In this paper we almost always work with \textit{two-particle generalized copula densities} $c$ (instead of $N$-particle generalized copula densities or their cumulative distribution functions $C$), and from now on refer to them for simplicity as copulas. 
Throughout this section we take the reference density to be the uniform density on the reference region $D$, $f_0\equiv {1}/{{\rm vol}(D)}$. 

\vspace{-.3cm}

\subsection{Independent electrons a.k.a. Hartree functional}

For independent electrons the pair density is given, in terms of the single-particle density $\rho$, by the mean field pair density
\begin{equation}
    \label{eq:MFrho2}
    \rho_2^{\rm MF} (\bfr,\bfr') = \tfrac12\rho(\bfr) \rho(\bfr').
\end{equation}
Whereas the mean field pair density depends on the density, the corresponding generalized copula obtained by plugging \eqref{eq:MFrho2} into  eq.~\eqref{defcopmulti} is just a  \textit{universal constant}: 
\begin{equation} \label{copMF}
   c^{\rm MF}(\bfr,\bfr') \equiv   \frac{N}{N\! -\! 1}  \frac{1}{(vol(D))^2}.
\end{equation}
Thus, just as in the classical 1D copula theory, also for generalized copulas with higher -dimensional marginals \textit{absence of correlation corresponds to a constant copula}. 
Note that, because the mean field pair density doesn't quite integrate to the correct number of pairs in the system, the mean field copula doesn't quite integrate to $1$. 

The mean field interaction energy is obtained by substituting expression \eqref{eq:MFrho2} into   \eqref{eq:vee_tot}, and is given by the Hartree energy 
\begin{equation} \label{hartree}
J[\rho]=\tfrac12 \int {\rm v}_{\rm ee}(|\bfr-\bfr'|)\rho(\bfr)\rho(\bfr') d\bfr \, d\bfr'.
\end{equation}
\vspace*{-5mm}

\subsection{Local Density Approximation} \label{sec:LDA}
We now discuss the simplest of the local density approximations, the exchange-only LDA.

{\bf Three dimensions.}~In the physical dimension $d=3$, the exchange-only LDA corresponds to the pair density\cite{Chen2015-gu}
\begin{align} 
\rho_2^{\rm LDA}(\bfr,\bfr') = &
 \tfrac12 \rho(\bfr)\rho(\bfr')  - \tfrac18 \rho(\bfr)^2 
   h\bigl((3\pi^2\rho(\bfr))^{1/3}|\bfr-\bfr'| \bigr) \nonumber \\
  & - \tfrac18 \rho(\bfr')^2 
   h\bigl((3\pi^2\rho(\bfr'))^{1/3}|\bfr-\bfr'| \bigr),  \label{rho2LDA3D}
\end{align}
with
\begin{equation}
    h(z) = \Bigl(\frac{3(\sin z - z \cos z)}{z^3}\Bigr)^2.
\end{equation}
So, from \eqref{defcopmulti}, the generalized copula is the following explicit functional of the density
$$
 c^{\rm LDA}(\bfr,\bfr') = \frac{2N}{N\! -\! 1} \;  \frac{\rho_2^{\rm LDA}\bigl(T^{-1}(\bfr),T^{-1}(\bfr')\bigr)}{\rho\bigl(T^{-1}(\bfr)\bigr) \, \rho\bigl(T^{-1}(\bfr'))},
$$
where $T$ is the Brenier map transporting $\rho/N$ to the uniform density $f_0\equiv 1/{\rm vol}(D)$ on $D$. Plugging \eqref{rho2LDA3D} into \eqref{VeeViaRho2} gives the Hartree energy plus Dirac exchange,
$$
   V_{\rm ee}^{\rm LDA}[\rho] = 
   J[\rho] - \tfrac{3}{4} 
   \bigl( \tfrac{3}{{\pi}} \bigr)^{1/3} 
   \!\!\int \rho^{4/3}.
$$

{\bf One dimension.}~To facilitate  comparison with our numerical results in section \ref{sec:cop_struct}, we also compute the (exchange-only) LDA copula in one dimension. 
For $N$ free electrons in a box $[-L,L]$ with periodic boundary conditions and homogeneous density $\rhobar=\tfrac{N}{2L}$, the exact pair density is\cite{Chen2015-gu}
$$
   \rho_2^{(N)}(x,y) = \tfrac12 \rhobar^2 - \rhobar^2 \frac{\sin^2\bigl( \tfrac{\pi}{2}\rhobar(x-y)\bigr)}{\bigl(N \sin (\tfrac{\pi}{N}\rhobar(x-y))\bigr)^2},
$$
and so its limit as $N$ goes to infinity is
\begin{equation} \label{rho2LDA}
   \rho_2^{\rm LDA}(x,y) =  \tfrac12 \rhobar^2 - \tfrac14 \rhobar^2 
   h\bigl(\tfrac{\pi}{2}\rhobar(x-y) \bigr)
\end{equation}
with
\begin{equation}
h(z)=\Bigl(\frac{\sin z}{z}\Bigr)^2.
\end{equation}
We remark that, just as in three dimensions, the pair density approaches that of statistically independent electrons (i.e. $\tfrac12\rhobar^2$) at long range, but depletes to half this value at short range, reflecting the physical fact that exchange only forbids \textit{same-spin} electrons to occupy identical positions. 

The LDA pair density for an \textit{inhomogeneous} system is given, analogously to the derivation of \eqref{rho2LDA3D}\cite{Chen2015-gu}, by replacing, in \eqref{rho2LDA}, the first term by $\tfrac12\rho(x)\rho(y)$ and the second term by its average value when substituting $\rhobar\to\rho(x)$ respectively $\rhobar\to\rho(y)$, that is to say
\begin{align} 
\rho_2^{\rm LDA}(x,y) = 
 \tfrac12 \rho(x)\rho(y) & - \tfrac18 \rho(x)^2 
   h\bigl(\tfrac{\pi}{2}\rho(x)(x-y) \bigr) \nonumber \\
  & - \tfrac18 \rho(y)^2 
   h\bigl(\tfrac{\pi}{2}\rho(y)(x-y) \bigr). \label{rho2LDAinhom}
\end{align}
Replacing the exact pair density in eq.~\eqref{defcopmulti} by \eqref{rho2LDAinhom} yields the LDA copula as an explicit functional of the density, 
$$
 c^{\rm LDA}(x,y) = \frac{2N}{N\! -\! 1} \;  \frac{\rho_2^{\rm LDA}\bigl(F^{-1}(x),F^{-1}(y)\bigr)}{\rho\bigl(F^{-1}(x)\bigr) \, \rho\bigl(F^{-1}(y)\bigr)},
$$
with $F$ defined by 
\begin{equation} \label{defCDF}
   F(x) = \int_{-\infty}^x \frac{\rho(t)}{N} dt.
\end{equation}
Figure \ref{fig:copula_fit} shows how the LDA copula changes as the interatomic distance in a two-electron system is changed from equilibrium to the dissociated state. One sees from the figure that  it does remarkably well qualitatively but also exhibits significant quantitative errors, e.g. by failing to fully deplete ionic configurations at dissociation. We also note that, just as the LDA pair density fails to give the correct single-particle density\footnote{its sometimes used nonsymmetric form obtained by changing the prefactor of the second term in \eqref{rho2LDAinhom} from  $\tfrac18$ to $\tfrac14$ and dropping the last term has the correct single-particle density over $x$ but -- by being nonsymmetric -- violates electron indistinguishability and still does not have the correct single-particle density over $y$}, the LDA copula fails to have uniform marginals. 

The LDA interaction energy is obtained by substituting expression \eqref{rho2LDAinhom} into \eqref{Vee}; by symmetry we can equivalently change the prefactor of the second term in \eqref{rho2LDAinhom} from $\frac18$ to $\tfrac14$ and drop the last term, and obtain
\begin{equation} \label{LDAen}
   V_{\rm ee}^{\rm LDA}[\rho] = J[\rho] + \int {\rm e}_x^{\rm LDA}(\rho(x)) \, dx 
\end{equation}
where $J$ is the Hartree energy \eqref{hartree} with $\bfr$ and $\bfr'$ replaced by $x$ and $y$, and the exchange energy density per unit volume is
\begin{equation} \label{ex}
  {\rm e}_x^{\rm LDA}(\rho(x)) = - \frac14 \rho(x)^2 \eta(\rho(x))
\end{equation}
with
\begin{equation} \label{ex2}
   \eta(a) = \int_{-\infty}^\infty h(\tfrac{\pi}{2}a z)\rmv_{\rm ee}(|z|)dz.
\end{equation}
In this derivation the interaction potential $\rmv_{ee}$ was arbitrary; in our numerical results in section \ref{sec:cop_struct} it will be taken to be the soft Coulomb potential \eqref{softcoul} from Wagner et al.\cite{Wagner2012-xc}. 

\subsection{Strictly correlated electrons}

\begin{figure*}
    \centering
    \subfigure[$N = 2$]{\includegraphics[width=0.2\linewidth]{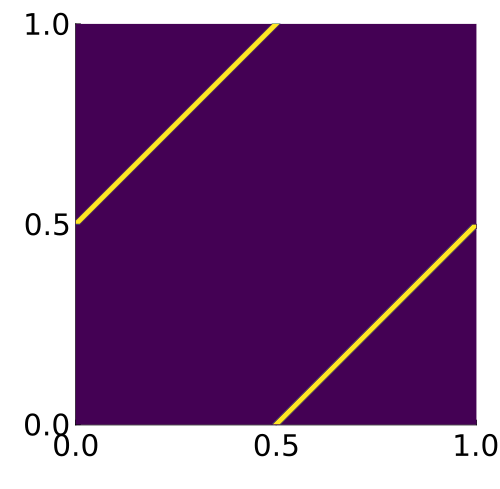}}
    \subfigure[$N = 3$]{
    \includegraphics[width=0.2\linewidth]{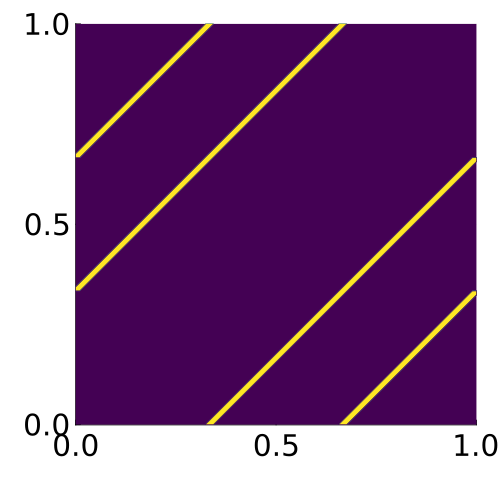}
    }
    \subfigure[$N = 4$]{
    \includegraphics[width=0.2\linewidth]{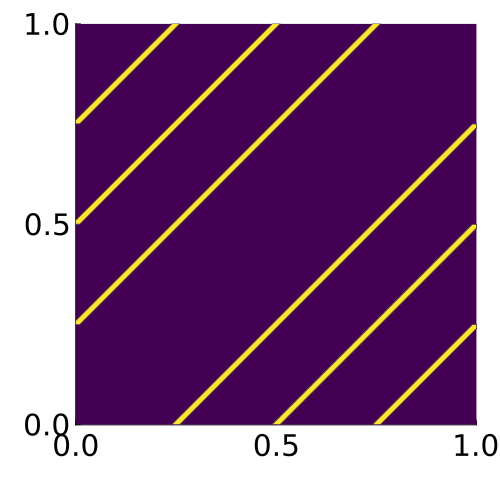}
    }
    \caption{Copula for strictly correlated electrons for two- to four-particle systems}
    \label{fig:seidl}
\end{figure*}

Strict correlations are an exteme case of strong correlations where $V_{ee}[\Psi]$ dominates over $T[\Psi]$. They arise 
for systems with lower and lower density (or, more mathematically, under the coordinate scaling $\rho \mapsto \rho_\lambda(\bfr)=\lambda^{d}\rho(\lambda\bfr)$ in the limit of $\lambda$ tending to zero). In this limit the exact Levy--Lieb functional $\min_{\Psi\mapsto\rho}(T[\Psi] + V_{ee}[\Psi])$ reduces to the strictly correlated electrons (SCE) functional
$$
   V_{\rm ee}^{\rm SCE}[\rho] = \min_{\gamma \mapsto \rho} \int_{\Omega^N} \sum_{1\le i<j \le N} {\rm v}_{\rm ee}(|\bfr_i-\bfr_j|) \gamma(\bfr_1,...,\bfr_N) d\bfr_1...d\bfr_N
$$
where the minimization is over $N$-particle densities $\gamma$ with single-particle density $\rho$.\cite{Seidl1999-ie, Friesecke2023} For one-dimensional systems the optimal $\gamma$ is known exactly\cite{Seidl1999-ie,Cotar13,Colombo2015-za}, 
\begin{equation}\label{SCENpt}
\gamma(x_1,...,x_N) = S_N \frac{\rho(x_1)}{N} \delta_{s_2(x_1)}(x_2) \cdots \delta_{s_N(x_1)}(x_N),
\end{equation}
where $S_N$ is the symmetrization operator and $s_2,..,s_N$ are the Seidl maps which give the positions of the other electrons as a function of the position of the first electron. These maps are 
characterized by the conditions that (i) they map the density $\rho$ to itself, that is to say 
\begin{equation} \label{Seidl}
   \rho(y) = \rho(s_i(y)) |s_i'(y)|,
\end{equation}
and (ii) they satisfy, when $s_1(y)<...<s_N(y)$ and with the convention $s_1(y)=y$, 
\begin{equation} \label{Seidl2}
  \int_{s_{i-1}(y)}^{s_i(y)} \rho = 1 \;\;\; \text{for all }i=2,...,N\text{ and all } y.
\end{equation}
Physically, \eqref{Seidl2} means that the electrons have an equal amount of density between them. By integrating out all but two electron coordinates it follows that
\begin{equation} \label{rho2SCE}
    \rho_2^{\rm SCE}(x,y) = \frac{\rho(x)}{2} \sum_{i=2}^N \delta_{s_i(x)}(y).
\end{equation}
It can be shown by investigating the interplay between the Seidl maps and the Brenier map (see below for a detailed proof) that the resulting copula in terms of the density is
\begin{equation} \label{copSCE1}
  c^{\rm SCE}(x,y) = \frac{1}{N-1} \sum_{i=2}^N \delta_{g_i(x)}(y)
\end{equation}
with the piecewise linear maps 
\begin{equation} \label{copSCE2}
  g_i(x) = x \, + \, \frac{i-1}{N} \!\!\!\!\!\mod{1}.
\end{equation}
Just like the mean field copula (which becomes accurate in a high-density limit) the SCE copula is \textit{universal}, that is, independent of the density $\rho$.
Figure \ref{fig:seidl} shows this copula for $N=2$, $3$, and $4$. 
We remark that the copulas for some realistic molecular systems, such as the right panel in the middle row of Figure \ref{fig:11dissociation}, show some similarity to a smeared-out and deformed version of the $N=2$ SCE copula.

To prove \eqref{copSCE1}--\eqref{copSCE2} we determine the $N$-point copula $c(y_1,...,y_N)$ from \eqref{c'} representing the $N$-point density \eqref{SCENpt}. We have, with $F$ given by \eqref{defCDF}, 
\begin{align*}
 c &= (F,...,F)_\sharp \gamma \\
   &= S_N (F,...,F)_\sharp (id,s_2,...,s_N)_\sharp \tfrac{\rho}{N} \\
   &= S_N (F,...,F)_\sharp (id,s_2,...,s_N)_\sharp F^{-1}{}_\sharp 1_{[0,1]} \\
   &= S_N(id,F\circ s_2\circ F^{-1},...,F\circ s_N \circ F^{-1})_\sharp 1_{[0,1]}.
\end{align*}
Now denote $g_i=F\circ s_i \circ F^{-1}$. We claim that the $g_i$ are the Seidl maps for the uniform density $1_{[0,1]}$. Indeed, assuming $g_{i-1}(y)<g_i(y)$ and using, in order of appearance, (i) the identity $1_{[0,1]}(y)=(F_\sharp \tfrac{\rho}{N})(y)=\tfrac{\rho}{N}(F^{-1}(y))/F'(F^{-1}(y))$, (ii) the change of variables $y=F(x)$, $dy=F'(x)dx$, and (iii) the fact that the $s_i=F^{-1}\circ g_i \circ F$ are the Seidl maps for $\tfrac{\rho}{N}$ and hence satisfy \eqref{Seidl2} gives
\begin{align*}
   \int_{g_{i-1}(y)}^{g_i(y)} 1_{[0,1]}(y)\, dy &= 
    \int_{g_{i-1}(y)}^{{g_i}(y)} (F_\sharp \tfrac{\rho}{N})(y)\, dy \\
    &= \int_{{F^{-1}(g_{i-1}(F(x)))}}^{{F^{-1}(g_i(F(x)))}} \tfrac{\rho}{N}(x)\, dx = \tfrac{1}{N}.   
\end{align*}
It follows that the $g_i$ are the Seidl maps for the uniform density $\rho_{\rm unif}=N \cdot 1_{[0,1]}$, which are -- by solving \eqref{Seidl}--\eqref{Seidl2} explicitly for the uniform density -- given by \eqref{copSCE2}. In total we have shown that $c(y_1,...,y_N) = S_N \delta_{g_2(y_1)}(y_2)\cdots \delta_{g_N(y_1)}(y_N)$. Formula \eqref{copSCE1} now follows by integrating out all but two coordinates.

\section{Dissociating systems}
\label{sec:dissociation_theory}

We now describe the correlation structure of dissociating systems in terms of generalized copulas. This is an important regime to investigate as standard density functionals struggle with it\cite{Cohen2012-xk}. We begin with
many-electron systems in one dimension,
where the theory of copulas can be applied directly. We observe the following very interesting phenomenon: when a system is dissociated into subsystems of arbitrary size, the exact copula is given by a simple, explicit, universal (i.e., density-independent) expression in terms of those of its constituents. 
We believe that an analogous explicit expression can be found for 3D systems. As a proof of concept, we derive such an expression in 3D for homonuclear dimers formed from zero angular momentum atoms (like H$_2$, Li$_2$,  Be$_2$, N$_2$), where the theoretical analysis greatly simplifies due to symmetry. General 3D systems lie beyond the scope of the present work.

\subsection{Wavefunction and pair density}
In this subsection we deal with many-electron systems in any region $\Omega$ in $d$-dimensional space $\R^d$. 

By a dissociating system we mean a family of systems with a density being a sum of two non-overlapping densities of integer mass, one of which is being translated to infinity, and which has minimum kinetic and interaction energy subject to the given density. More precisely, we assume that the density is of the form
\begin{equation} \label{dis1}
   \rho(\bfr) = \rho^A(\bfr) + \rho^B(\bfr-\bfR)
\end{equation}
where $\bfR$ is some vector in $\R^d$, 
\begin{equation} \label{dis2}
   \rho^A(\bfr) \rho^B(\bfr-R) = 0 \; \mbox{for all }\bfr,
\end{equation}
and
\begin{equation} \label{dis3}
   \int_\Omega \rho^A(\bfr)\, d\bfr = N_A, \;\;\; \int_\Omega \rho^B(\bfr) \, d\bfr = N_B,
\end{equation}
for some positive integers (physically: subsystem particle numbers) $N_A$, $N_B$; and the underlying wavefunction is a minimizer of the Levy--Lieb variational principle 
\begin{align} \label{LL1} 
   & \text{Minimize }T[\Psi]+V_{ee}[\Psi] \mbox{ over }\Psi \mbox{ in }H^1((\Omega\times\Z_2)^N)  \\
   & \text{subject to } \Psi \mbox{ antisymmetric, } \Psi\mapsto\rho, \label{LL2}
\end{align}
where $H^1$ denotes the usual space of square-integrable functions with square-integrable gradient. 
By the Hohenberg--Kohn theorem~\cite{Hohenberg1964-da}, ground state wavefunctions for any external potential $V_{ext}$ will always satisfy \eqref{LL1}--\eqref{LL2} {\it exactly}, with $\rho$ being their own density $\rho^\Psi$; and they are expected to  satisfy \eqref{dis1}--\eqref{dis3} {\it asymptotically} in the limit of large nuclear separation, e.g., in the case of diatomic molecules
$$
   V_{\rm ext}(\bfr) = - \frac{Z_A}{|\bfr-\bfR_A|} - \frac{Z_B}{|r-\bfR_B|}, \;\; Z_A, \, Z_B\in\N,
$$
in the limit where $\bfR_A$ is held fixed and 
$$
      |\bfR|=|\bfR_B-\bfR_A|\to\infty.
$$

It is well known that in the limit of large nuclear separation, the ground state wavefunction $\Psi$ of the joint system  factorizes, up to antisymmetrization, into an $N_A$-particle wavefunction for subsystem A and an $N_B$-particle wavefunction for subsystem B, 
\begin{align*} 
  \Psi\approx & \;\Psi^A \otimes_a \Psi^B,
\end{align*}
with the antisymmetric tensor product $\Psi^A\otimes_a\Psi_B$ defined as
\begin{align}\label{ATP}
  (\Psi_A\otimes \Psi_B)(\bfx_1,...,\bfx_N) = & \;\frac{1}{\sqrt{N!\,k!\,(N\! - \!  k)!}}\sum_{\sigma\in {\cal S}_N}
   \epsilon(\sigma) \cdot \nonumber  \\
   &\hspace*{-20mm} \Psi^A(\bfx_{\sigma(1)},\ldots,\bfx_{\sigma(k)})
   \Psi^B(\bfx_{\sigma(k+1)},\ldots,\bfx_{\sigma(N)}).
\end{align}
Here ${\cal S}_N$ denotes the group of permutations of the $N$ electron coordinates, and the $\bfx_i$ are space-spin coordinates. 
The prefactor guarantees that, when the densities $\rho^A$ and $\rho^B$ of $\Psi^A$ and $\Psi^B$ have disjoint support, normalization is preserved, that is to say if $||\Psi^A||=||\Psi^B||=1$ then $||\Psi^A\otimes_a \Psi^B||=1$. A more rigorous discussion of this asymptotic factorization in the context of Hohenberg--Kohn DFT would be desirable but lies beyond our scope. 

Understanding the behaviour of correlations then reduces to understanding the correlation structure of antisymmetrized tensor products of disjointly supported fewer-particle wavefunctions. The pair density of such states is described by an interesting exact expression involving only the subsystem densities and subsystem pair densities.
\begin{theorem}[Pair density of dissociated systems] \label{thm:L:paird} 
Let $N_A$, $N_B$ be positive integers, let $\Psi^A$ be an antisymmetric normalized $N_A$-particle wavefunction in $L^2_a((\Omega\times\Z_2)^{N_A})$ and $\Psi^B$ an antisymmetric normalized $N_B$-particle wavefunction in $L^2_a((\Omega\times\Z_2)^{N_B})$, and assume that the densities $\rho^A$ and $\rho^B$ of these wavefunctions have disjoint support. Then the density $\rho$ and the pair density $\rho_2$ of the antisymmetric tensor product $\Psi=\Psi^A\otimes_a\Psi^B$ are given by the following formulae:
\begin{eqnarray}
   \rho & = & \rho^A + \rho^B, \label{L:d} \\
   \rho_2 & = & \rho_2^A + \rho_2^B + \tfrac{1}{2}\bigl( \rho^A\otimes\rho^B + \rho^B\otimes\rho^A \bigr), \label{L:paird}
\end{eqnarray}
where $\rho_2^A$ and $\rho_2^B$ are the pair densities of $\Psi^A$ and $\Psi^B$. Here we use the convention that the pair density of a single-particle wavefunction is zero. 
\end{theorem}
Because $\rho^A$ and $\rho^B$ have disjoint support (denoted respectively $\Omega_A$ and $\Omega_B$), the four terms appearing in \eqref{L:paird} are supported in mutually disjoint spatial regions and our  formula for the pair density $\rho_2=\rho_2(\bfr,\bfr')$ can be summarized in Table~\ref{T:1}.

\begin{table}[http!]
\begin{center}
\begin{tabular}{c|c|c|}
    &   $\bfr'\in\Omega_A$ & 
    $\bfr'\in\Omega_B$ \\[2mm]
    \hline 
$\bfr\in\Omega_A$
&  
$\rho_2^A$
 & 
 $\frac12 \rho^A\otimes\rho^B$\\[2mm]
  \hline 
$\bfr\in\Omega_B$ & 
$\frac12 \rho^B\otimes\rho^A$
 & $\rho_2^B$  \\[2mm]
\hline
\end{tabular}
\caption{Pair density of an antisymmetric tensor product whose tensor factors have densities supported in disjoint regions $\Omega_A$ and $\Omega_B$.}
\label{T:1}
\end{center}
\end{table}

These formulas have the important physical meaning that
\begin{itemize}
    \item correlations between electrons in the same subsystem are unaffected by the presence of the other subsystem
    \item correlations between electrons in different subsystems are of mean field type, agreeing with those produced by the Hartree term $\rho_2^{\rm Hartree}=\frac12 \rho\otimes\rho = \frac12(\rho_A\otimes\rho_A + \rho_A\otimes\rho_B+\rho_B\otimes\rho_A+\rho_B\otimes\rho_B)$.  
\end{itemize}

While the resulting expressions are rather simple, deriving them requires some careful bookkeeping, 
due to the sum over coordinate permutations present in \eqref{ATP}, which 
-- when plugged into the formula \eqref{eq:rho2} for the pair density -- becomes a double sum and
leads to many possibilities of the numerous position coordinates belonging to the regions $\Omega_A$ or $\Omega_B$. The main technical reason leading to the final result is that when the densities $\rho^A$ and $\rho^B$ of the tensor factors $\Psi^A$ and $\Psi^B$ have disjoint support, we have
\begin{align}  
  &\Psi^A(\bfr_1,s_1\ldots,\bfr_k,s_k)\Big|_{\bfr_i=\bfa} \!\Psi^B(\bfr'_1,s'_1,\ldots,\bfr'_{N-k},s_{N-k})\Big|_{\bfr_j' \!\!\!= \bfa} = 0 \nonumber \\
  & \hspace*{5.5cm}\mbox{ for all } i,j,\bfa. \label{TO}
\end{align}  
This implies that many of the $N!^2$ terms in the double sum vanish. The details are provided in Appendix A.

\subsection{Copula for 1D systems} 

One of our main results for one-dimensional systems is the following.
\begin{theorem}[Copula of dissociated systems] \label{T:cop}
Let $\Omega_A$, $\Omega_B$ be disjoint one-dimensional intervals of the form $\Omega_A=(-\infty,l]$, $\Omega_B=(l,\infty)$ for some $l\in\R$. Let $N_A$, $N_B$ be positive integers, let $\Psi^A$ be an antisymmetric normalized $N_A$-particle wavefunction in $L^2_a((\Omega\times\Z_2)^{N_A})$ and $\Psi^B$ an antisymmetric normalized $N_B$-particle wavefunction in  $L^2_a((\Omega\times\Z_2)^{N_B})$, and assume that the densities $\rho^A$ and $\rho^B$ of these wavefunctions are supported in $\Omega_A$ respectively $\Omega_B$. Then the copula $c=c(x,y)$ of the antisymmetric tensor product $\Psi=\Psi^A\otimes_a\Psi^B$ is given in terms of the copulas $c^A$ and $c^B$ of $\Psi^A$ and $\Psi^B$ as indicated in the following table:
\begin{small}
\begin{center}
\begin{tabular}{c|c|c|}
    &  $y<\frac{\; N_A}{N}$ & $y \ge \frac{\; N_A}{N}$ 
    \\[2mm]
    \hline 
    $x < \frac{\; N_A}{N}$
 & $\frac{N_A-1}{N_A}\frac{N}{N-1} \, c^A\Bigl(\tfrac{N}{\; N_A}(x,y)\Bigr)$   &  $\frac{N}{N-1}$  
 \\[2mm]
   \hline
   $x\ge \frac{\; N_A}{N}$
 & $\frac{N}{N-1}$ & $\frac{N_B-1}{N_B}\frac{N}{N-1} \, c^B\Bigl(\tfrac{N}{\; N_B}(x-\tfrac{\;\; N_A}{N}, \, y - \tfrac{\;\; N_A}{N})\Bigr)$ \\[2mm]
   \hline
\end{tabular}
\end{center}
\end{small}

\end{theorem}

Here we have used the convention that the copula of a single-particle wavefunction is zero. 

Thus in the region $x,y<\frac{\; N_A}{N}$, the copula equals that of subsystem $A$, re-scaled to a domain of size $\frac{\; N_A}{N} \times \frac{\; N_A}{N}$; in the opposite region $x,y\ge \frac{\; N_A}{N}$, it equals that of subsystem $B$, translated by $\frac{\; N_A}{N}$ in both the $x$ and $y$ direction and re-scaled to a domain of size $\frac{\; N_B}{N}\times \frac{\; N_B}{N}$.
The key advantage of working with copulas (as compared to pair densities as in Theorem \ref{thm:L:paird}) is that the dependence on the density has now disappeared and we have a {\it universal} expression for the copula of the joint system in terms of the subsystem copulas.

Also, we find it quite interesting that the value of the joint copula in the `off-diagonal' region is constant and independent of the size of the subsystems. This constant equals 2 when $N=2$, 1.5 when $N=3$ and $N_A=2$, $N_B=1$, and 4/3 when $N=4$ and either $N_A=3$, $N_B=1$ or $N_A=N_B=2$, 
in agreement with our numerical simulations below.
\\[2mm]
{\bf Proof of Theorem \ref{T:cop}} This follows by computing the CDF of $\rho/N$ (eq.~\eqref{defCDF}) and its generalized inverse and applying \eqref{defcopmulti} and~\eqref{L:paird}. Since ${\rm supp}\, \rho_A \subseteq(-\infty,l)$,  ${\rm supp}\, \rho_B\subseteq [l,\infty)$, and $\rho=\rho_A+\rho_B$, we have, denoting the CDFs of $\rho_A/N_A$ and $\rho_B/N_B$ by $F_A$ respectively $F_B$, 
$$
   F(x) = \int_{-\infty}^x \frac{\rho^A(t)+\rho^B(t)}{N} \, dt = 
   \begin{cases} \frac{\; N_A}{N} F_A(x), & x<l \\
                 \frac{\; N_A}{N} + \frac{\; N_B}{N} \, F_B(x), & x \ge l \end{cases}
$$
and consequently 
\begin{equation} \label{Finv}
    F^{-1}(y) = \begin{cases} 
    F_A^{-1}\Bigl( \tfrac{N}{N_A} y\Bigr), & y < \frac{\; N_A}{N} \\
    F_B^{-1}\Bigl( \tfrac{N}{N_B}(y - \tfrac{\; N_B}{N})\Bigr), & y\ge \frac{\; N_A}{N}. 
    \end{cases}
\end{equation}
To compute the copula $c(x,y)$ of $\Psi=\Psi^A\otimes_a\Psi^B$ we distinguish four cases, according to whether $x$ and $y$ are $<\tfrac{\; N_A}{N}$ or $\ge \tfrac{\; N_A}{N}$. 

First, suppose $x,y\ge \tfrac{\; N_A}{N}$. By \eqref{defcopmulti}, Table \ref{T:1}, and \eqref{Finv}, 
\begin{align*}
 c(x,y) &= \frac{{N\choose 2}^{-1} \rho_2^A\Bigl( F_A^{-1}(\frac{N}{N_A} x), \, F_A^{-1}(\frac{N}{N_A}y)\Bigr)}
 {\frac{\rho^A}{N}\Bigl( F_A^{-1}(\tfrac{N}{N_A}x)\Bigr) \; 
 \frac{\rho^A}{N}\Bigl( F_A^{-1}(\tfrac{N}{N_A}y)\Bigr)} 
 \\
 & =  \frac{{N_A\choose 2} {N\choose 2}^{-1}}{\Bigl(\frac{N_A}{N}\Bigr)^2} \; c^A\Bigl(\tfrac{N}{N_A}(x,y)\Bigr).
\end{align*}
The prefactor in front of $c_A$ equals $\frac{N_A-1}{N_A}\frac{N}{N-1}$, yielding the asserted expression for the copula. 

In the region $x,y\ge \tfrac{\; N_A}{N}$, the second alternative in \eqref{Finv} becomes active and we obtain analogously 
\begin{equation*}
 c(x,y) =  \frac{{N_B-1}}{N_B} \, \frac{N}{N-1} \;  c^B\Bigl(\tfrac{N}{N_B}(x-\tfrac{\; N_A}{N},y-\tfrac{\; N_A}{N})\Bigr).
\end{equation*}
Next, suppose $x<\frac{\; N_A}{N}$, $y\ge \frac{\; N_A}{N}$. In this case Table \ref{T:1} gives
\begin{align*}
      c(x,y) &=  \frac{{N\choose 2}^{-1} \tfrac{1}{2} \, \rho^A\Bigl( F_A^{-1}(\frac{N}{N_A} x)\Bigr) \,  \rho^B\Bigl( F_B^{-1}(\frac{N}{N_B}(y-\frac{\; N_A}{N}))\Bigr)}
 {\frac{\rho^A}{N}\Bigl( F_A^{-1}(\tfrac{N}{N_A}x)\Bigr) \; 
 \frac{\rho^B}{N}\Bigl(F_B^{-1}(\frac{N}{N_B}(y-\frac{\; N_A}{N}))\Bigr)} 
 \\
 & = \frac{N}{N-1}.
\end{align*}
Analogously, in the final case $x\ge\frac{\; N_A}{N}$, $y< \frac{\; N_A}{N}$ we also obtain $c(x,y)=\tfrac{N}{N-1}$. This completes the proof.  

\subsection{Generalized copula for a class of 3D systems}

We now determine the generalized copula in the dissociated limit for diatomic molecules composed of identical zero-angular-momentum atoms.
As a reference density we choose the uniform density on any bounded three-dimensional reference region $D$ which is symmetric with respect to reflection at a plane perpendicular to the molecular axis.

\begin{theorem}[Generalized copula of dissociated systems] \label{T:cop_3D}
Let $\Omega_A$, $\Omega_B$ be disjoint balls in $\R^3$ of the form $\Omega_A=B(-\mathbf{R}, R )$, $\Omega_B=B(\mathbf{R}, R)$ denoting $\mathbf{R} = (R,0,0)$ for some $R\in\R_+$. Let $\Omega=\Omega_A\cup\Omega_B$, let $N$ be an even positive integer, let $\Psi^A$ be an antisymmetric normalized $N/2$-particle wavefunction in $L^2_a((\Omega\times\Z_2)^{3N})$ and $\Psi^B$ the antisymmetric normalized $N/2$-particle wavefunction in  $L^2_a((\Omega\times\Z_2)^{3N})$ defined as the image of $\Psi^A$ by reflection of the spatial variables $\bfr_i=(x_i,y_i,z_i)$ with respect to the plane $x_i=0$, and assume that the density $\rho^A$ of 
 $\Psi^A$ is supported in $\Omega_A$, so that the density $\rho^B$ of $\Psi^B$ is supported in $\Omega_B$. Then the generalized copula $c=c(\bfrtilde,\bfrtilde')$ of the antisymmetric tensor product $\Psi=\Psi^A\otimes_a\Psi^B$ is given as indicated in the following table, where we denote $\bfrtilde=(u,v,w)$ and $\bfrtilde'=(u',v',w')$: 

\begin{small}
\begin{center}
\begin{tabular}{c|c|c|}
    &  $u'< 0$ & $u' \ge 0$ 
    \\[2mm]
    \hline 
    $u < 0$
 & $\frac{N-2}{N-1}\; c^A\Bigl(
 \varphi_A^{-1}(\bfrtilde),\varphi_A^{-1}(\bfrtilde')
 \Bigr)$   &  $\frac{N}{N-1}{\rm vol}(D)^{-2}$  
 \\[2mm]
   \hline
   $u\ge 0$
 & $\frac{N}{N-1}{\rm vol}(D)^{-2}$   & $\frac{N-2}{N-1}\; c^B\Bigl(
 \varphi_B^{-1}(\bfrtilde),\varphi_B^{-1}(\bfrtilde')
 \Bigr)$ \\[2mm]
   \hline
\end{tabular}
\end{center}
\end{small}
\noindent
where $c^A$ and $c^B$ are the generalized copulas of $\Psi^A$, respectively, $\Psi^B$, and $\varphi_A$ and $\varphi_B$ are invertible mappings defined in the proof.
\end{theorem}
\vspace*{4mm}

\begin{figure}[h!]
    \centering
    \includegraphics[width=0.25\textwidth]{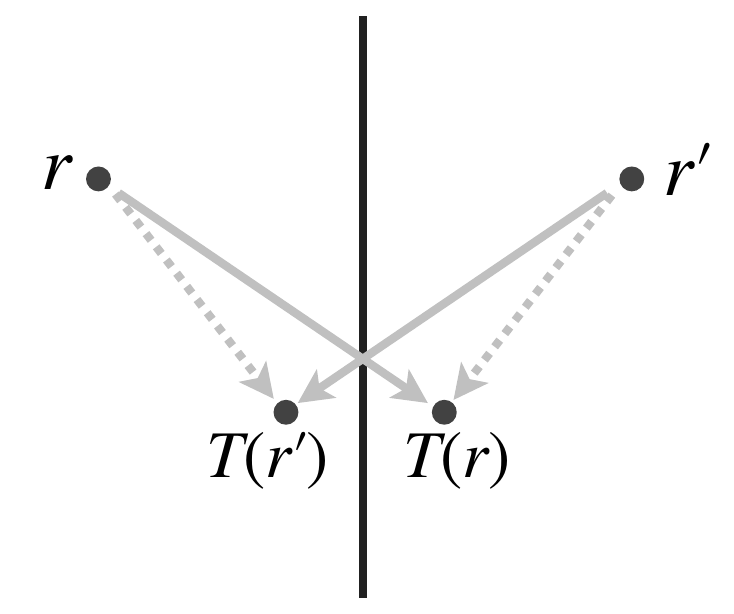}
    \vspace*{-2mm}
    \caption{Hypothetical behavior of the Brenier map for two symmetric points with respect to the reflection plane.}
    \label{fig:reflection}
\end{figure}

{\bf Proof of Theorem \ref{T:cop_3D}}
The proof combines a symmetry argument with a variational argument using the monotonicity of the Brenier map. 
Starting point is the expression~\eqref{c'multidim2} for the generalized copula. The main point is to show that the Brenier map $T$ from $\rho_A+\rho_B$ to the uniform density on the reference region $D$  maps the physical region $\Omega_A$  to the left half of $D$,  
$D^-=\{(u,v,w) \, | \, u < 0 \} \cap D$, and $\Omega_B$ to the right half $D^+$. To prove this, let $\bfr=(x,y,z)$ be any point in  $\Omega_A$ to the left of the plane $x=0$, i.e. $x<0$, with $\rho_A(\bfr)>0$. Denote its reflection at the plane $x=0$ by $\bfr'=(-x,y,z)$, and its image under $T$ by $T(\bfr)=(u,v,w)$. 
By a symmetry property of Brenier maps\cite{Dusson2023-vn},  $T(\bfr')=(-u,v,w)$. See Figure \ref{fig:reflection}.

 Now suppose that, contrary to our claim, $\bfr$ is mapped to the right half of $B(0,1)$, that is, $u>0$, as in Fig.~\ref{fig:reflection}. Then
\begin{align*} |\bfr-T(\bfr)|^2 &=
   |x-u|^2 \;\;\;\;\;\;\, + |(y,z)-(v,w)|^2 \\
   &> |x-(-u)|^2 + |(y,z)-(v,w)|^2 = |\bfr-T(\bfr')|^2, 
\end{align*}
which means that the solid arrow emanating from $\bfr$ in the figure is longer than the dashed arrow.  
Analogously $|\bfr'-T(\bfr')|^2 > |\bfr'-T(\bfr)|^2$. Adding these two inequalities, expanding the squares, cancelling terms, and rearranging gives
$$
    0 > (T(\bfr)-T(\bfr'))\cdot (\bfr-\bfr'),
$$
contradicting the monotonicity of the Brenier map $T$. 

Using~\eqref{L:paird}, we conclude that the copula is constant on the subdomains $D^+\times D^-$ and $D^-\times D^+$. Next, let us derive the expression in the above table on $D^+\times D^+$. 
Let $T_A$ be the Brenier map from $\rho_A$ to the uniform measure on $D$.
We have that for $\bfr\in\Omega_A$, $T(\bfr) = \varphi_A (T_A(\bfr))$ for some invertible map $\varphi_A:D\rightarrow D^-$. (Since $T$ depends on the joint density $\rho$, and $T_A$ depends on the subsystem density $\rho_A$, this map depends on $\rho$ and $\rho_A$, but not on any pair density.) Then for $\bfrtilde\in D^-$,
\[
    T^{-1}(\bfrtilde) = T_A^{-1}(\varphi_A^{-1}(\bfrtilde)).
\]
Now using~\eqref{defcopmulti}, for $\bfrtilde,\bfrtilde'\in D^-$,
\begin{align*}
 c(\bfrtilde,\bfrtilde') &= \frac{{N\choose 2}^{-1} \rho_2^A\Bigl( T_A^{-1}(\varphi_A^{-1} (\bfrtilde)), \, T_A^{-1}(\varphi_A^{-1} (\bfrtilde'))\Bigr)}
 {\frac{\rho^A}{N}\Bigl( T_A^{-1}(\varphi_A^{-1} (\bfrtilde))\Bigr) \; 
 \frac{\rho^A}{N}\Bigl( T_A^{-1}(\varphi_A^{-1} (\bfrtilde'))\Bigr)} 
 \\
 & = \frac{N-2}{N-1}\; c^A\Bigl(
 \varphi_A^{-1}(\bfrtilde),\varphi_A^{-1}(\bfrtilde')
 \Bigr).
\end{align*}
The copula on the domain $D^+\times D^+$ is obtained similarly, completing the proof.

\section{Representability constraints on the copula} \label{sec:representability}
$N$-representability for pair densities translates as follows into an analogous constraint on the associated two-particle generalized copula densities \eqref{defcopmulti}. 
\begin{theorem} \label{thm:rep} Let $f_0$ be the uniform reference probability density on any bounded reference region $D\subset\R^d$. The following are equivalent: \\[1mm]
{\rm (1)} $c$ is the two-particle generalized copula density \eqref{defcopmulti} of some $N$-representable pair density $\rho_2$. 
\\[1mm]
{\rm (2)} $c$ is the pair density of some $N$-particle wavefunction on $D$ with uniform single-particle density (i.e., of an $N$-particle homogeneous electron gas in $D$). 
\end{theorem}
{\bf Proof} We prove $(1)\Longrightarrow(2)$, the proof of the other direction being analogous. If $\rho_2$ is an $N$-representable pair density on $\Omega\times\Omega\subseteq\R^d\times\R^d$, then by definition there exists an $N$-electron wavefunction $\Psi$ on $\Omega^N$ such that $\rho_2=\rho_2^\Psi$ where $\rho_2^\Psi$ is given by eq.~\eqref{eq:rho2}. Now let $\rho$ be the density of $\Psi$ and let $T$ be the Brenier map which pushes the normalized density $\rho/N$ of $\Psi$ forward to the reference density $f_0$, and introduce the following $N$-electron wavefunction $\Phi$ on $D^N$: 
\begin{align*}
   \Phi(\bfr'_1,s_1,...,\bfr'_N,s_N) = &\Psi\bigl(T^{-1}(\bfr'_1),s_1,...,T^{-1}(\bfr'_N),s_N\bigr) \\
   &\cdot \prod_{i=1}^N \Bigl(\frac{1}{\rho(T^{-1}(\bfr'_i))/N}\Bigr)^{1/2} \cdot \frac{1}{{\rm vol}(D)^{N/2}}.
\end{align*}
Squaring, summing over spins, and integrating over $\bfr'_3,...,\bfr'_N$ (using the change-of-variables formula) gives that 
$$
   \rho_2^{\Phi}(\bfr'_1,\bfr'_2) =  \, \frac{2N}{N-1} \; \frac{\rho_2^\Psi(T^{-1}(\bfr'_1),T^{-1}(\bfr'_2))}{\rho(T^{-1}(\bfr'_1)) \rho(T^{-1}(\bfr'_2))} \cdot \frac{1}{{\rm vol}(D)^2}.
$$
Hence the pair density $\rho_2^\Phi$ of $\Phi$ is precisely the two-particle generalized copula density of $\rho_2$ defined in eq.~\eqref{defcopmulti}. 

As a consequence, known necessary representability constraints on the pair density yield corresponding constraints on the copula. 
\\[2mm]
{\bf Example} (obtained by combining Theorem \ref{thm:rep} with Corollary IV.1 in Friesecke et al.\cite{Friesecke2013-tp}) For $f_0$ and $D$ as above, and any  partitioning of $D$ into two disjoint regions $\Omega_A$, $\Omega_B$, the probabilities $c_{XY}=\int_{\Omega_X\times \Omega_Y}\! c$ satisfy the following: \\[1mm]
{\rm (a)} If $c$ is $3$-representable, then 
\[
 c_{AB}+c_{BA} \le 2(c_{AA}+c_{BB}).
\]
{\rm (b)} If $c$ is $N$-representable for all $N$, then 
\[
c_{AB}+c_{BA} \le 2 (c_{AA}\cdot c_{BB})^{1/2}.
\]
An analogous condition to (a) for arbitrary $N$ (which  approaches (b) when $N$ gets large) is also known\cite{Friesecke2013-tp}. Interestingly, in the dissociation limit of three-particle systems into a two-particle and a one-particle system (see Theorem~\ref{T:cop} and Figure~\ref{fig:21dissociation}), the constraint (a) is exactly saturated for the regions associated with the subsystems, and hence of physical relevance.

Analogously to the use of certain exact constraints in the design of exchange-correlation functionals\cite{scan}, such representability constraints could be systematically incorporated into future machine-learned copulas.

\section{Numerically computed copulas of quantum-mechanical ground states}
\label{sec:cop_struct}

In this section we numerically compute the exact copulas \eqref{defcopmulti} corresponding to ground states 
of the fundamental quantum mechanical many-electron energy \eqref{eq:energy}. Since the accurate computation of Brenier maps in 3D is somewhat involved, we confine ourselves to 1D many-electron systems, leaving 3D simulations to a forthcoming paper.
Our goal is to observe the copulas for different numbers of electrons and different positions of the atomic nuclei. We are particularly interested in how the copula changes when the nuclei are pulled apart.

For systems with two to four electrons, we plot the chosen external potential, the electronic density~\eqref{eq:rho}, the pair density~\eqref{eq:rho2}, the mean field pair density defined by~\eqref{eq:MFrho2}
where $\rho$ is the electronic density, 
and the copula~\eqref{c'}. The code for reproducing the plots can be found at \url{https://github.com/dussong/copula_pair_density}.
In all the simulations, the interactions between the electrons is based on the soft Coulomb potential proposed by Wagner et al.\cite{Wagner2012-xc}, that is
\begin{equation}\label{softcoul}
   \rmv_{\rm ee}(x) = \frac{1}{\sqrt{1+x^2}}.
\end{equation}
For a nucleus with a charge $z$ and position $R$, the nucleus-electron interaction takes the form
\begin{equation} \label{eq:Vext0}
   V_{z,R}(x) = -z V_{\rm ee}(x-R).
\end{equation}
Then for a molecule having $N$ nuclei with charges  $(z_1,\ldots,z_N)$ and positions $(R_1,\ldots,R_N)$, the external potential is defined as
\begin{equation}
    \label{eq:Vext} 
    V_{\rm ext}(x) = \sum_{i=1}^N V_{z_i,R_i}(x).
\end{equation}
We discretize the eigenvalue equation~\eqref{eq:HpsiEpsi} using piecewise linear finite elements. We therefore start by computing the ground-state wavefunction of the system by solving the Schr\"odinger equation~\eqref{eq:HpsiEpsi}, and then compute derived quantities such as the density and pair density.

\paragraph{Two-particle system.}

\begin{figure*}[!t]
    \centering
    \subfigure[External potential $V_{\rm ext}$ defined in~\eqref{eq:Vext} and electronic density~\eqref{eq:rho}]{
    \begin{tabular}{@{}c@{}}
         \includegraphics[width=0.23\textwidth]{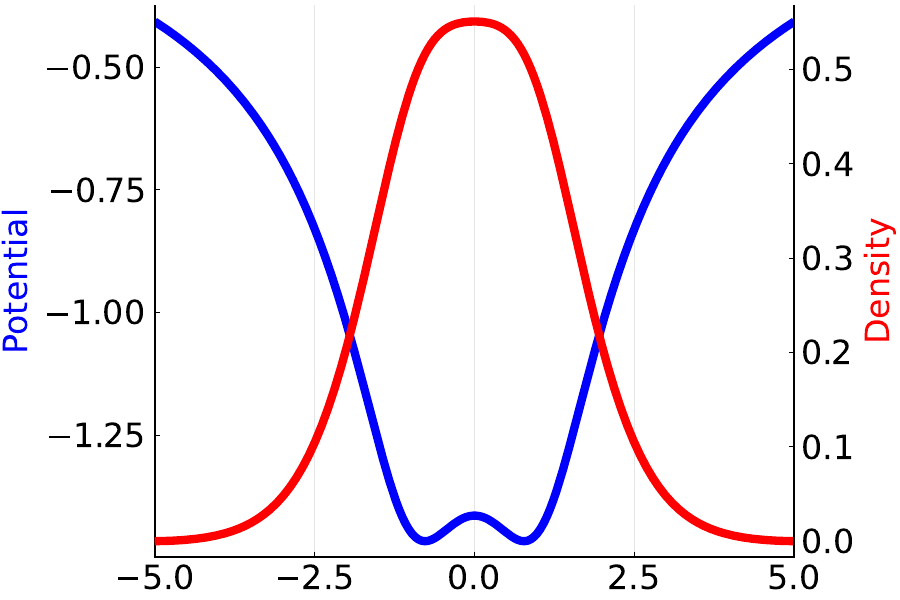} \\
         \includegraphics[width=0.23\textwidth]{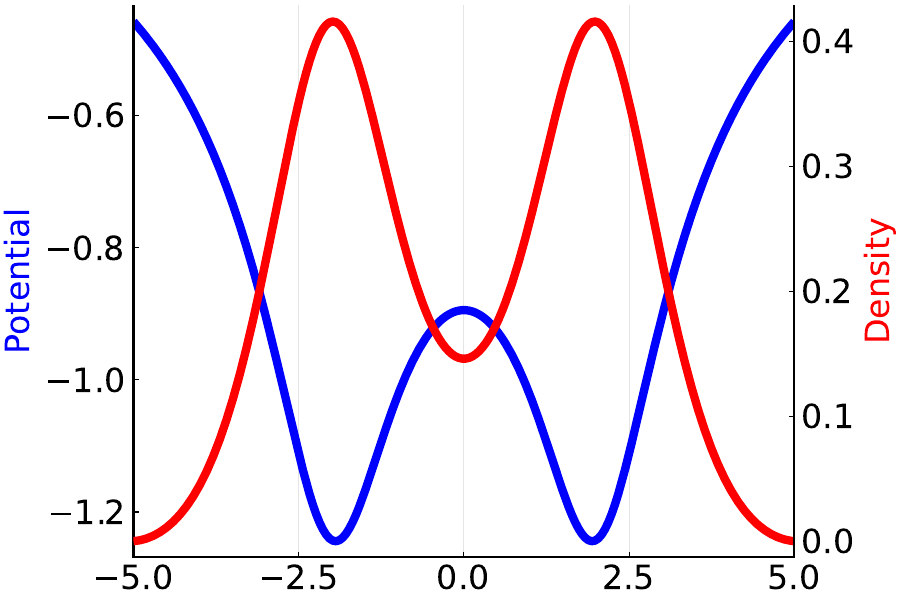} \\
         \includegraphics[width=0.23\textwidth]{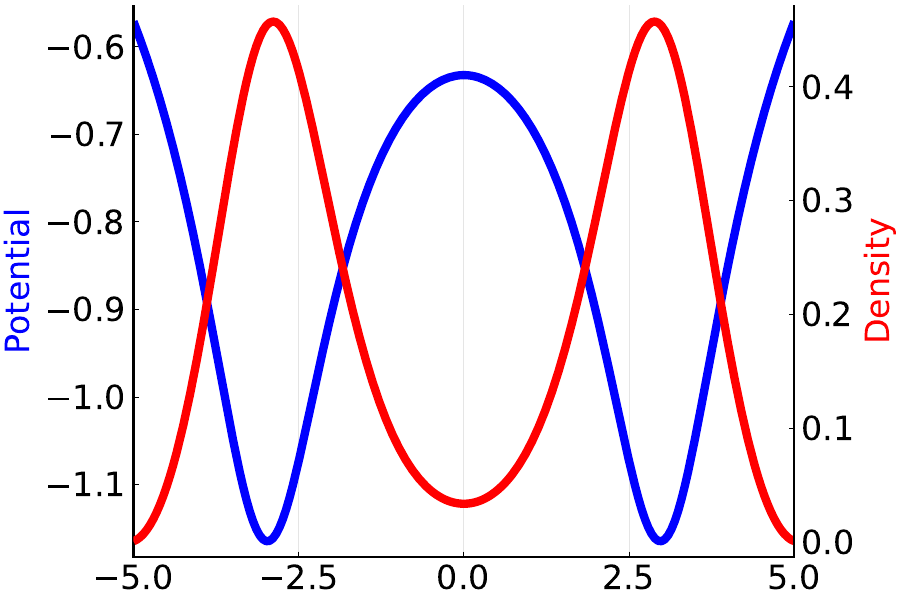}
    \end{tabular}
    }
    \subfigure[Pair density~\eqref{eq:rho2}]{
    \begin{tabular}{@{}c@{}}
    \includegraphics[width=0.23\textwidth]{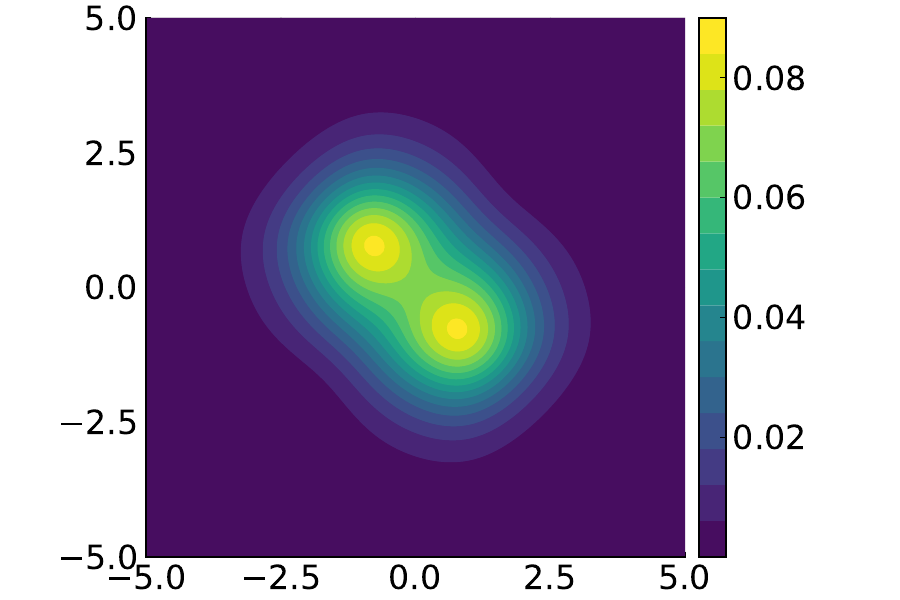} \\
    \includegraphics[width=0.23\textwidth]{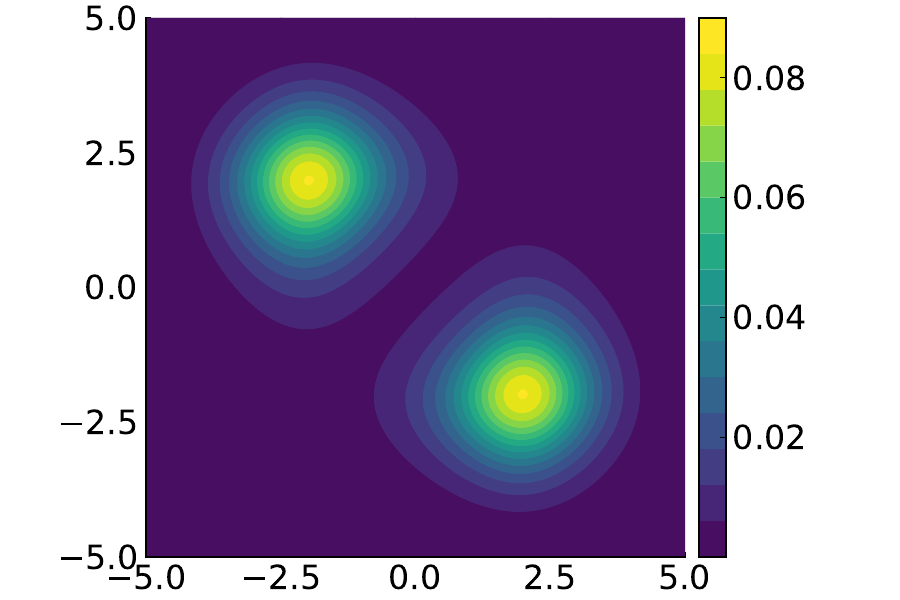} \\
        \includegraphics[width=0.23\textwidth]{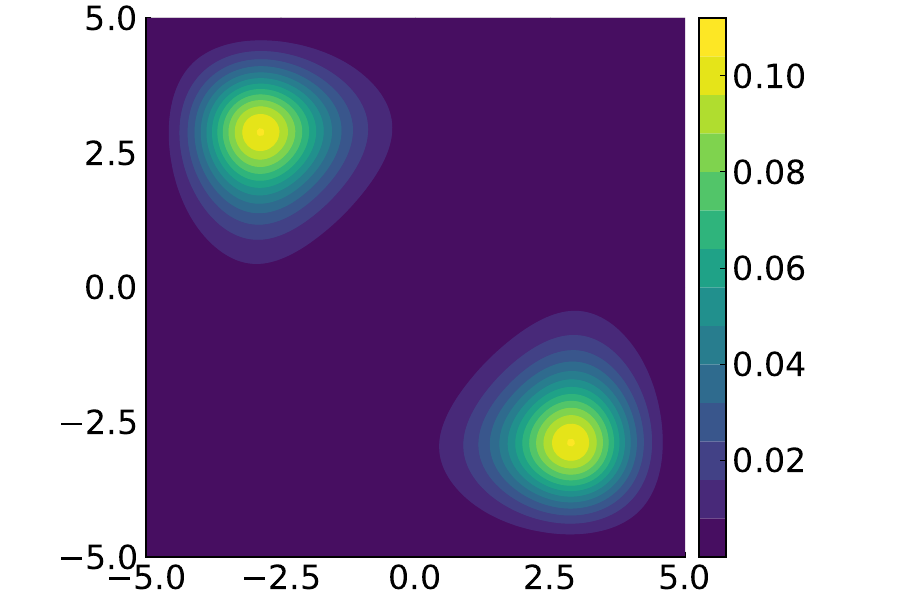} \\
    \end{tabular}
    }
    \subfigure[Mean field pair density~\eqref{eq:MFrho2}]{
    \begin{tabular}{@{}c@{}}
        \includegraphics[width=0.23\textwidth]{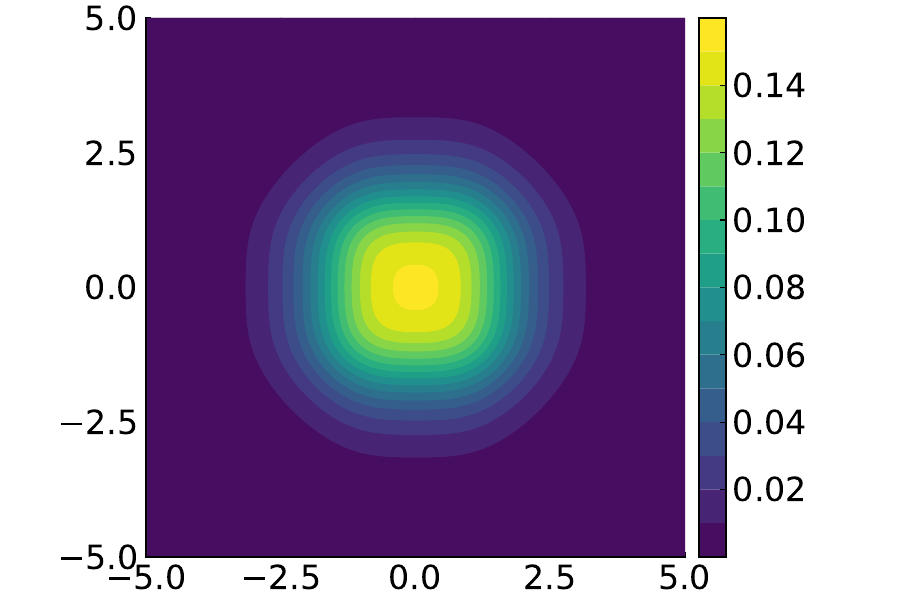} \\
        \includegraphics[width=0.23\textwidth]{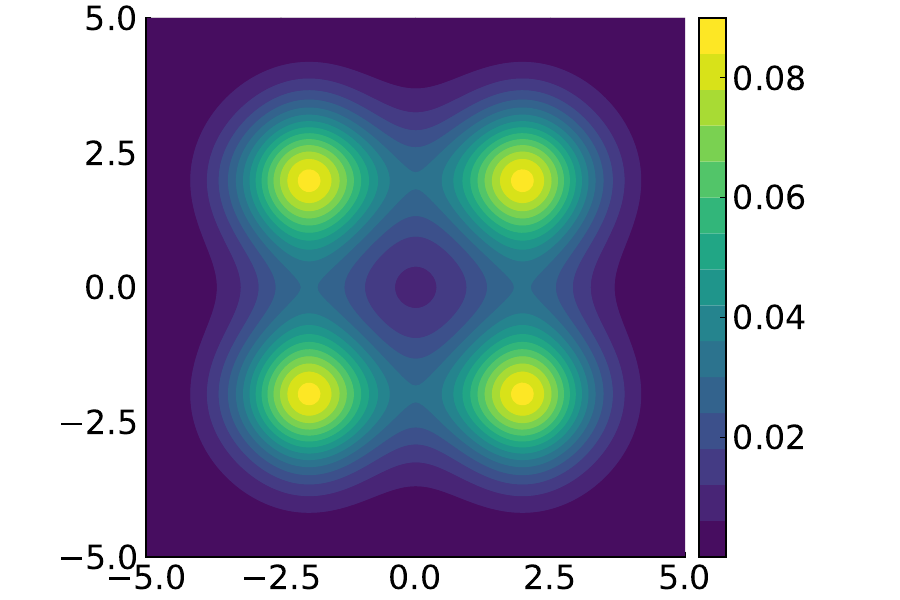} \\
        \includegraphics[width=0.23\textwidth]{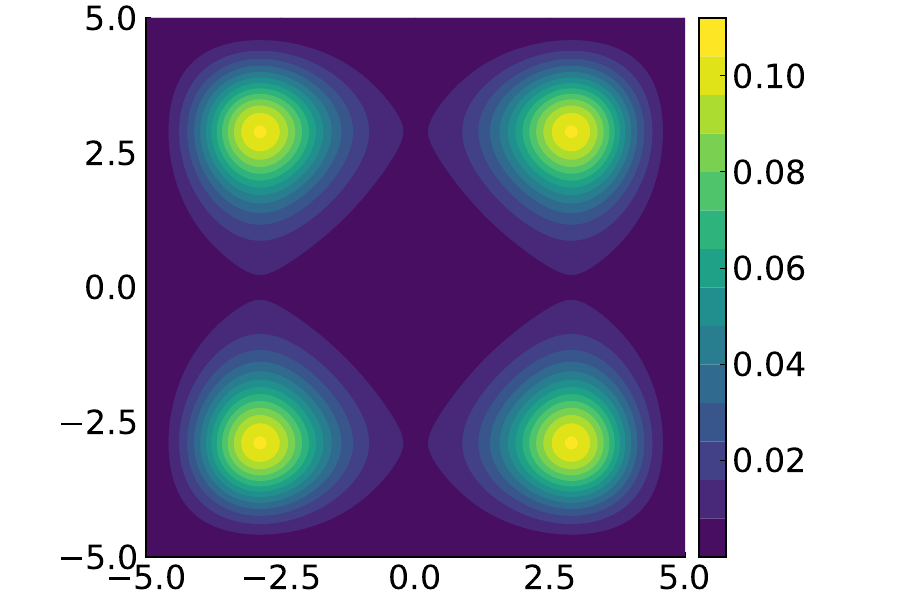} 
    \end{tabular}
    }
    \subfigure[Copula~\eqref{defcopmulti}]{
    \begin{tabular}{@{}c@{}}
        \includegraphics[width=0.23\textwidth]{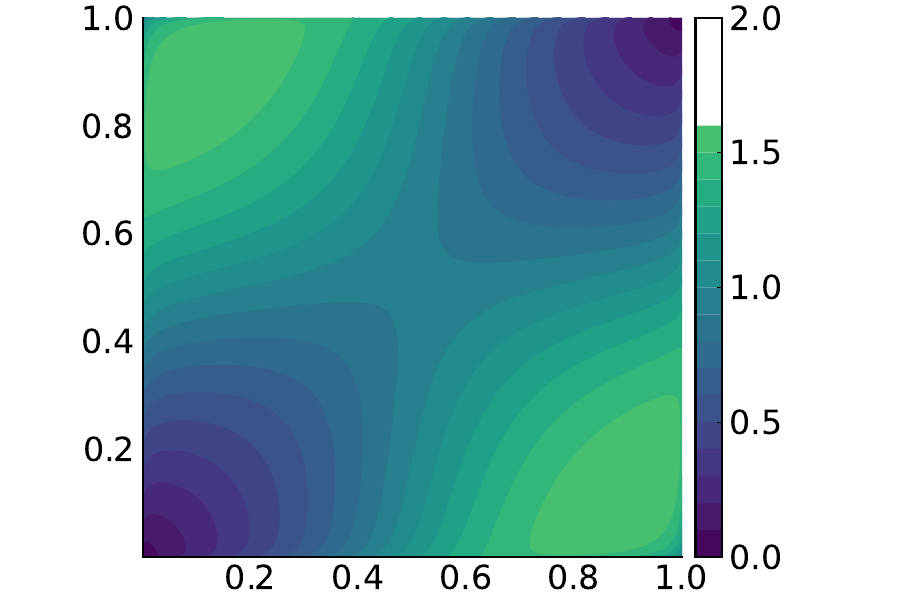} \\
        \includegraphics[width=0.23\textwidth]{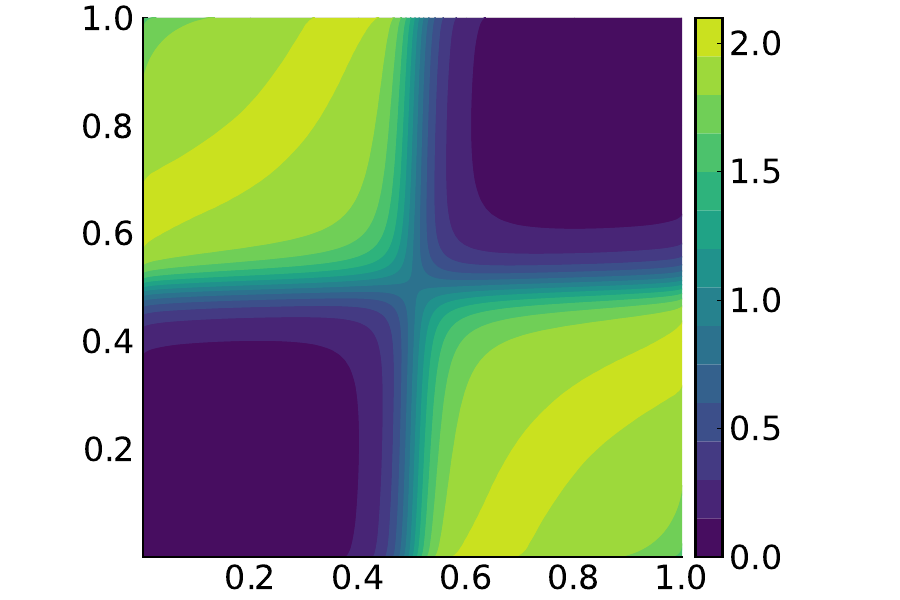} \\
        \includegraphics[width=0.23\textwidth]{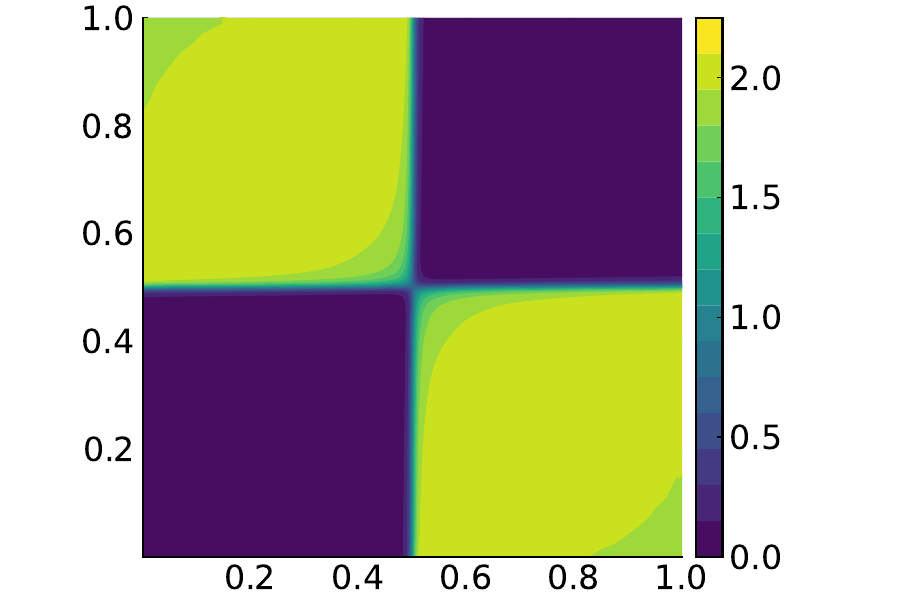} 
    \end{tabular}
    }
    \caption{Right column: exact copula of ground state for a  two-particle system dissociating into two one-electron densities.  The other columns show related quantities.
    Nuclei positions for top row: $(-1,1)$, second row: $(-2,2)$, bottom row: $(-3,3)$. 
    }
    \label{fig:11dissociation}
\end{figure*}

We first consider two nuclei  with equal charges $z_1=z_2=1$ at  positions $(R_1,R_2) = (-a,a)$ for some $a\ge 0$. We simulate the system on the interval $[-5,5]$ and with a $150$-point  mesh in each dimension.
The results for the different parameter values $a=1, 2, 3,$ are shown in  Figure~\ref{fig:11dissociation}. 
The left column represents the external potential for the different values of $a$, together with the corresponding ground state densities. 
As expected from the functional form \eqref{eq:Vext0}--\eqref{eq:Vext}, the potential shows two wells that are separating when the interatomic distance grows, while the electronic density changes from a one-peak density to a two-peaks density, which is already almost dissociated for $a=3$.
The second column represents the pair density. For a two-particle system, the pair density is just the modulus square of the wavefunction summed over the four possible spin combinations.
From left to right, we observe one peak splitting into two peaks that are symmetric with respect to the axis $y=x$. 
Moreover the position of the two peaks indicates anti-correlation between the electrons. When an electron is situated in one of the peaks, the other has to be in the other peak, since the probability of finding the two electrons at the same value is zero. This can be clearly understood by looking at the mean field electronic density, which is shown on the third column. Finally, in the last column we plot the copula. We observe that when the system dissociates, the copula takes the form of a checkerboard with four squares, two going to the value of 2, two going to zero. This is exactly what is predicted by Theorem~\ref{T:cop}.

\paragraph{Three-particle systems.}

\begin{figure*}[t!]
    \centering
    \subfigure[External potential $V_{\rm ext}$ defined in~\eqref{eq:Vext} and electronic density~\eqref{eq:rho}]{
    \begin{tabular}{@{}c@{}}
         \includegraphics[width=0.23\textwidth]{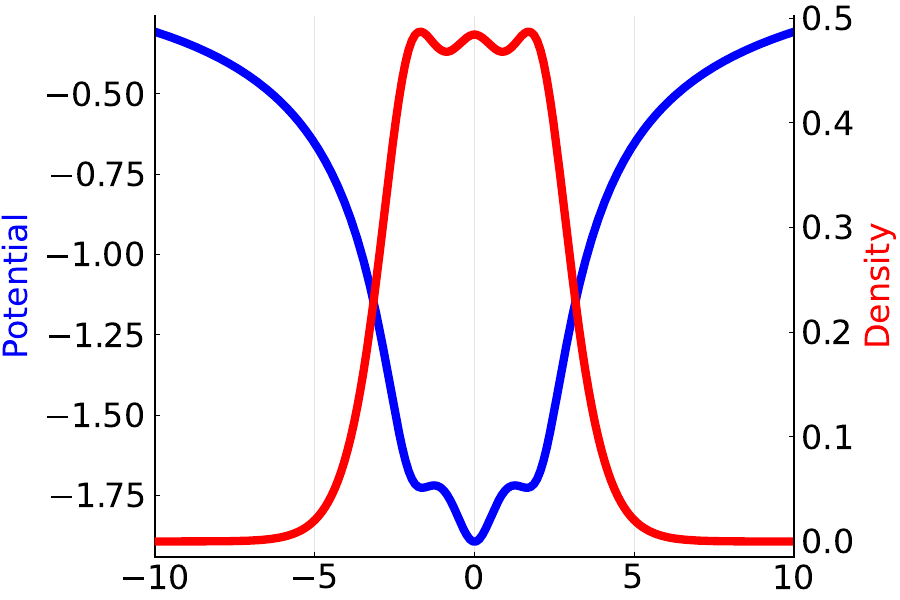} \\
         \includegraphics[width=0.23\textwidth]{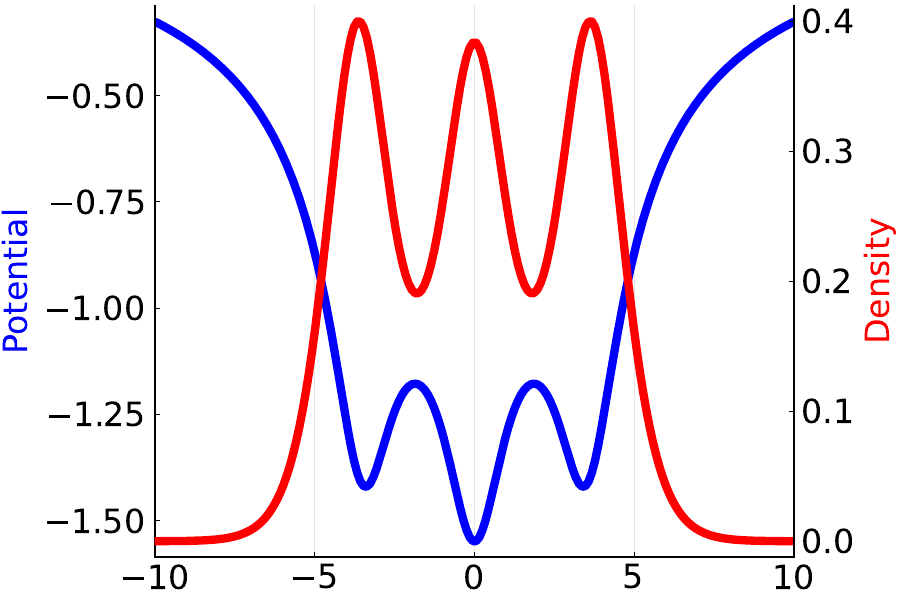} \\
         \includegraphics[width=0.23\textwidth]{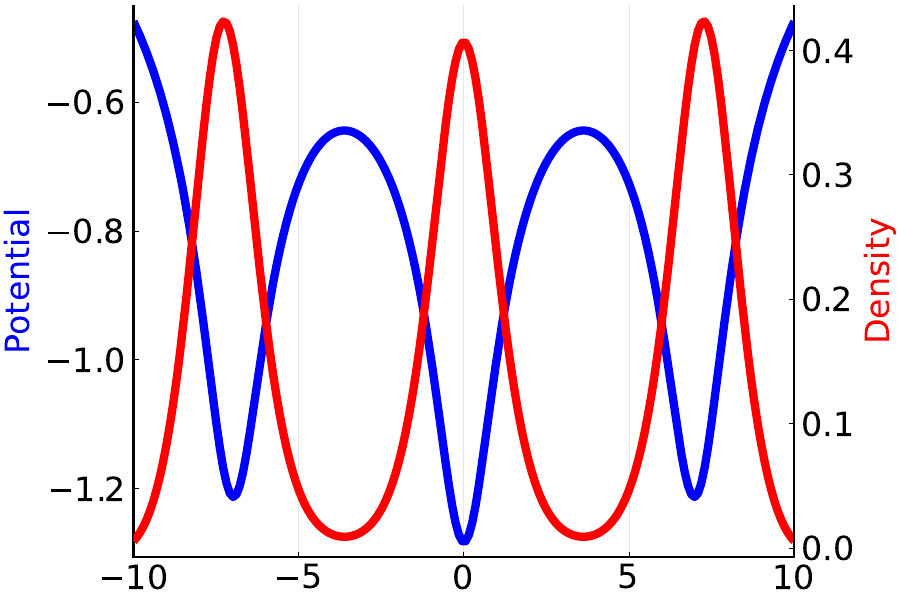}
    \end{tabular}
    }
    \subfigure[Pair density~\eqref{eq:rho2}]{
    \begin{tabular}{@{}c@{}}
    \includegraphics[width=0.23\textwidth]{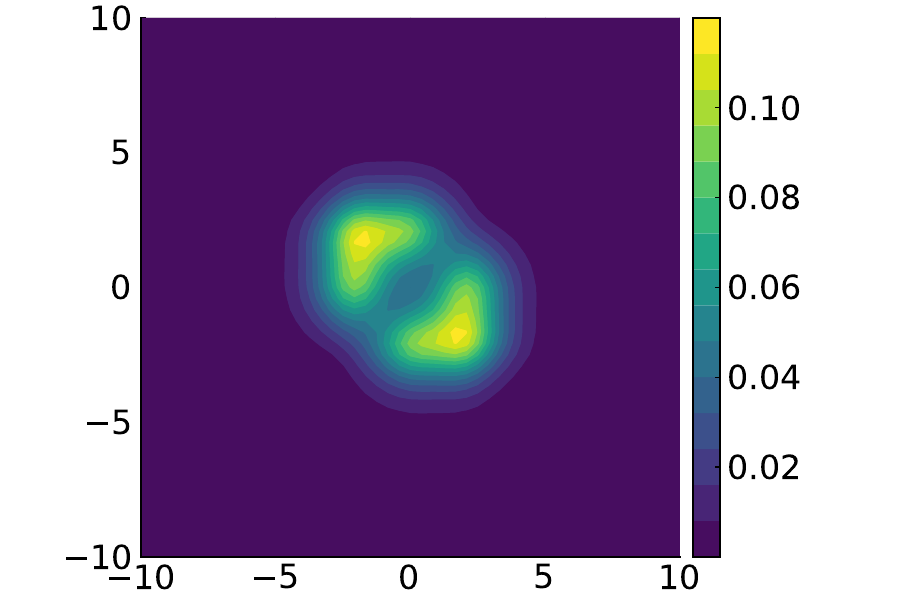} \\
    \includegraphics[width=0.23\textwidth]{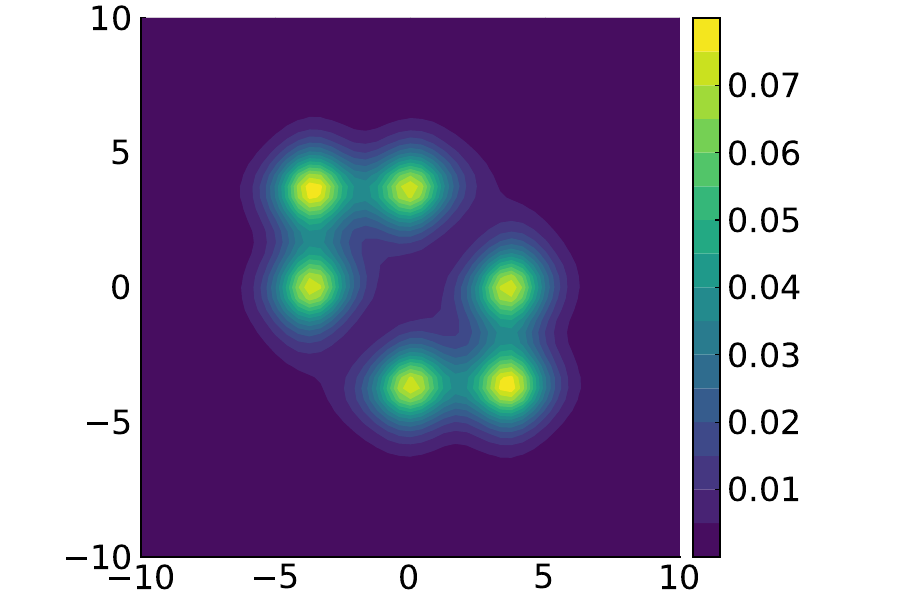} \\
        \includegraphics[width=0.23\textwidth]{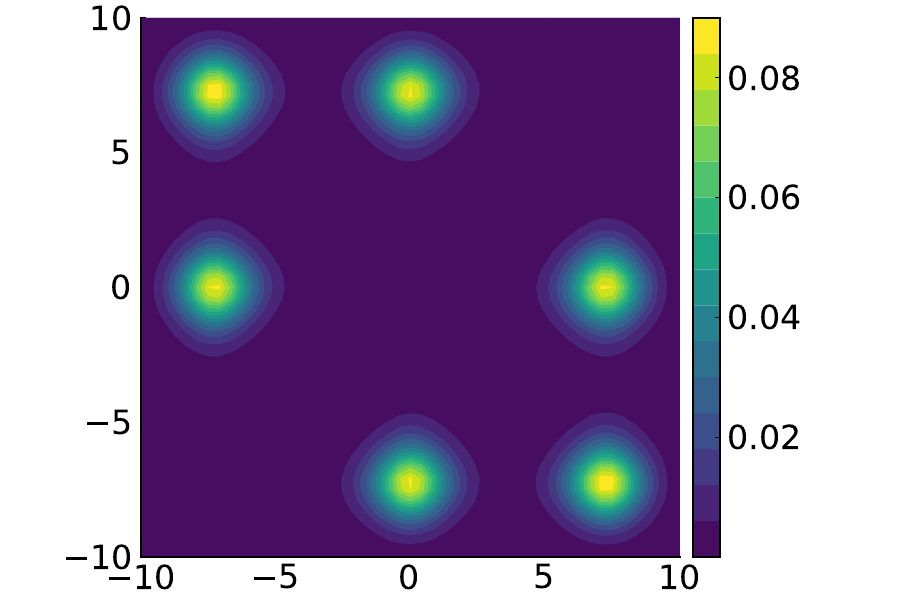} \\
    \end{tabular}
    }
    \subfigure[Mean field pair density~\eqref{eq:MFrho2}]{
    \begin{tabular}{@{}c@{}}
        \includegraphics[width=0.23\textwidth]{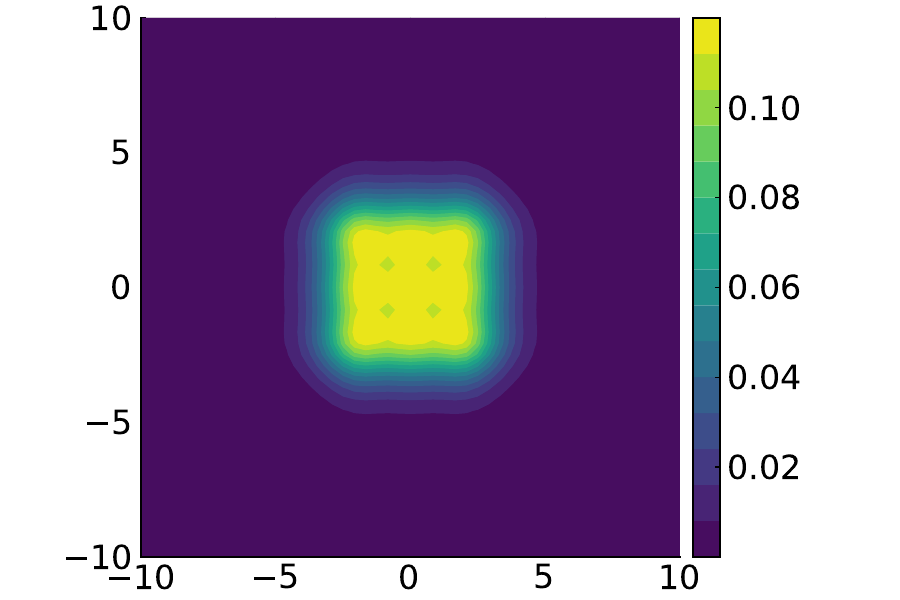} \\
        \includegraphics[width=0.23\textwidth]{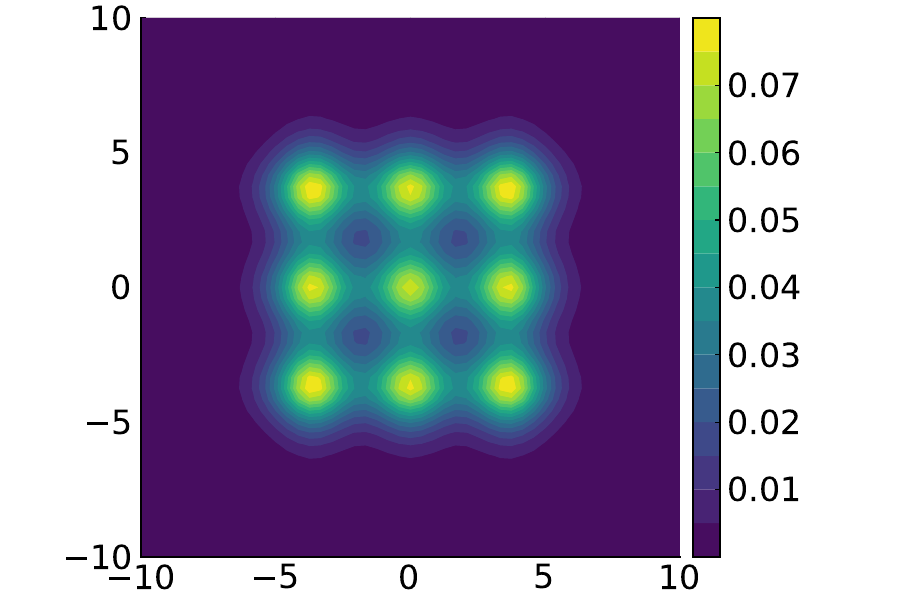} \\
        \includegraphics[width=0.23\textwidth]{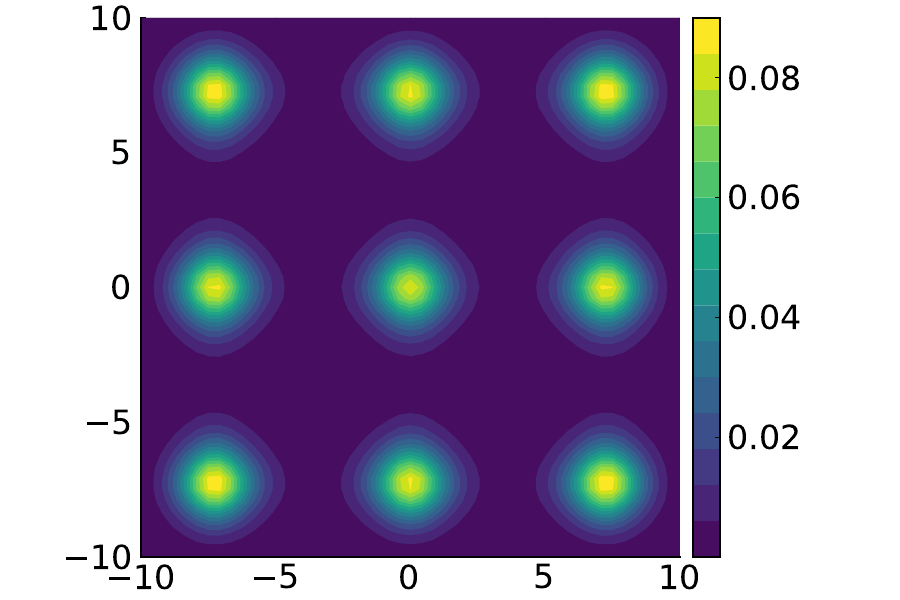} 
    \end{tabular}
    }
    \subfigure[Copula~\eqref{defcopmulti}]{
    \begin{tabular}{@{}c@{}}
        \includegraphics[width=0.23\textwidth]{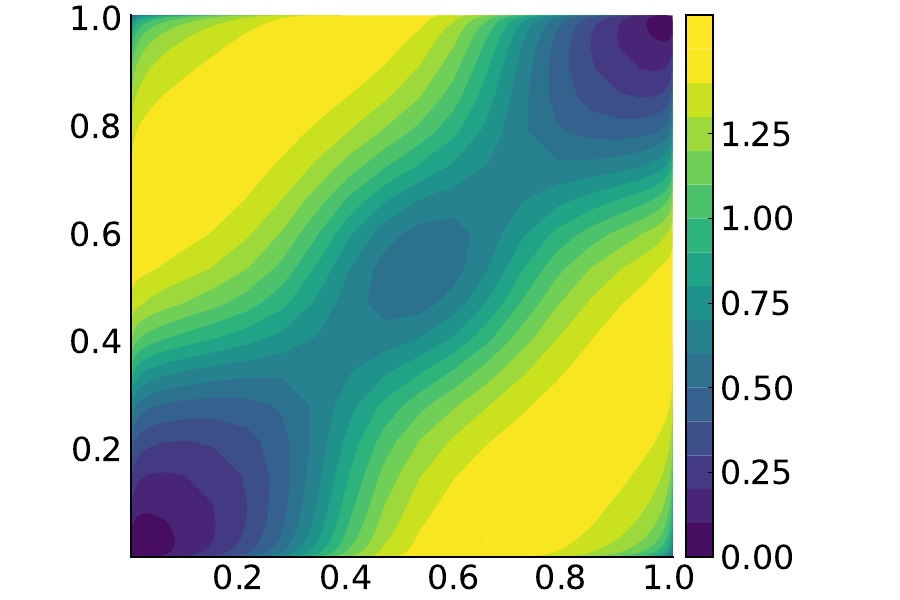} \\
        \includegraphics[width=0.23\textwidth]{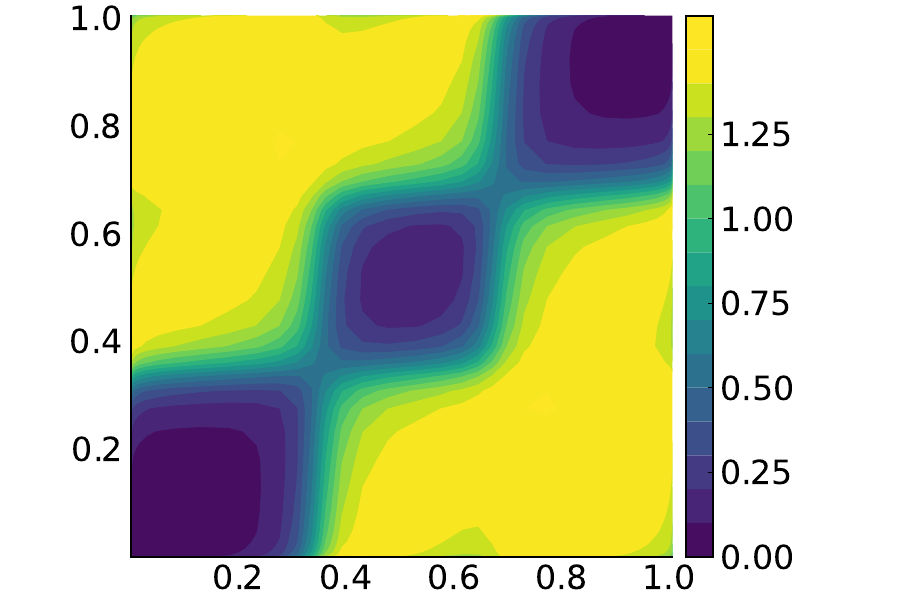} \\
        \includegraphics[width=0.23\textwidth]{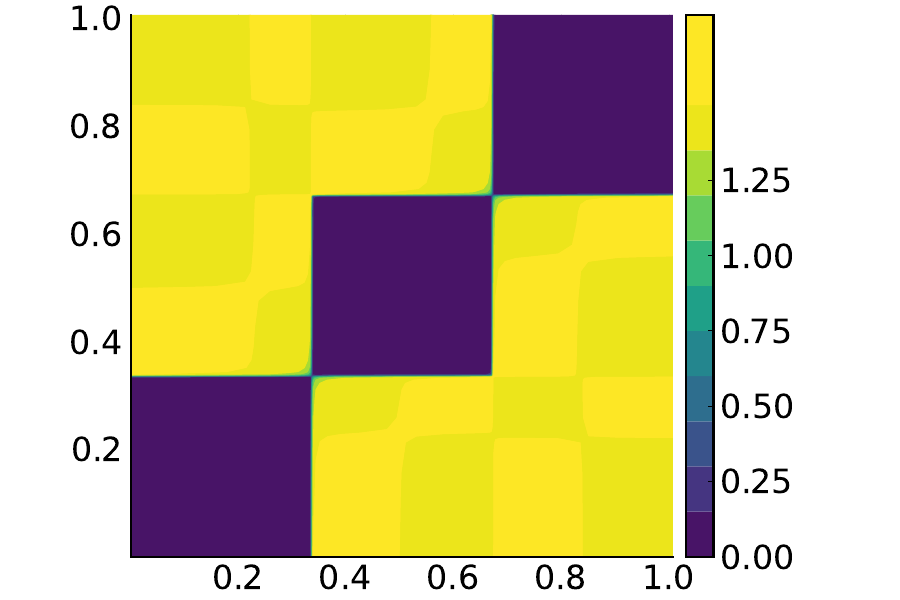} 
    \end{tabular}
    }
    \caption{
    Right column: exact copula of ground state for a  three-particle system dissociating into three one-electron densities.  
    The other columns show related quantities.
    Nuclei positions for top row: $(-2, 0, 2)$, second row: $(-3.5, 0, 3.5)$, bottom row: $(-7, 0, 7)$. 
    }
    \label{fig:111dissociation}
\end{figure*}

We now consider three charges $z_1=z_2=z_3=1$ for the nuclei with positions $(R_1,R_2,R_3)$. We consider a spatial domain $[-10,10]$, and discretize it with $50$ points. We compare here different dissociation paths. We start by observing in Figure~\ref{fig:111dissociation} the dissociation of the system into three separate systems with charge 1 each, mimicking the dissociation H$_3 \to $H+H+H. To do so, we choose for the positions of the nuclei $(-a,0,a)$ with $a = 2,3.5$ and $7$.
The potential then shows three wells separating when $a$ increases.
The electronic density also separates into three peaks with one-electron charge each. 
And as in the case of two particles, the copula structure is in accordance with the theory, that is when the particles dissociate, the copula converges to a checkerboard, here with nine squares, six of which are 3/2 and three of which are zero.

\begin{figure*}[t!]
    \centering
    \subfigure[External potential $V_{\rm ext}$ defined in~\eqref{eq:Vext} and electronic density~\eqref{eq:rho}]{
    \begin{tabular}{@{}c@{}}
         \includegraphics[width=0.23\textwidth]{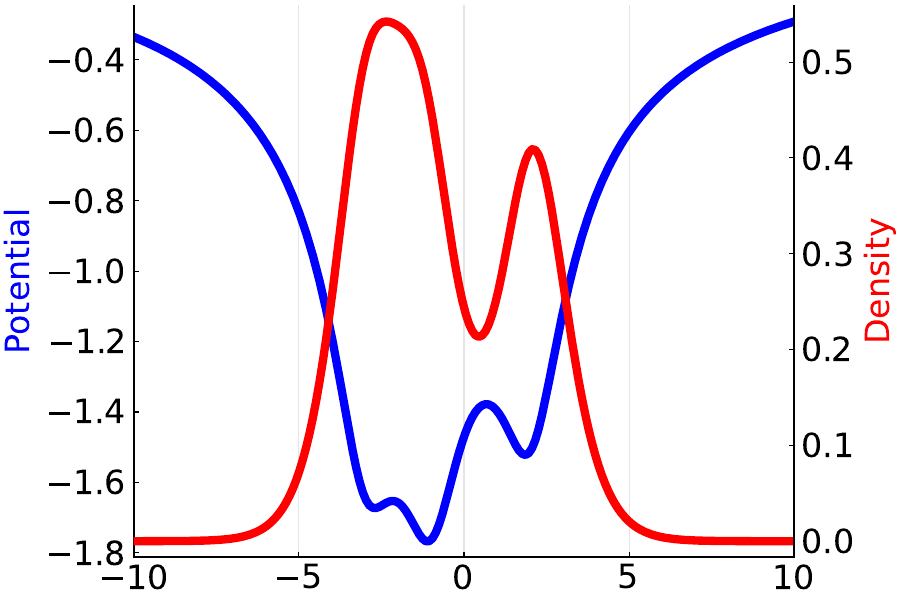} \\
         \includegraphics[width=0.23\textwidth]{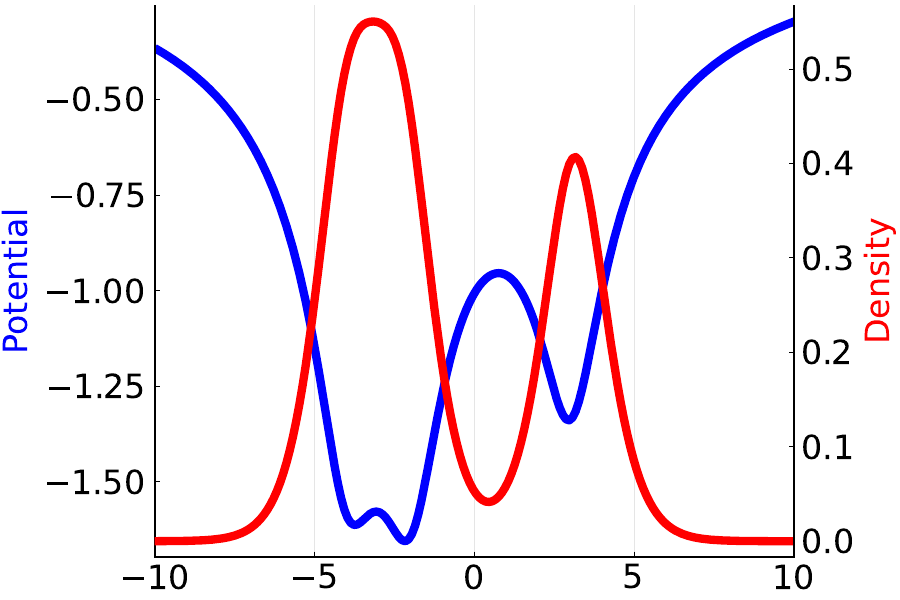} \\
         \includegraphics[width=0.23\textwidth]{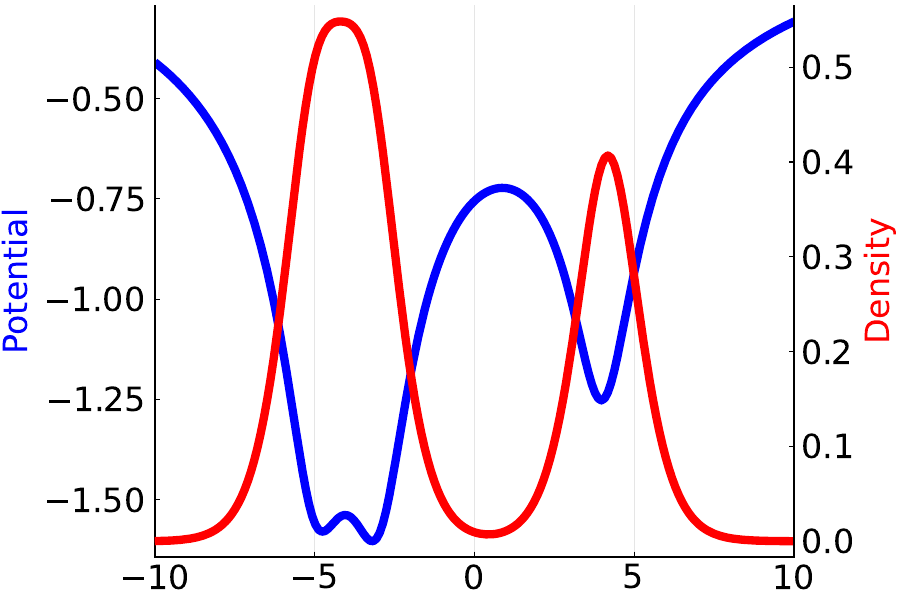}
    \end{tabular}
    }
    \subfigure[Pair density~\eqref{eq:rho2}]{
    \begin{tabular}{@{}c@{}}
    \includegraphics[width=0.23\textwidth]{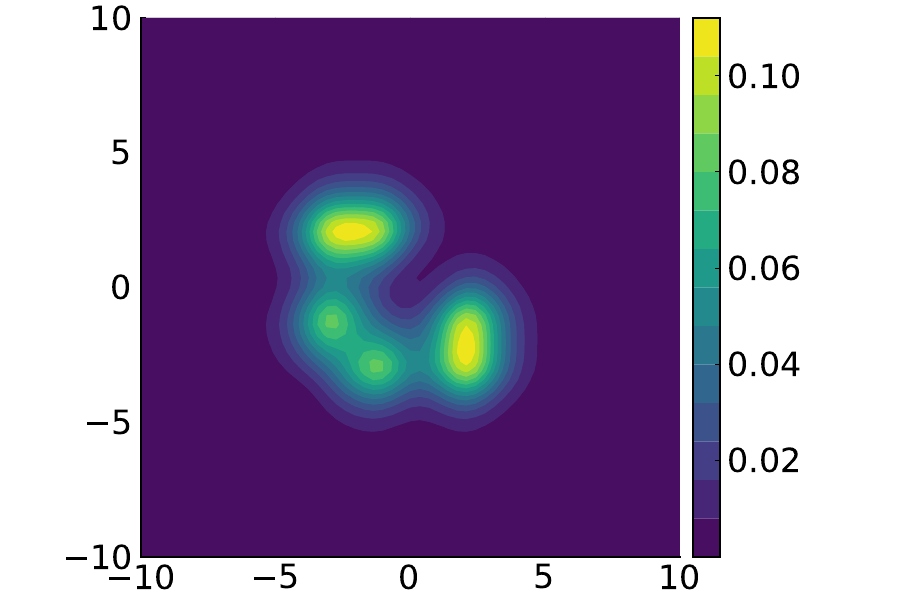} \\
    \includegraphics[width=0.23\textwidth]{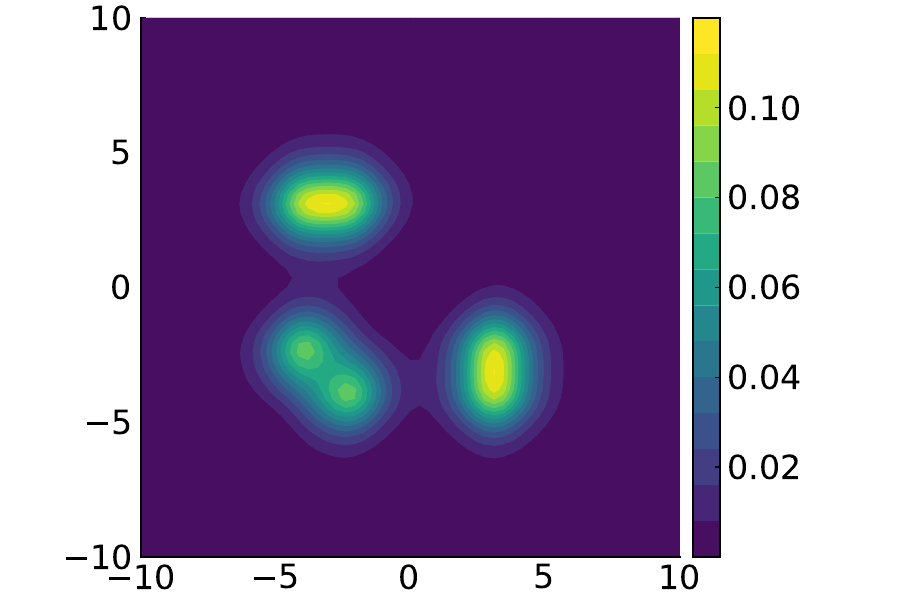} \\
        \includegraphics[width=0.23\textwidth]{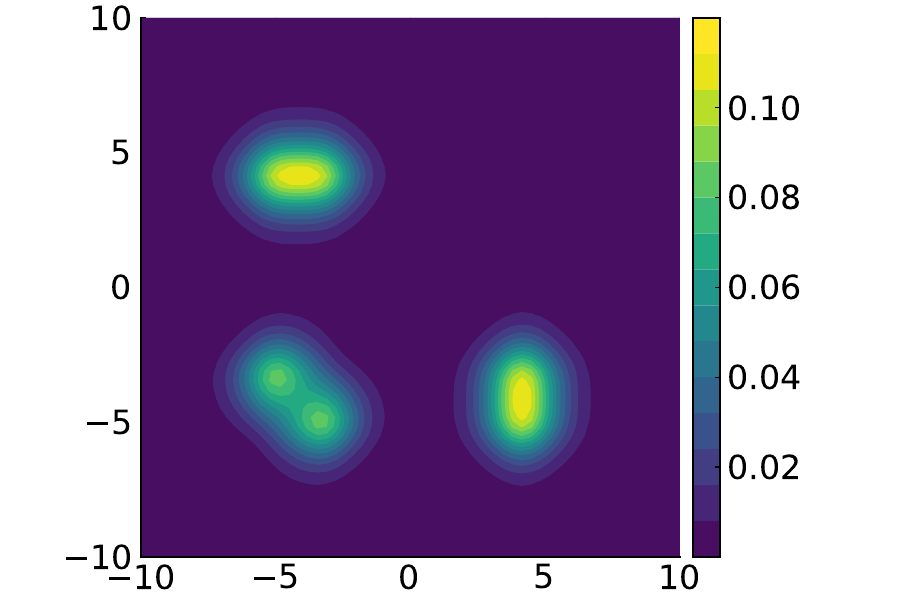} \\
    \end{tabular}
    }
    \subfigure[Mean field pair density~\eqref{eq:MFrho2}]{
    \begin{tabular}{@{}c@{}}
        \includegraphics[width=0.23\textwidth]{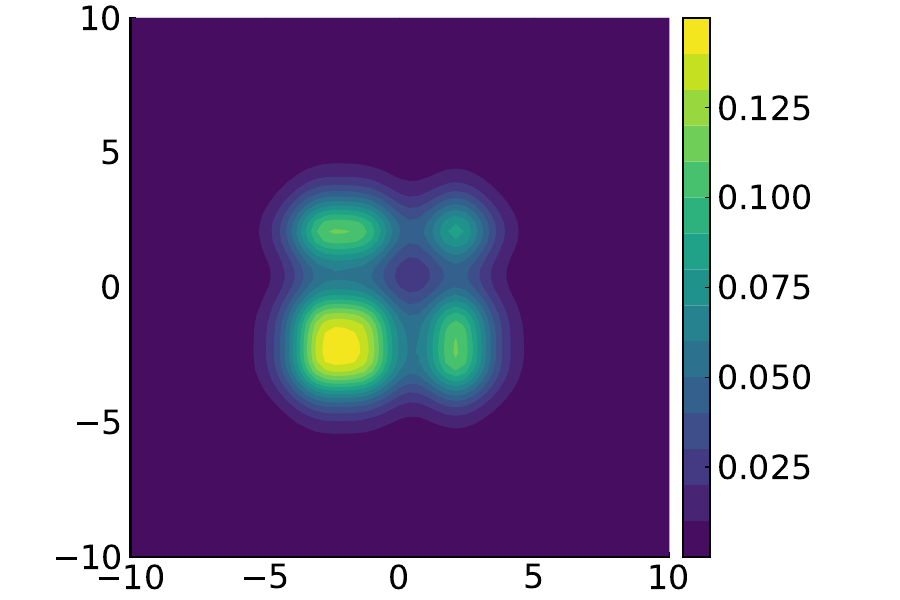} \\
        \includegraphics[width=0.23\textwidth]{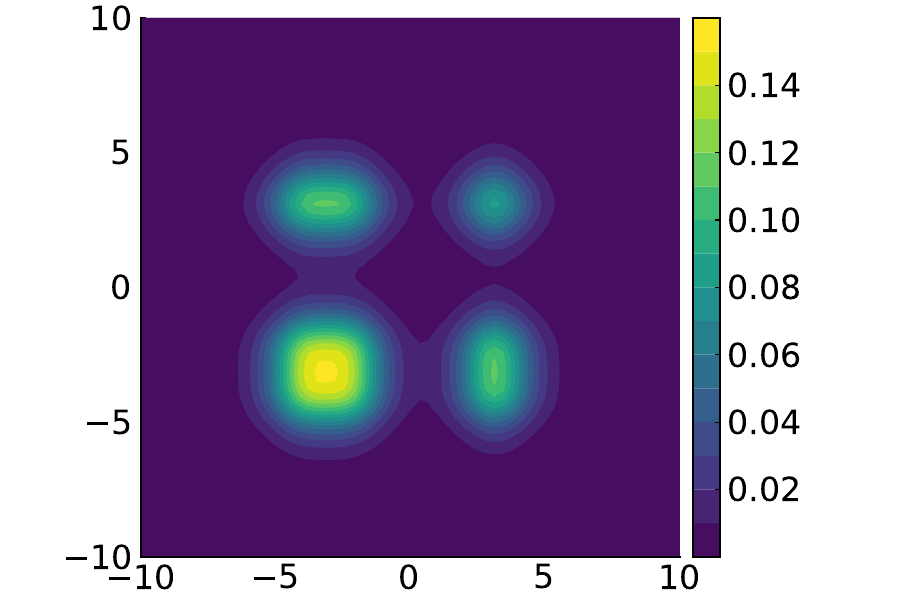} \\
        \includegraphics[width=0.23\textwidth]{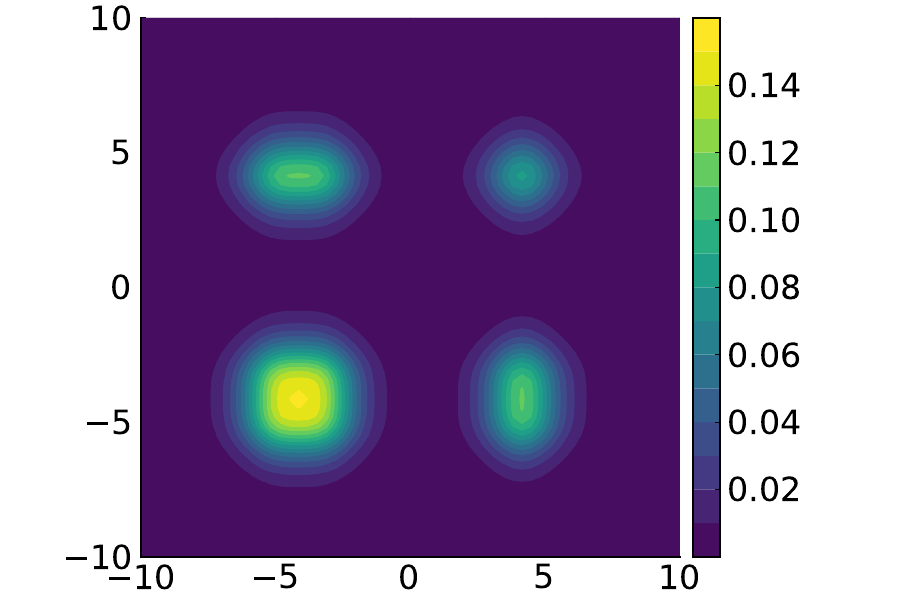} 
    \end{tabular}
    }
    \subfigure[Copula~\eqref{defcopmulti}]{
    \begin{tabular}{@{}c@{}}
        \includegraphics[width=0.23\textwidth]{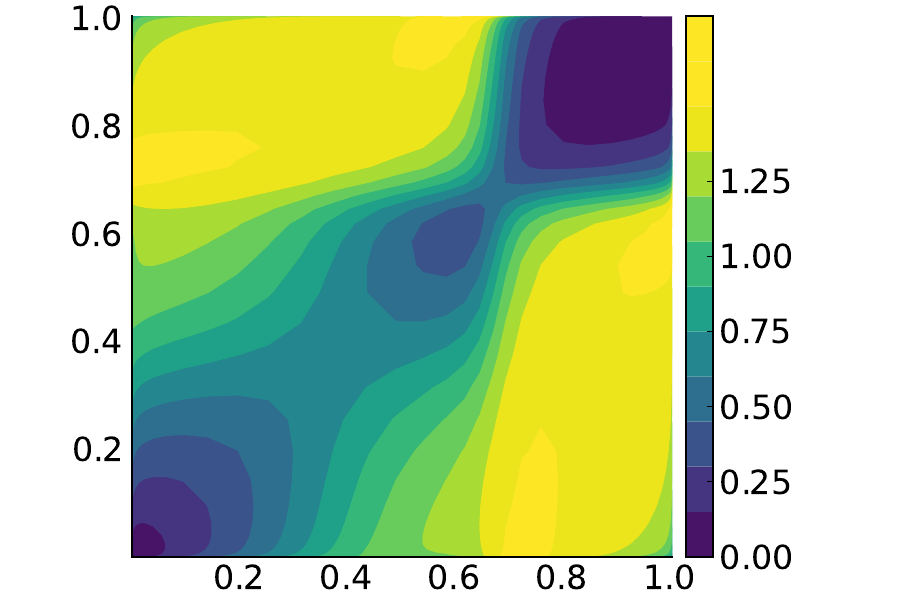} \\
        \includegraphics[width=0.23\textwidth]{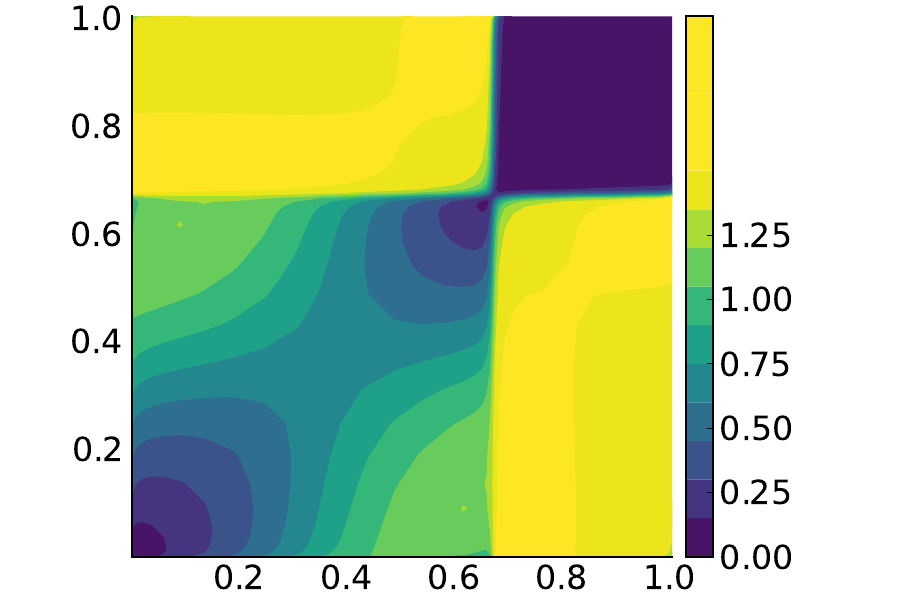} \\
        \includegraphics[width=0.23\textwidth]{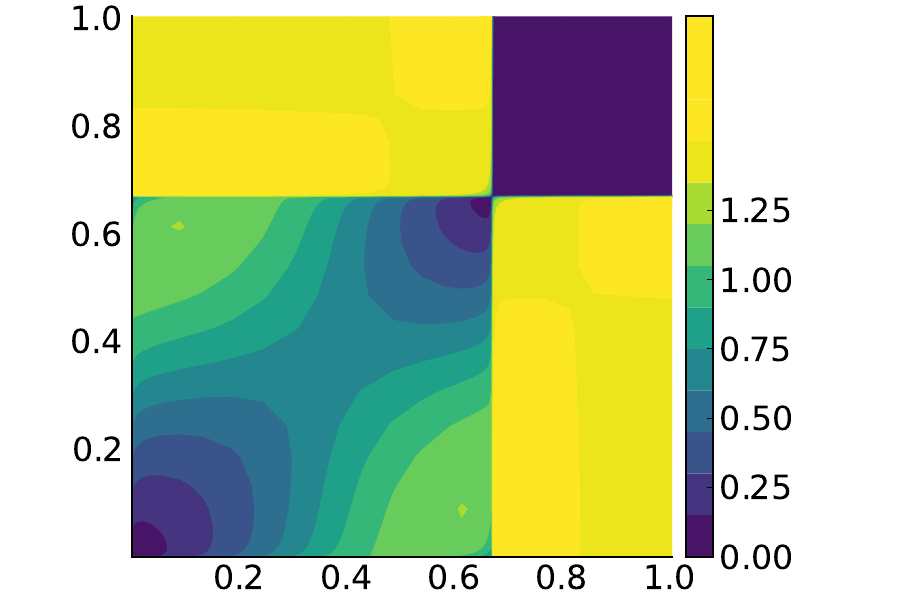} 
    \end{tabular}
    }
    \caption{
    Right column: exact copula of ground state for a  three-particle system dissociating  into a one-electron density and a two-electron density.  
    The other columns show related quantities.
    Nuclei positions for top row: $(-3, -1, 2)$, second row: $(-4, -2, 3)$, bottom row: $(-5, -3, 4)$. 
    }
    \label{fig:21dissociation}
\end{figure*}

\begin{figure*}[t!]
    \centering
    \subfigure[External potential $V_{\rm ext}$ defined in~\eqref{eq:Vext} and electronic density~\eqref{eq:rho}]{
    \begin{tabular}{@{}c@{}}
         \includegraphics[width=0.23\textwidth]{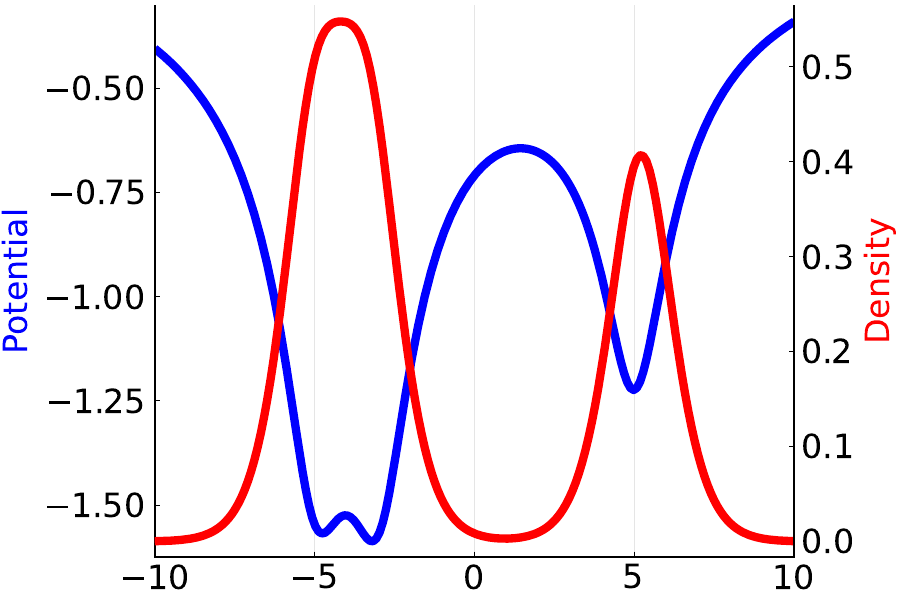} \\
         \includegraphics[width=0.23\textwidth]{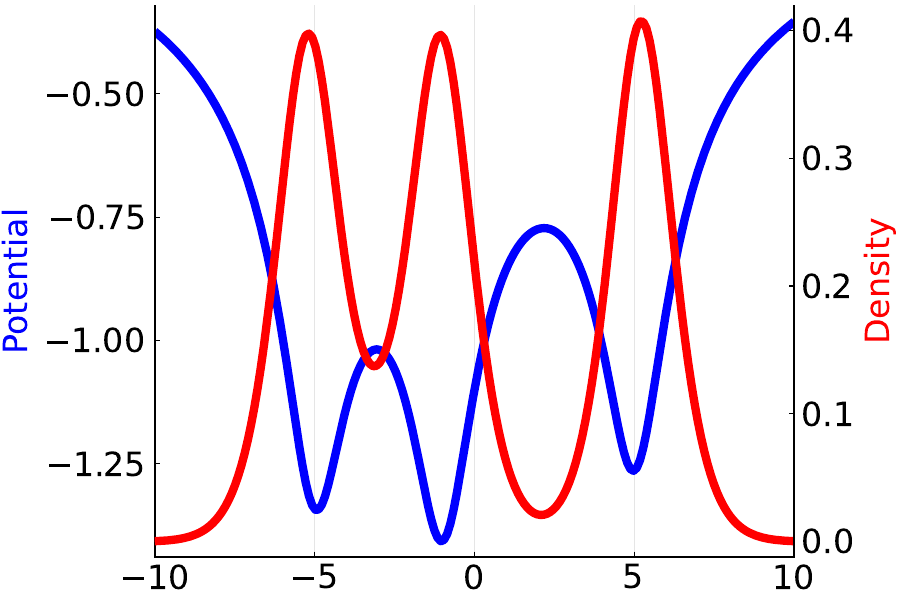} \\
         \includegraphics[width=0.23\textwidth]{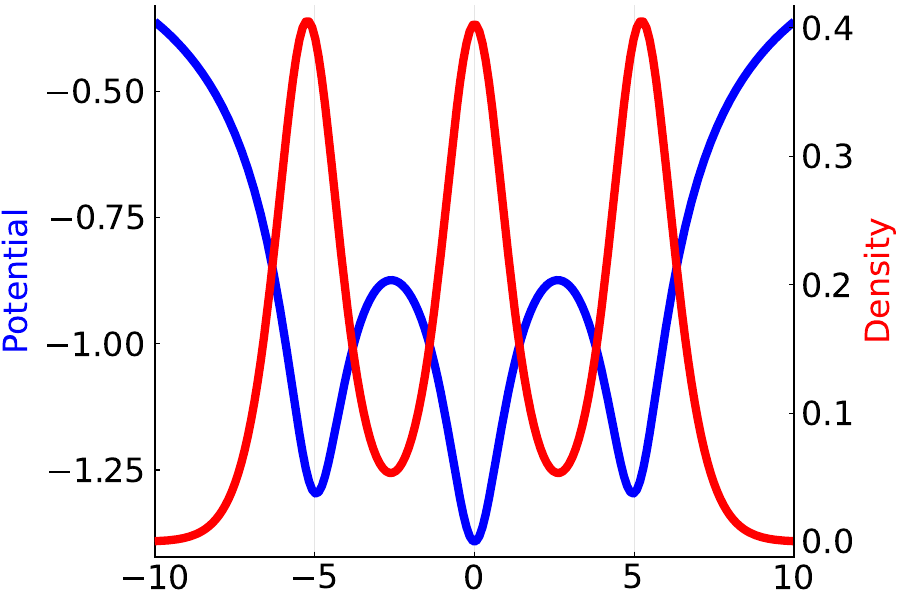}
    \end{tabular}
    }
    \subfigure[Pair density~\eqref{eq:rho2}]{
    \begin{tabular}{@{}c@{}}
    \includegraphics[width=0.23\textwidth]{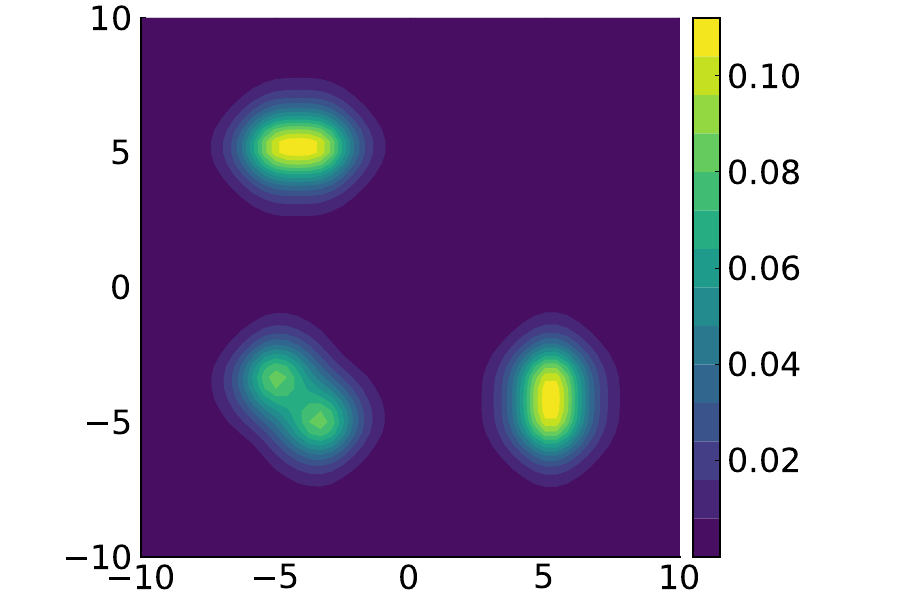} \\
    \includegraphics[width=0.23\textwidth]{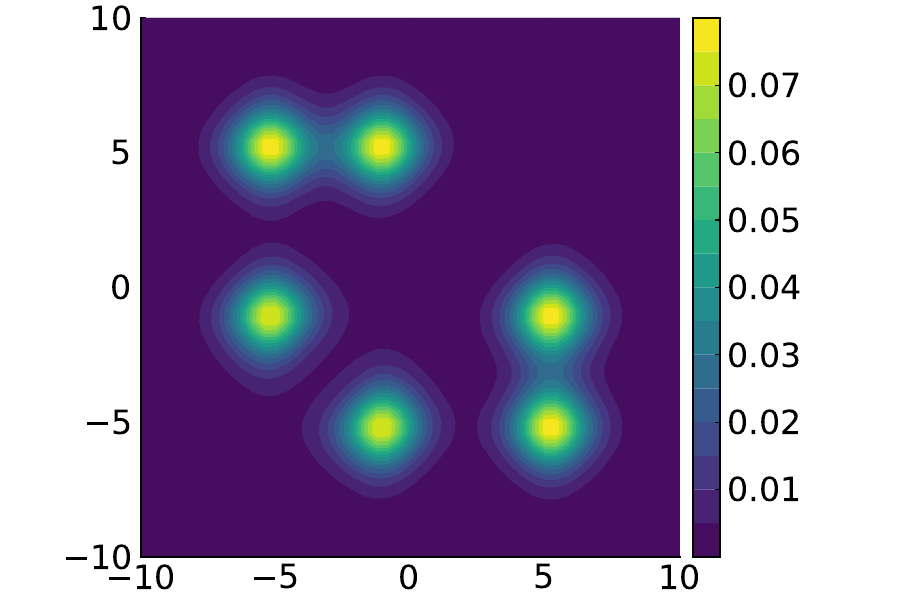} \\
        \includegraphics[width=0.23\textwidth]{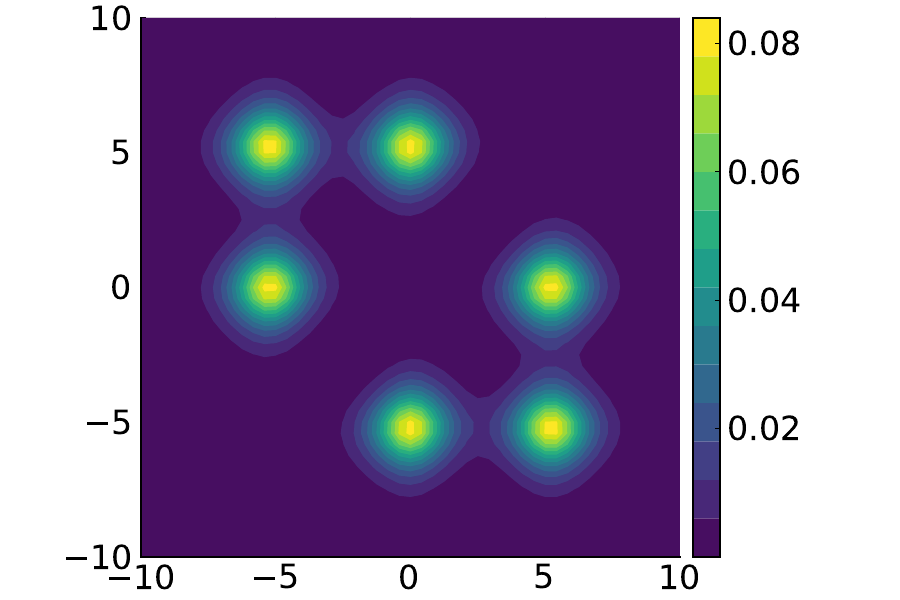} \\
    \end{tabular}
    }
    \subfigure[Mean field pair density~\eqref{eq:MFrho2}]{
    \begin{tabular}{@{}c@{}}
        \includegraphics[width=0.23\textwidth]{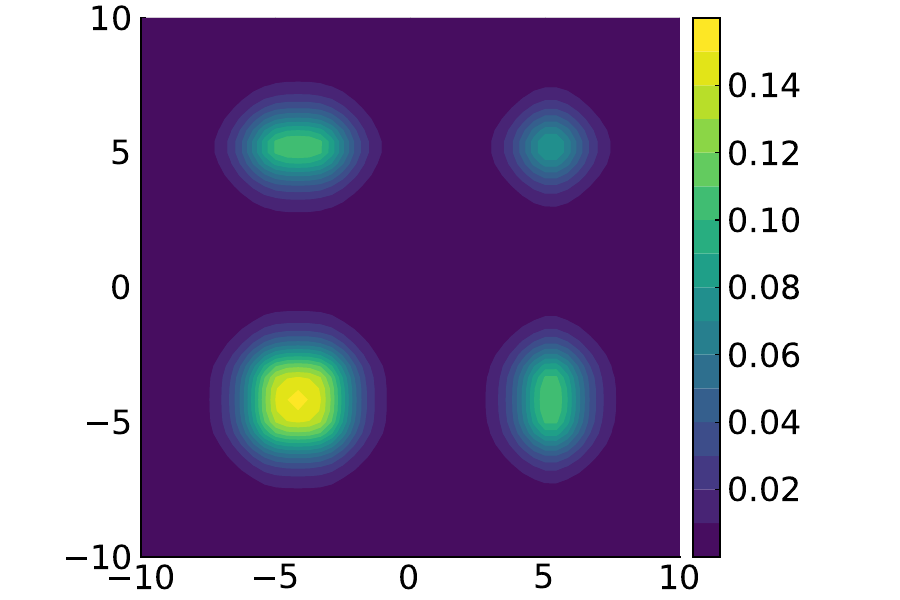} \\
        \includegraphics[width=0.23\textwidth]{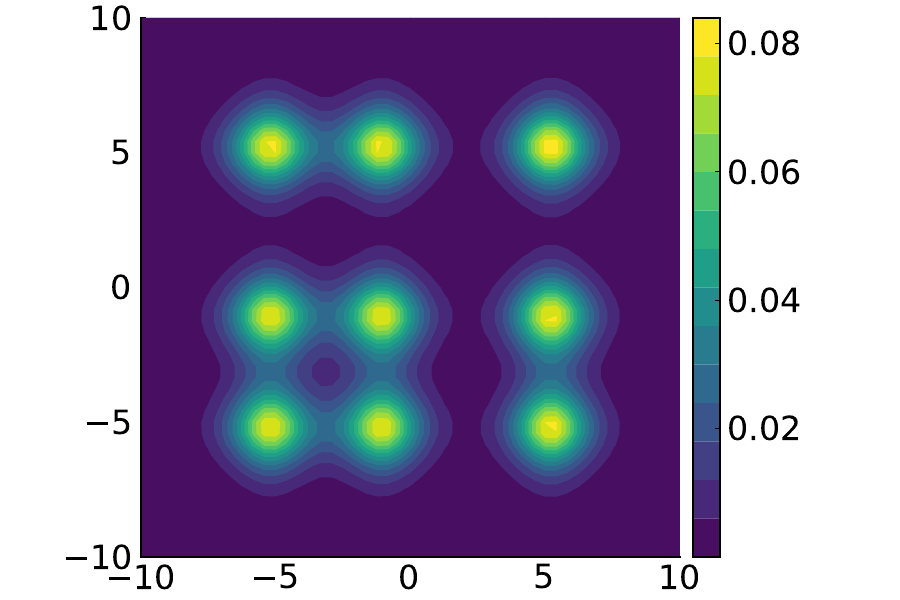} \\
        \includegraphics[width=0.23\textwidth]{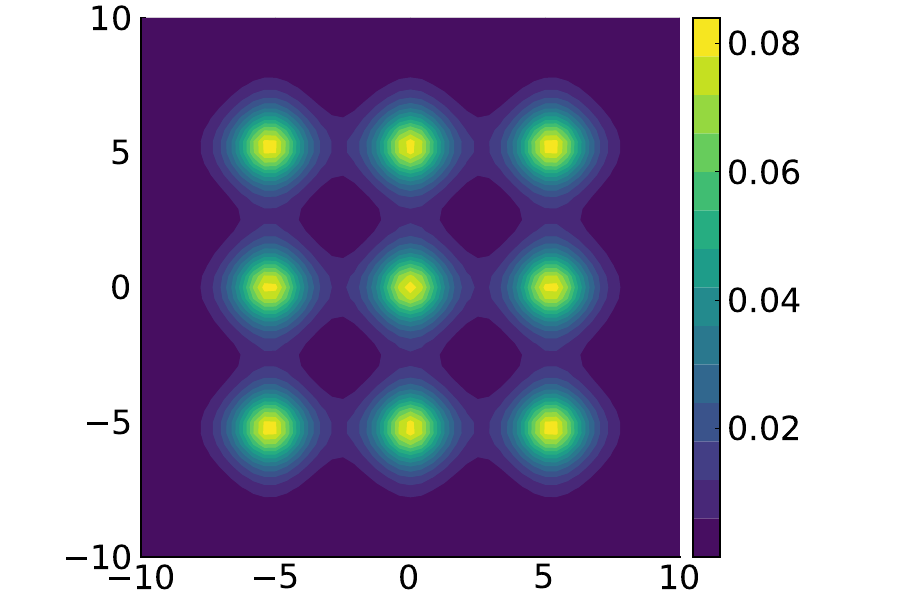} 
    \end{tabular}
    }
    \subfigure[Copula~\eqref{defcopmulti}]{
    \begin{tabular}{@{}c@{}}
        \includegraphics[width=0.23\textwidth]{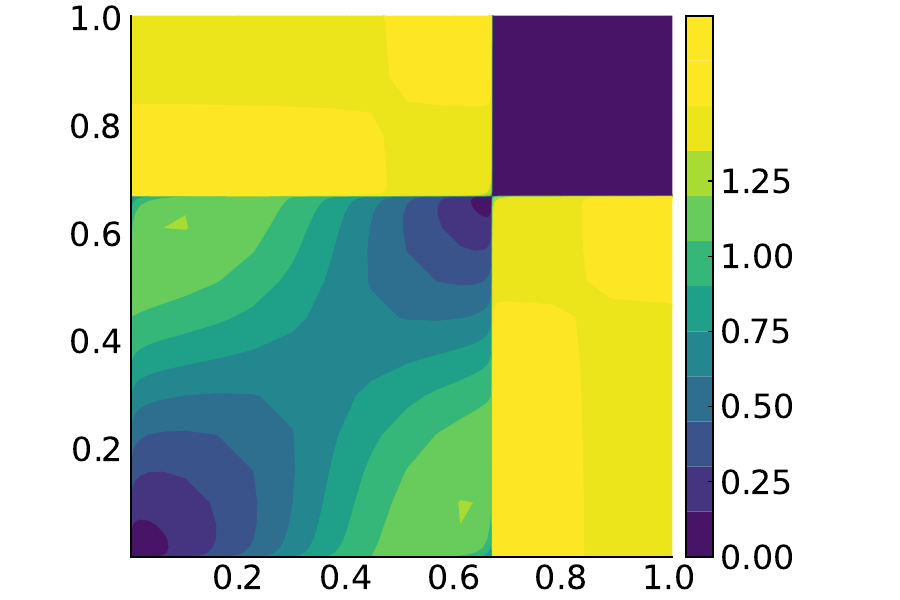} \\
        \includegraphics[width=0.23\textwidth]{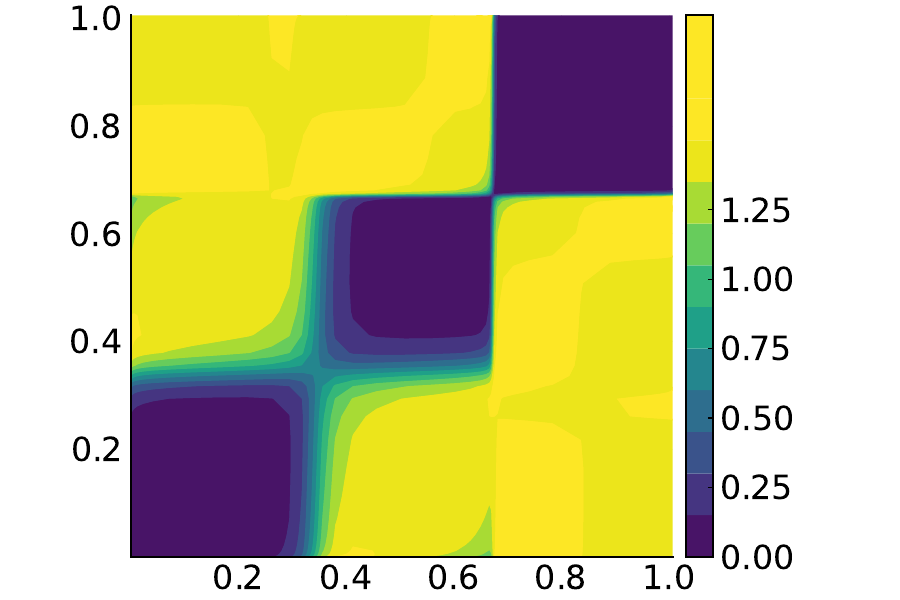} \\
        \includegraphics[width=0.23\textwidth]{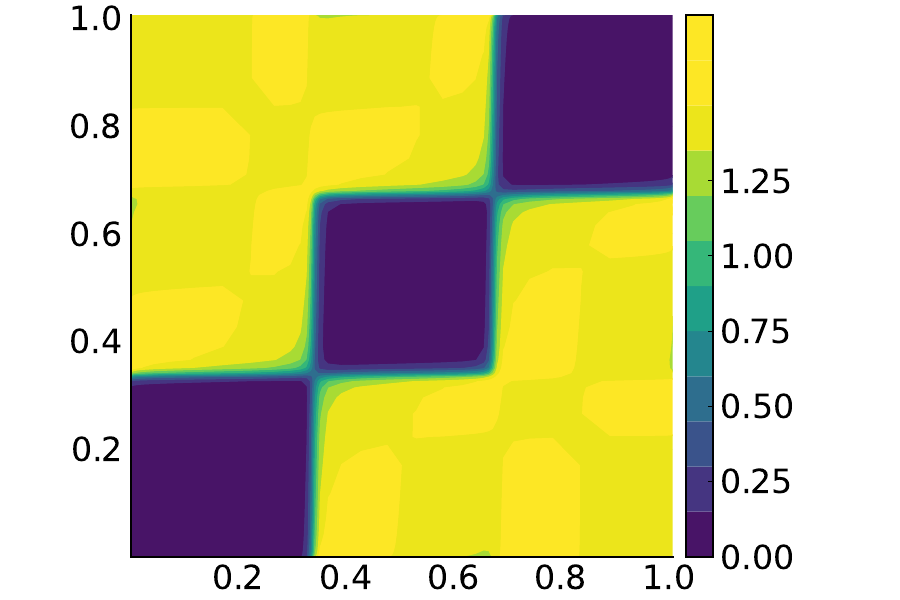} 
    \end{tabular}
    }
    \caption{
        Right column: exact copula of ground state for a  three-particle system dissociating  into three one-electron densities.  
    The other columns show related quantities. 
    Nuclei positions for top row: $(-5, -3, 5)$, second row: $(-5, -1, 5)$, bottom row: $(-5, 0, 5)$. 
    }
    \label{fig:21-111dissociation}
\end{figure*}

In Figure~\ref{fig:21-111dissociation}, we consider a different scenario, namely the dissociation of the density into one part with total charge 1 and another part with total charge 2, mimicking the dissociation H$_3\to $H+H$_2$. 
To do so, we take for the positions of the nuclei $(-3,-1,2),$ $(-4,-2,3)$ and $(-5,-3,4)$,
that is, two nuclei are kept at the same small distance as in the left part of Figure~\ref{fig:11dissociation}, while the third one is placed further and further away. 
The potential naturally separates into two wells, one deeper than the other. As a consequence, the density separates into two parts, one with total charge 1, the other with total charge 2.
In this case, the copula converges in a checkerboard as well, but with only four rectangles, three of them having at least one side of size 1/3 with constant values 3/2, 3/2 and 0, which corresponds to the one dissociated electron. The last one
looks precisely like the copula for the least dissociated two-particle system (i.e. Figure~\ref{fig:11dissociation}, (a)), as predicted by theory (Theorem \ref{T:cop}). 

\begin{figure*}[t!]
    \centering
    \subfigure[External potential $V_{\rm ext}$ defined in~\eqref{eq:Vext} and electronic density~\eqref{eq:rho}]{
    \begin{tabular}{@{}c@{}}
         \includegraphics[width=0.23\textwidth]{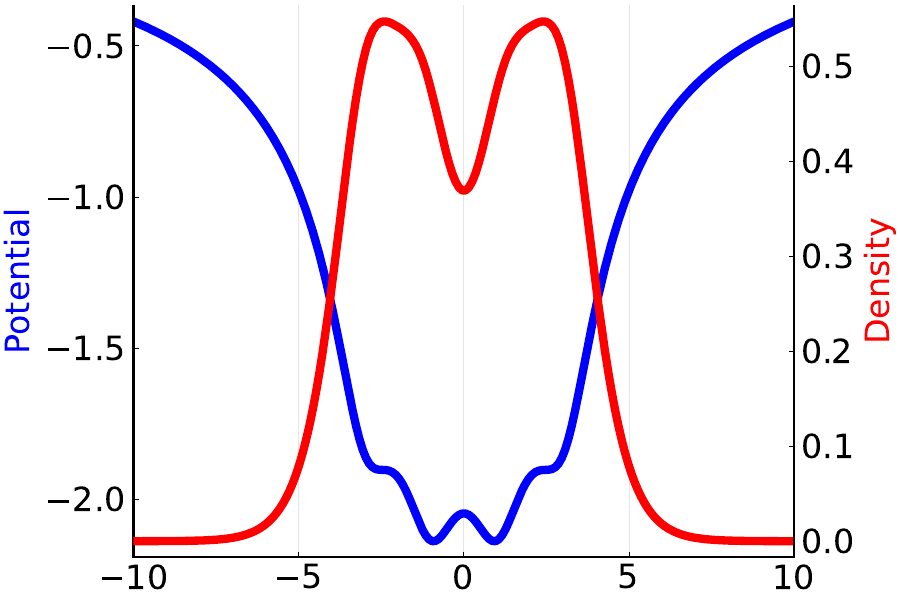} \\
         \includegraphics[width=0.23\textwidth]{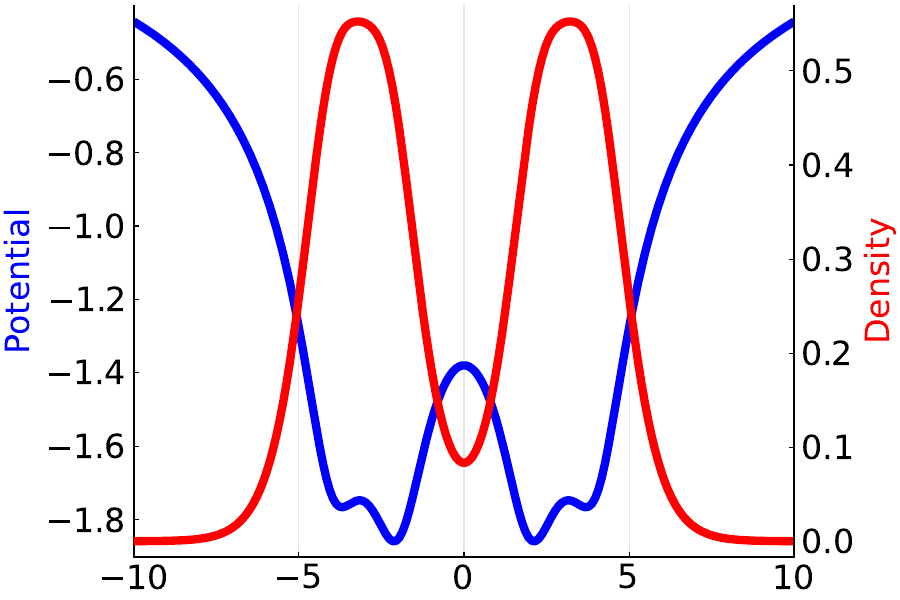} \\
         \includegraphics[width=0.23\textwidth]{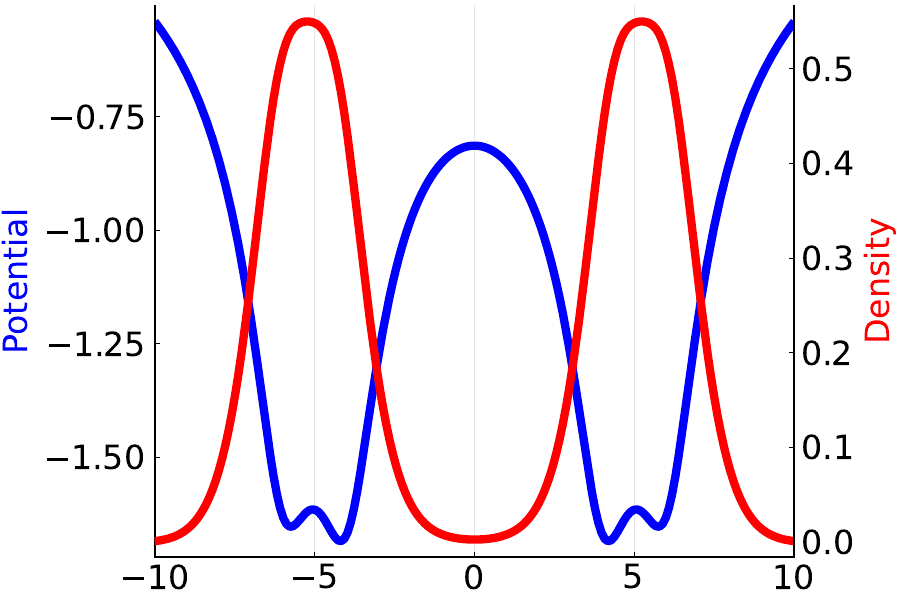}
    \end{tabular}
    }
    \subfigure[Pair density~\eqref{eq:rho2}]{
    \begin{tabular}{@{}c@{}}
    \includegraphics[width=0.23\textwidth]{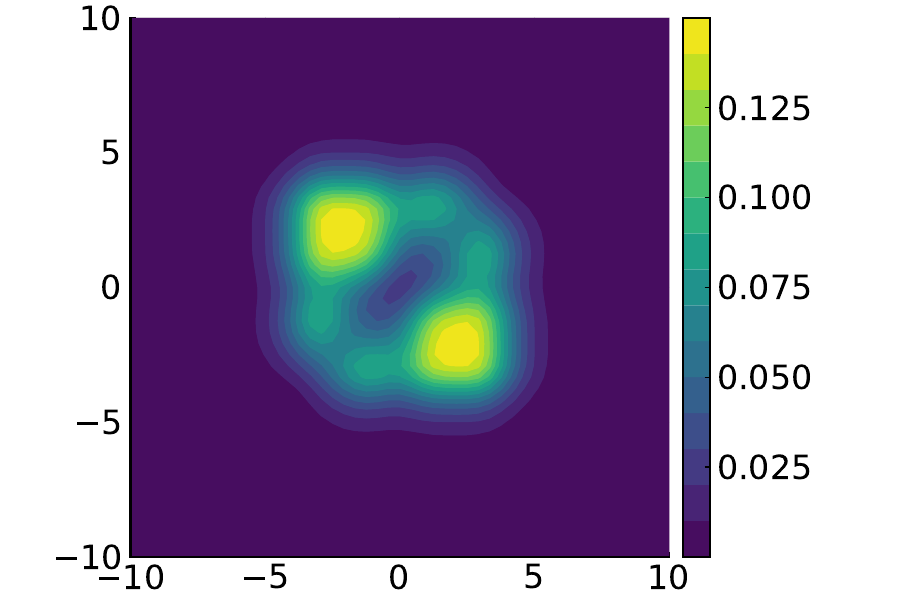} \\
    \includegraphics[width=0.23\textwidth]{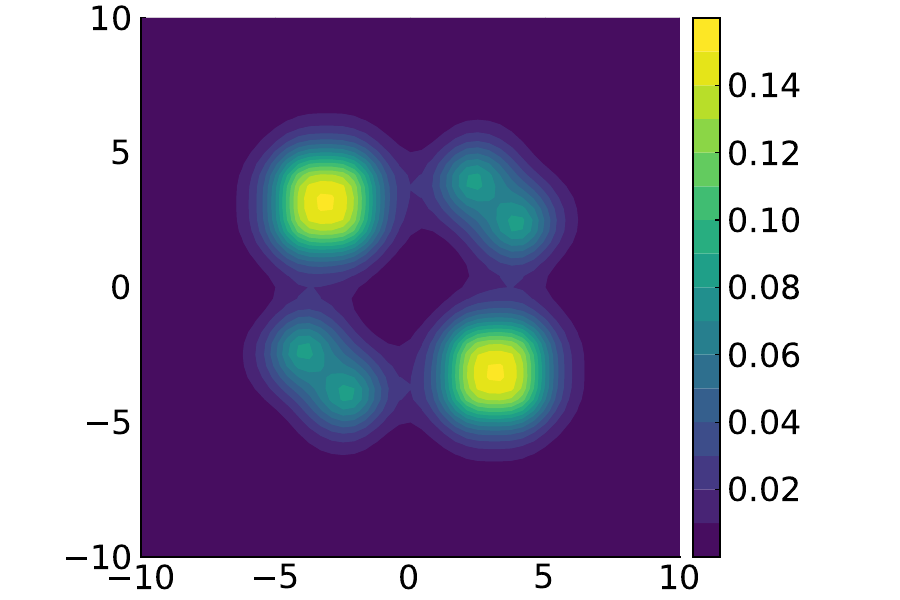} \\
        \includegraphics[width=0.23\textwidth]{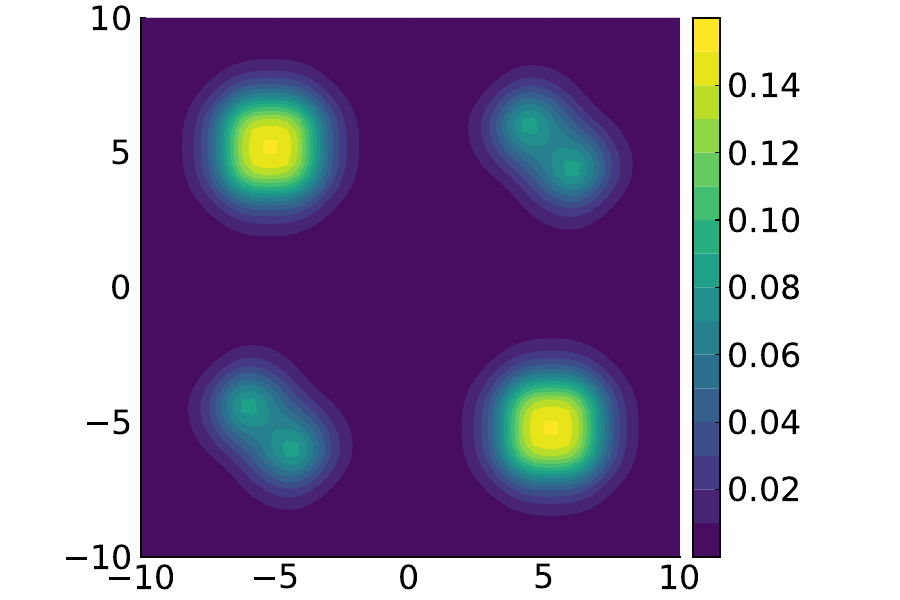} \\
    \end{tabular}
    }
    \subfigure[Mean field pair density~\eqref{eq:MFrho2}]{
    \begin{tabular}{@{}c@{}}
        \includegraphics[width=0.23\textwidth]{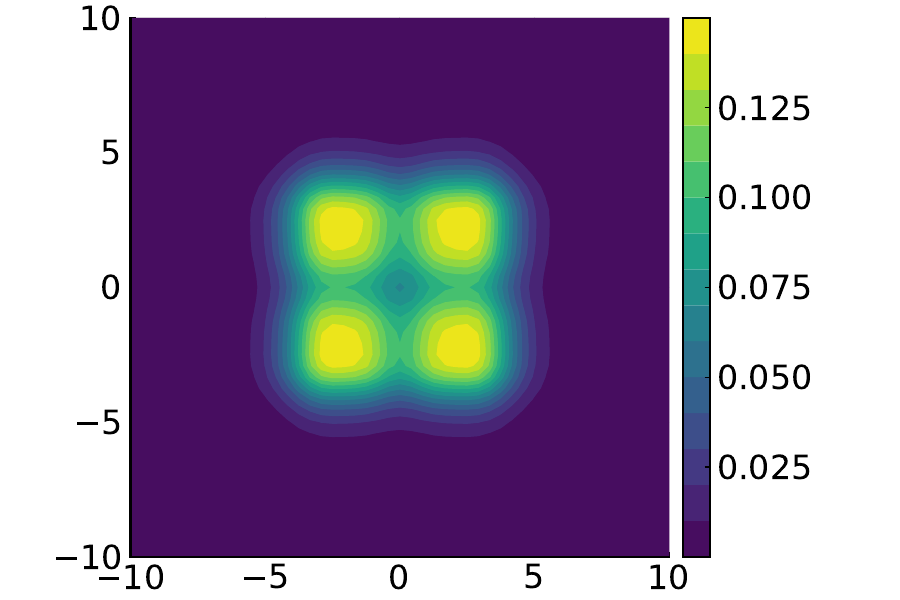} \\
        \includegraphics[width=0.23\textwidth]{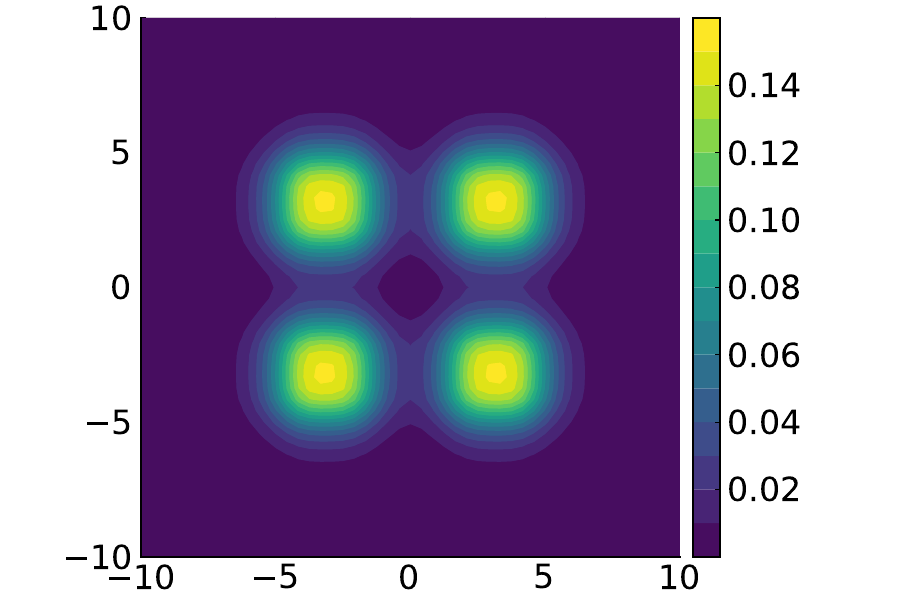} \\
        \includegraphics[width=0.23\textwidth]{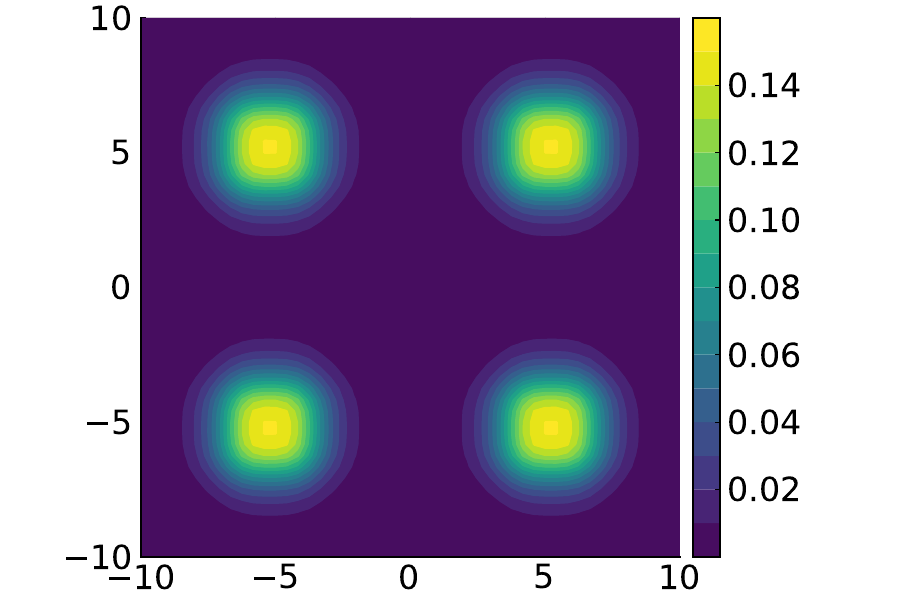} 
    \end{tabular}
    }
    \subfigure[Copula~\eqref{defcopmulti}]{
    \begin{tabular}{@{}c@{}}
        \includegraphics[width=0.23\textwidth]{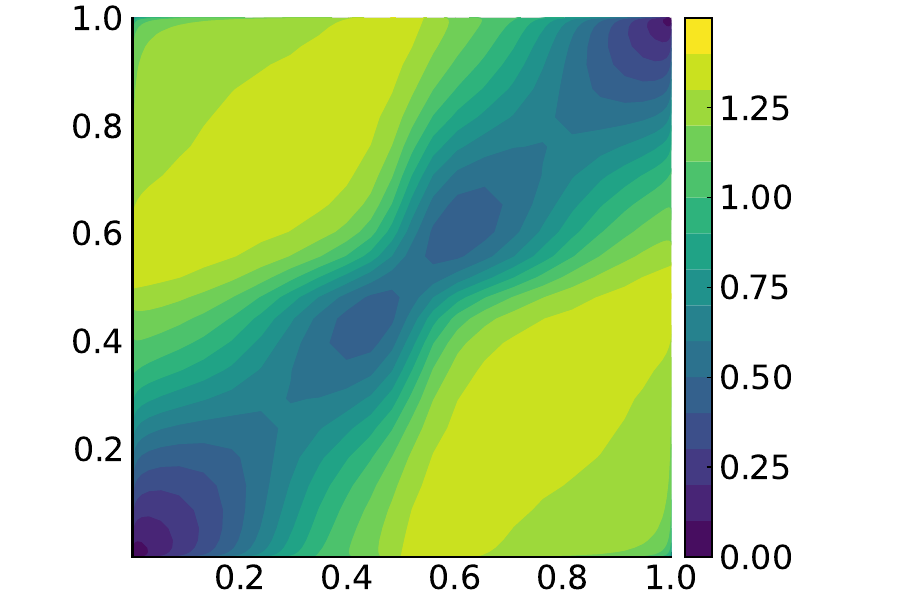} \\
        \includegraphics[width=0.23\textwidth]{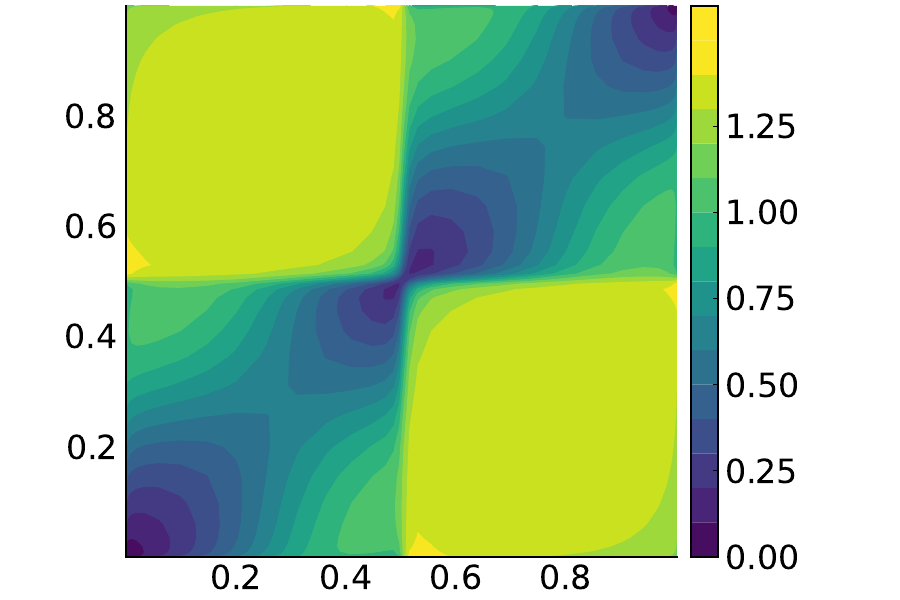} \\
        \includegraphics[width=0.23\textwidth]{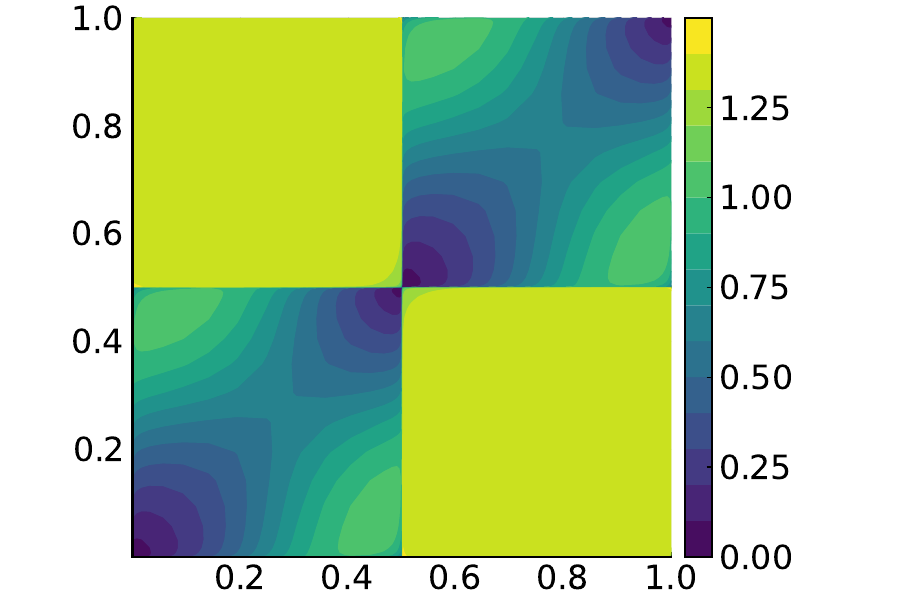} 
    \end{tabular}
    }
    \caption{
    Right column: exact copula of ground state for a  four-particle system dissociating into  two two-electron densities. 
    The other columns show related quantities.
    Parameter $a$ for top row: $a=1$, second row: $a=2$, bottom row: $a=3.$
    }
    \label{fig:22dissociation}
\end{figure*}

This feature of the copula is even more visible on the additional example provided in Figure~\ref{fig:21-111dissociation}. Here we start from a three-particle system, dissociated into one part with charge 2 and one part with charge 1, and we observe the dissociation of the part with charge 2 into two parts with charge 1, mimicking the dissociation H$_2$+H $\to$ H+H+H. To do so, we choose an external potential with nuclei positions $(-5,a,5)$ with $a = -3, -1, 0$. 
The potential shows three wells, where the two wells on the left progressively separate. 
In Figure~\ref{fig:21-111dissociation} (a), 
the corresponding ground-state density is the sum of two dissociated densities, one with total charge 2, the other with total charge 1. This corresponds to the dissociated case of Figure~\ref{fig:21dissociation} (c). 
In the partially dissociated case (b), we clearly see the copula structure as predicted by theory which has the form of a checkerboard, with four rectangles, three of them being constant. Those parts do not involve nontrivial electron-electron correlations. The copula on the last square is nontrivial, and has the structure of the two-particle copula, as in (b) of Figure~\ref{fig:11dissociation}. The most dissociated case looks exactly as in (c) of Figure~\ref{fig:111dissociation}, that is the copula has the form of a checkerboard with nine squares, all of them being constant, with value 4/3 for six of them and 0 for three of them.

\paragraph{Four-particle systems.}

To conclude this section, we illustrate the copula structure for a four-particle system, dissociated into two systems with two particles. In this case, the number of discretization points per dimension is 50, and we consider an interval $[-10,10]$. Thus, the nuclei positions are $(-a-1.5,-a+1.5,a-1.5,a+1.5)$ with $a = 3,4,5$. Naturally, the density separates into two main parts of charge 2 each. Figure \ref{fig:22dissociation} shows that the copula exhibits a checkerboard structure with four squares. The copula is constant on two squares, whereas in the two others it agrees with the copula of the two-particle system of (b) in Figure~\ref{fig:11dissociation}, as predicted by theory.

\section{Fitting the copula for two-particle systems}
\label{sec:fitting}

In this section, we turn to the approximation of the copula and the pair density for two-particle systems. More precisely, we introduce several simple models for interpolating the copula from knowledge of its equilibrium and dissociated shape, and demonstrate that these models already provide accurate approximations of the copula in the intermediate regime, and also of the pair density and the interaction energy. An intuitive explanation of the observed favorable approximability properties of the copula is that for uncorrelated systems it is a uniform constant (see \eqref{copMF}) and one needs to fit deviations of order one; by contrast the pair density is small in important regions, and one needs to fit small corrections to it. 

\begin{figure*}
    \centering
    \subfigure[$a=1$]{
    \begin{tabular}{@{}c@{}}
        \includegraphics[width=0.13\textwidth]{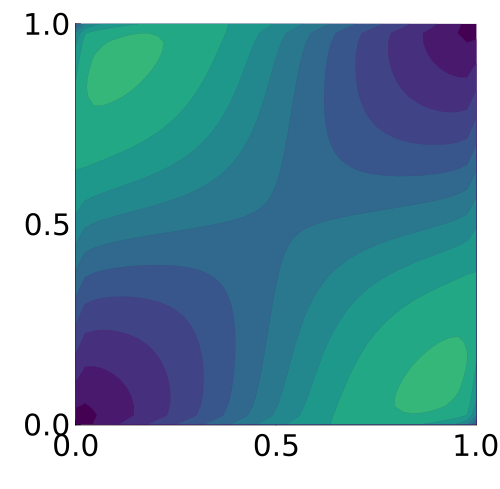} \\
         {
        \setlength{\fboxsep}{0pt}%
        \setlength{\fboxrule}{1.5pt}%
    \fcolorbox{white}{white}{\includegraphics[width=0.13\textwidth]{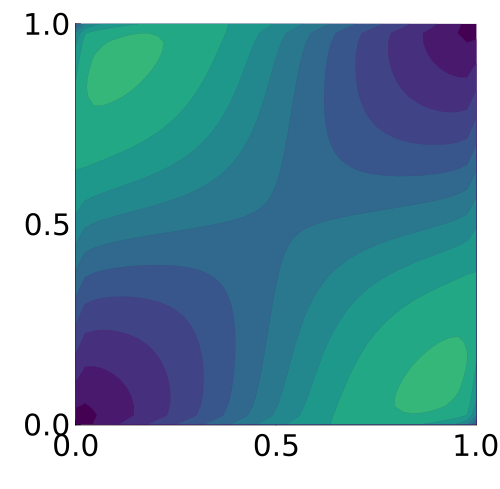}}} \\
          {
        \setlength{\fboxsep}{0pt}%
        \setlength{\fboxrule}{1.5pt}%
    \fcolorbox{white}{white}{\includegraphics[width=0.13\textwidth]{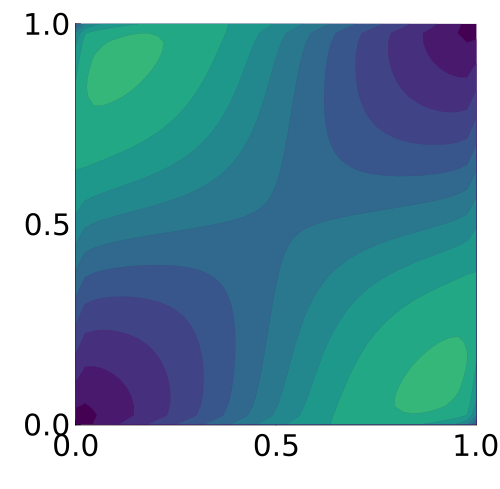}}} \\
         {
        \setlength{\fboxsep}{0pt}%
        \setlength{\fboxrule}{1.5pt}%
    \fcolorbox{purple}{white}{ \includegraphics[width=0.13\textwidth]{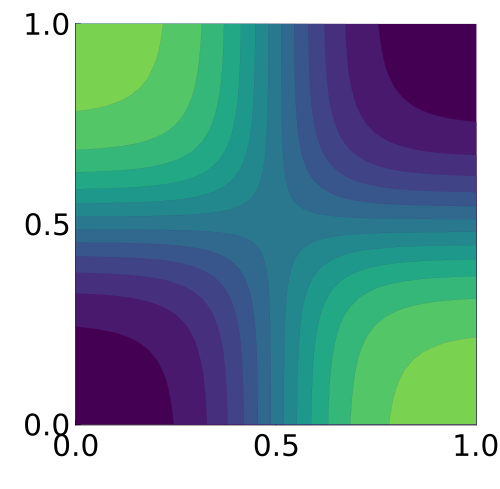}}} \\
         {
         \setlength{\fboxsep}{0pt}%
        \setlength{\fboxrule}{1.5pt}%
    \fcolorbox{lightgray}{white}{
         \includegraphics[width=0.13\textwidth]{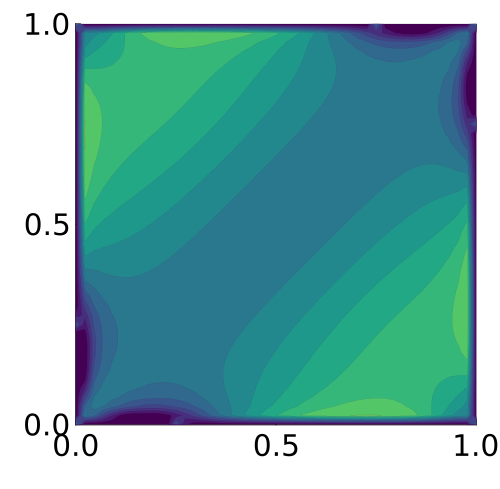}}}
    \end{tabular}
    }
    \subfigure[$a=1.5$]{
    \begin{tabular}{@{}c@{}}
         \includegraphics[width=0.13\textwidth]{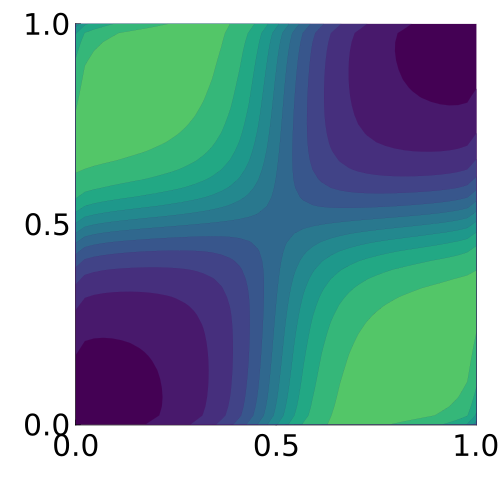} \\
             {
         \setlength{\fboxsep}{0pt}%
        \setlength{\fboxrule}{1.5pt}%
    \fcolorbox{purple}{white}{\includegraphics[width=0.13\textwidth]{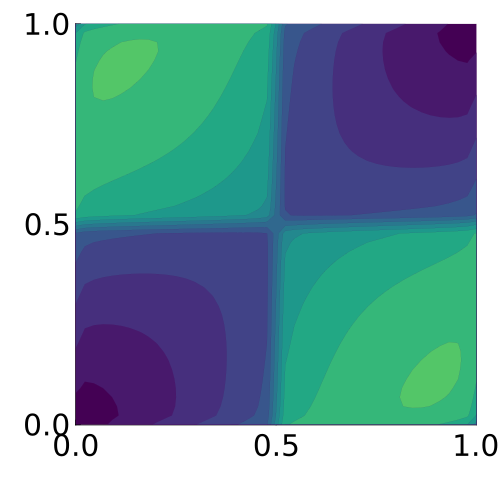}
    }} \\
         {
         \setlength{\fboxsep}{0pt}%
        \setlength{\fboxrule}{1.5pt}%
    \fcolorbox{purple}{white}{\includegraphics[width=0.13\textwidth]{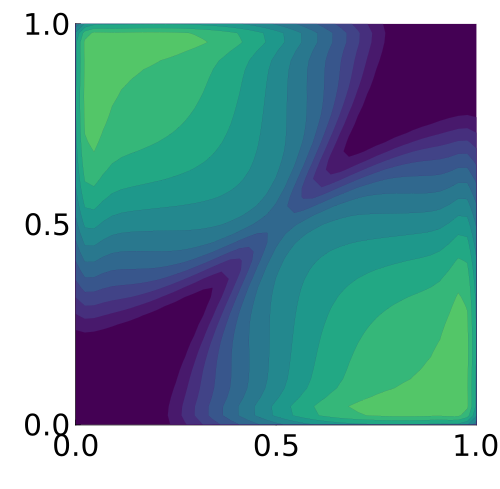}}} \\
    {
         \setlength{\fboxsep}{0pt}%
        \setlength{\fboxrule}{1.5pt}%
    \fcolorbox{purple}{white}{
         \includegraphics[width=0.13\textwidth]{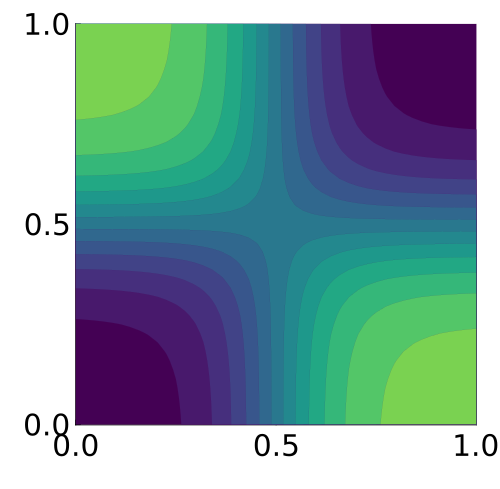}}} \\
          {
         \setlength{\fboxsep}{0pt}%
        \setlength{\fboxrule}{1.5pt}%
    \fcolorbox{lightgray}{white}{\includegraphics[width=0.13\textwidth]{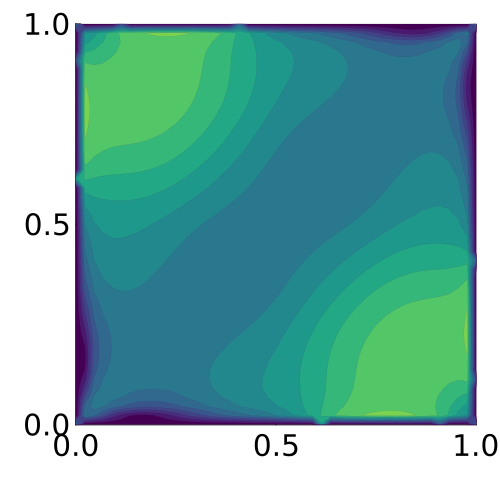}}}
    \end{tabular}
    }
    \subfigure[$a=2$]{
    \begin{tabular}{@{}c@{}}
         \includegraphics[width=0.13\textwidth]{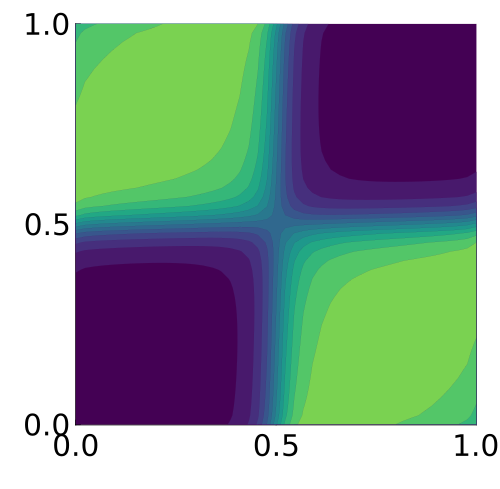} \\
         {
         \setlength{\fboxsep}{0pt}%
        \setlength{\fboxrule}{1.5pt}%
    \fcolorbox{purple}{white}{
     \includegraphics[width=0.13\textwidth]{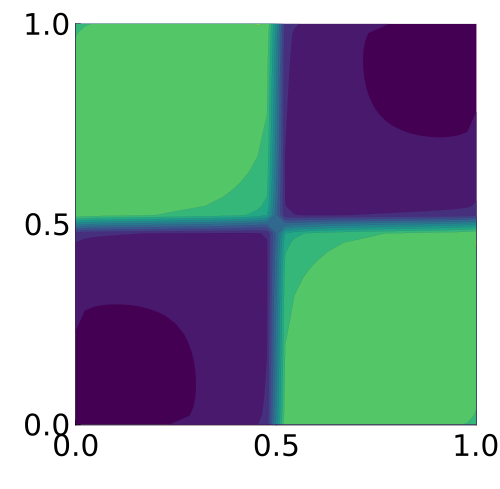}}} \\
         {
         \setlength{\fboxsep}{0pt}%
        \setlength{\fboxrule}{1.5pt}%
    \fcolorbox{purple}{white}{\includegraphics[width=0.13\textwidth]{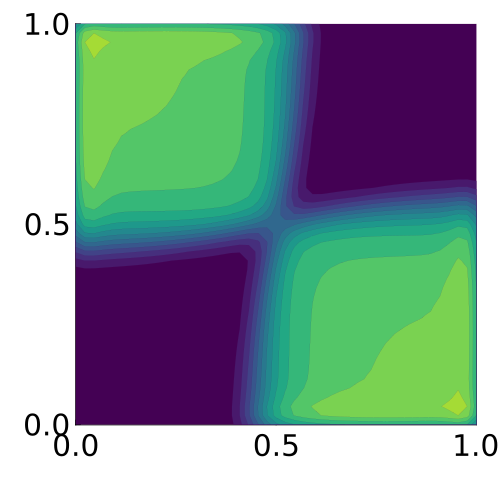}}} \\
         {
         \setlength{\fboxsep}{0pt}%
        \setlength{\fboxrule}{1.5pt}%
    \fcolorbox{purple}{white}{\includegraphics[width=0.13\textwidth]{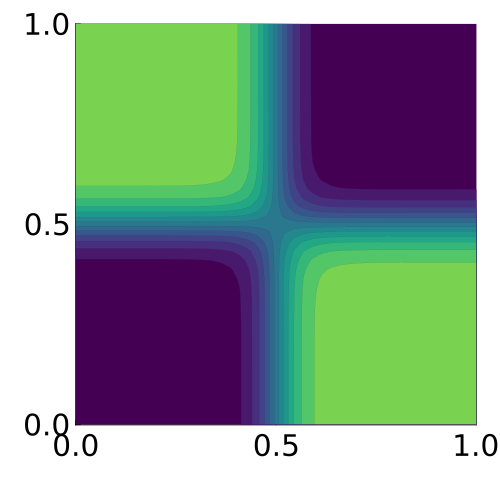}}} \\
    {
        \setlength{\fboxsep}{0pt}%
        \setlength{\fboxrule}{1.5pt}%
    \fcolorbox{lightgray}{white}{ \includegraphics[width=0.13\textwidth]{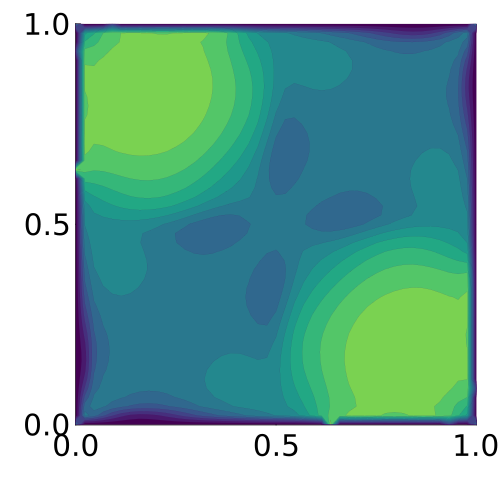}}}
    \end{tabular}
    }
    \subfigure[$a=2.5$]{
    \begin{tabular}{@{}c@{}}
         \includegraphics[width=0.13\textwidth]{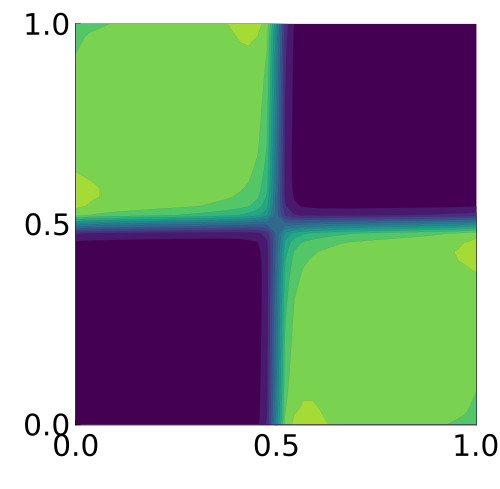} \\    
         {
         \setlength{\fboxsep}{0pt}%
        \setlength{\fboxrule}{1.5pt}%
    \fcolorbox{purple}{white}{\includegraphics[width=0.13\textwidth]{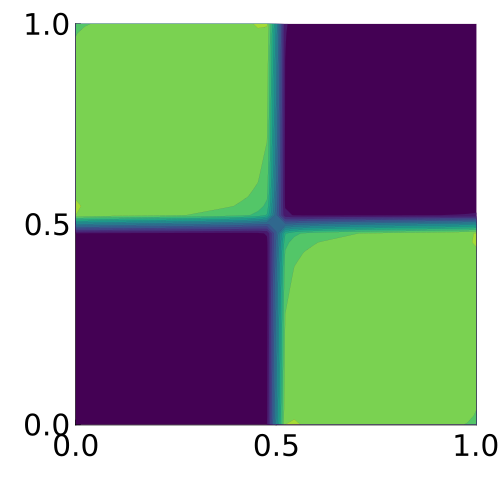}}} \\
        {
         \setlength{\fboxsep}{0pt}%
        \setlength{\fboxrule}{1.5pt}%
    \fcolorbox{purple}{white}{\includegraphics[width=0.13\textwidth]{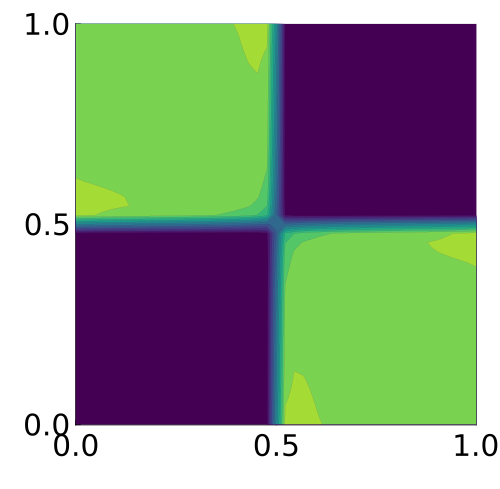}}} \\
         {
         \setlength{\fboxsep}{0pt}%
        \setlength{\fboxrule}{1.5pt}%
    \fcolorbox{purple}{white}{\includegraphics[width=0.13\textwidth]{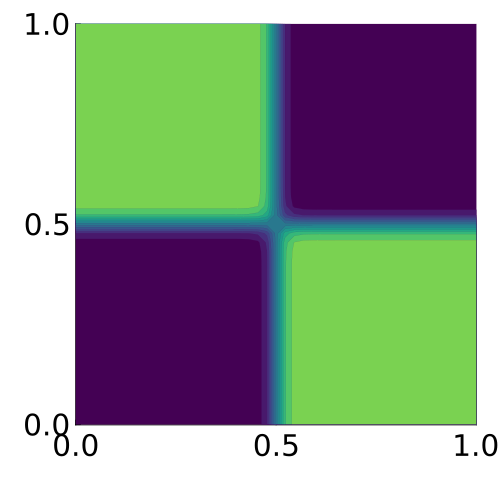}}} \\
          {
        \setlength{\fboxsep}{0pt}%
        \setlength{\fboxrule}{1.5pt}%
    \fcolorbox{lightgray}{white}{\includegraphics[width=0.13\textwidth]{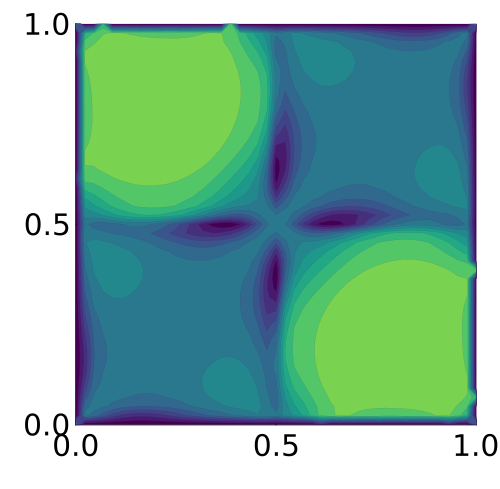}}}
    \end{tabular}
    }
    \subfigure[$a=3$]{
    \begin{tabular}{@{}c@{}}
         \includegraphics[width=0.13\textwidth]{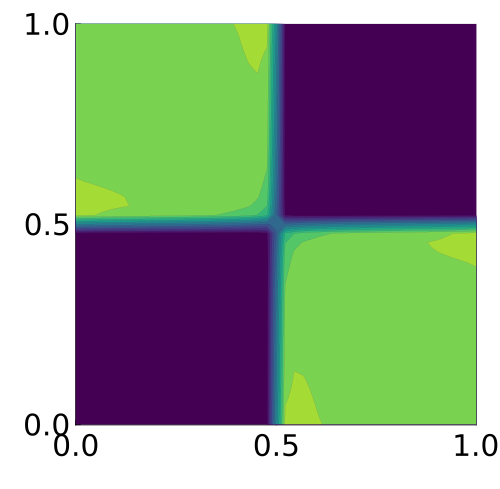} \\
         {
        \setlength{\fboxsep}{0pt}%
        \setlength{\fboxrule}{1.5pt}%
    \fcolorbox{white}{white}{\includegraphics[width=0.13\textwidth]{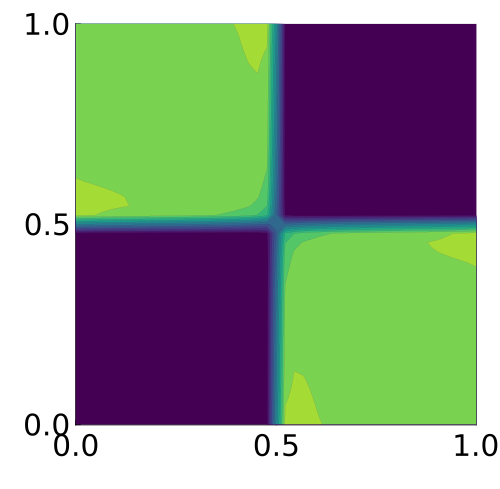}}} \\
         {
        \setlength{\fboxsep}{0pt}%
        \setlength{\fboxrule}{1.5pt}%
    \fcolorbox{white}{white}{ \includegraphics[width=0.13\textwidth]{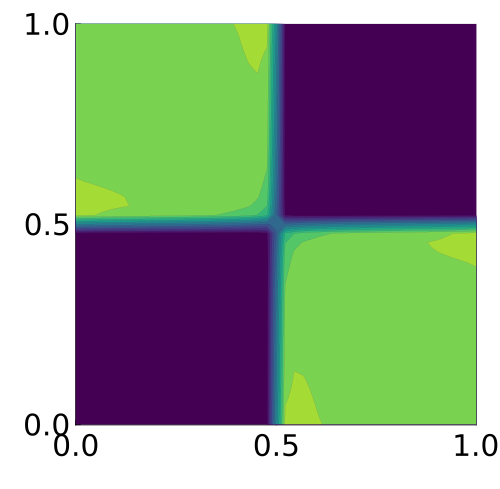}}} \\
          {
        \setlength{\fboxsep}{0pt}%
        \setlength{\fboxrule}{1.5pt}%
    \fcolorbox{purple}{white}{\includegraphics[width=0.13\textwidth]{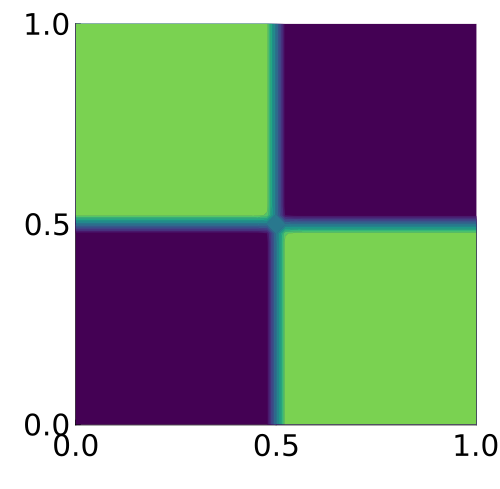}}} \\
         {
        \setlength{\fboxsep}{0pt}%
        \setlength{\fboxrule}{1.5pt}%
    \fcolorbox{lightgray}{white}{ \includegraphics[width=0.13\textwidth]{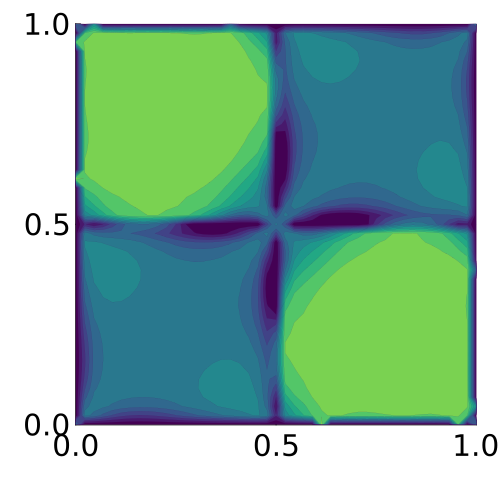}}}
    \end{tabular}
    }
    \caption{Copulas for two-particle systems obtained from different models. Exact (top row), Linear interpolation (second row), Wasserstein barycenter (third row), one-parameter neural net (fourth row), LDA (bottom row). Pictures further to the right correspond to larger internuclear distances. The color limits are (0,2.5).
    Red boxes correspond to fitted copulas. Gray boxes correspond to the LDA copulas. Other images show exact copulas.
    }
    \label{fig:copula_fit}
\end{figure*}

To simplify the setting, we consider the systems with nuclei positions $(-a,a)$ for $a = 1,1.5,2,2.5,3$ for which the copulas are shown in the top row of Figure~\ref{fig:copula_fit}. As in Figure~\ref{fig:11dissociation}, we progressively see the checkerboard structure appear.

The aim here is to approximate the intermediate copulas, i.e. for $a=1.5,2,2.5$, only from the knowledge of the copulas for $a=1$ and $a=3.$ To facilitate notation, we denote by $c_a$ the copula with parameter $a$.

\textit{Linear interpolation.}  
A basic interpolation scheme is to perform a linear combination of the two copulas $c_1$ and $c_3.$
We therefore define the approximation
\[
{c}^{\rm lin}_a = (1-t^*) c_1 + t^* c_3,
\]
where $t^*$ is found by minimizing the error in the $L^2$ norm, 
\[
    \displaystyle t^* \in \underset{t\in [0,1]}{\rm argmin} \;  \| c_a - (1-t^*) c_1 - t^* c_3 \|_{L^2}.
\]
Note that the minimization over $t$'s allows to find the best possible approximation in the one-parameter set generated by all convex combinations between the copulas $c_1$ and $c_3$. We observe in the second row of Figure~\ref{fig:copula_fit} that the approximation does not match the exact copula so well, especially for $a=1.5$, where we already see the square structure of the copula of dissociated systems while the exact copula does not show this behaviour.

\textit{Wasserstein barycenter.} Another natural interpolation scheme in this setting is to compute Wasserstein barycenters between the copulas for $a=1$ and $a=3.$ Physically, Wasserstein barycenters linearly interpolate the location of lumps of mass instead of their amplitudes. Mathematically, they  correspond to replacing the $L^2$-norm in the previous convex combination by the Wasserstein distance.  
More precisely, we compute an approximation of the copula as
\[
{c}^{W_2}_a = {\rm Bar}^{W_2}_{t*} (c_1,c_3) := u^*(t^*),
\]
where 
\[
    u^*(t) =  \underset{u}{\rm argmin}
    (1-t) W_2(u,c_1)^2 + t  W_2(u,c_3)^2,
\]
$W_2$ is the Wasserstein distance, the minimization is over arbitrary probability densities $u$ on $[0,1]^2$, and 
\[
    t^* = \underset{t\in [0,1]}{\rm argmin} \; \; 
    \underset{u}{\rm min}
    (1-t) W_2(u,c_1)^2 + t  W_2(u,c_3)^2.
\]
Note that the optimization over $t$'s allows to find the best possible approximation in the one-parameter set generated by all Wasserstein barycenters between the copulas $c_1$ and $c_3$.
These barycenters are computed using a Sinkhorn algorithm from the Python Optimal Transport library\cite{flamary2021pot} with a regularization parameter of $10^{-3}$ and 1000 iterations. The results are shown in the third row of Figure~\ref{fig:copula_fit}. We observe that the copulas for $a=2$ and $a=2.5$ are reasonably well approximated, while the approximation for $a=1.5$ still seems a bit poor.

\textit{Sigmoid/neural network interpolation}.
Another approach is to choose a parametrized functional form for the copula and perform parameter optimization. 
As an example, we consider a simple functional form based on a sigmoid which respects the exact marginal constraints, is consistent with our asymptotic results for dissociated systems, and whose parameter is adapted to match the target copula. 
For a parameter $\lambda>0$, the sigmoid function is defined as
\[
    \sigma_\lambda(x) = \frac{1}{1+e^{-\lambda x}},
\]
and our approximation of the copula is
\begin{align} 
    \widetilde\sigma_\lambda(x,y) =  2\Big[\sigma_\lambda(x-1/2)\big(1-\sigma_\lambda(y-1/2)\big) \nonumber \\
   \;\; + \big(1-\sigma_\lambda(x-1/2)\big)\sigma_\lambda(y-1/2)
\Big]. \label{sigmoid}
\end{align}

For each value of the atomic distance parameter $a$, we find the best possible approximation of the copula in  $L^2$-norm, that is
\begin{equation} \label{training}
{c}^{\sigma}_a = \widetilde \sigma_{\lambda^*}, 
\quad 
    \lambda^* = \underset{\lambda}{\rm argmin} \| c_a - \widetilde\sigma_\lambda \|_{L^2}.
\end{equation}
In practice, we solve the above  minimization problem for $\lambda\in [10,1000]$. The results are plotted on the fourth row of Figure~\ref{fig:copula_fit}. We see that the fit, while not perfect, exhibits really nice approximation features, including for $a=1.5$.
Moreover, once the parametrization is decided, computing the copula for a new parameter is extremely cheap, which is not the case when computing Wasserstein barycenters, which can be very costly, especially in higher dimension.

\begin{figure} 
\includegraphics[width=0.35\textwidth]{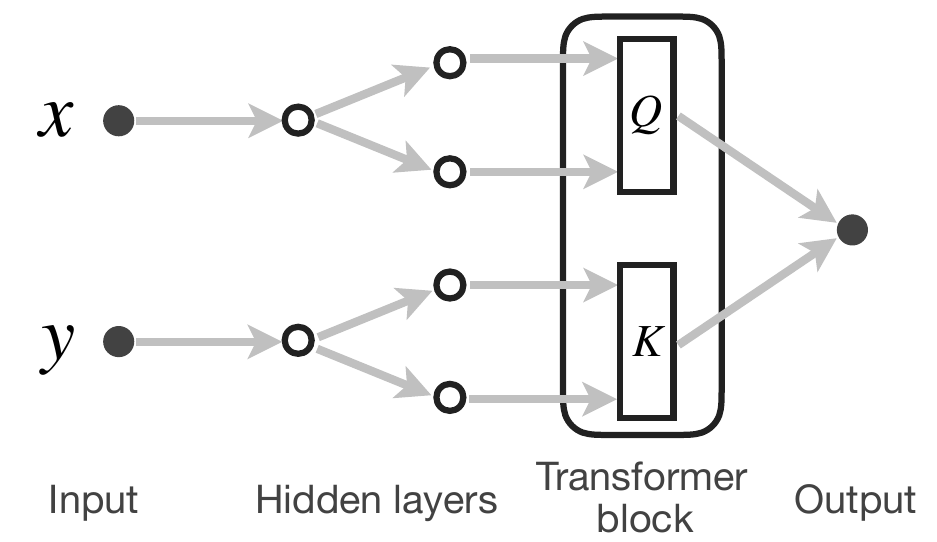}
\caption{The copula \eqref{sigmoid} interpreted as a neural network}
\label{fig:NN}
\end{figure}

The sigmoid copula \eqref{sigmoid} and the parameter optimization \eqref{training} is a basic example of a neural network and its training, and is therefore amenable both to systematic refinement and to generalization to systems with an arbitrary number of particles. 
More precisely, our ansatz \eqref{sigmoid} corresponds to the neural network in Figure \ref{fig:NN}, with all parameters pre-chosen except for the weight in the first layer. One obtains precisely the function \eqref{sigmoid} if 
(i) the first layer is a standard layer $z_i\mapsto \sigma(w_i z_i + b_i)$ with weights $w_i=\lambda$, biases $b_i=-\lambda/2$, and activation function $\sigma(z)=1/(1+e^{-z})$, 
(ii) the second layer is a linear layer $z_i\mapsto w'_iz_i + b'_i$ with weights $w'_1=w'_4=-1$ and $w'_2=w'_3=1$ and biases $b'_1=b'_4=1$ and $b'_2=b'_3=0$, and 
(iii) the third layer is a transformer block with dot-product self attention\cite{Vaswani2017-ps} (as underlying the successes of  large language models), $z\mapsto {\rm v} \; K^T Q$ with ${\rm v}=2$, query and key vectors
$$
  Q = W_Q\begin{pmatrix} z_1 \\ z_2\end{pmatrix}, \;\;\;
  K = W_K\begin{pmatrix} z_3 \\
  z_4\end{pmatrix},
$$
and weight matrices $W_Q=W_K=id_{2\times 2}$. Note that the coupling between the electron positions $x$ and $y$ only occurs through a single dot-product self-attention term, whose input is not however given by the bare electron positions but by the positions pre-processed by neural nets.

\begin{figure*}
    \centering
    \subfigure[$a=1.$]{
    \begin{tabular}{@{}c@{}}
         \includegraphics[width=0.13\textwidth]{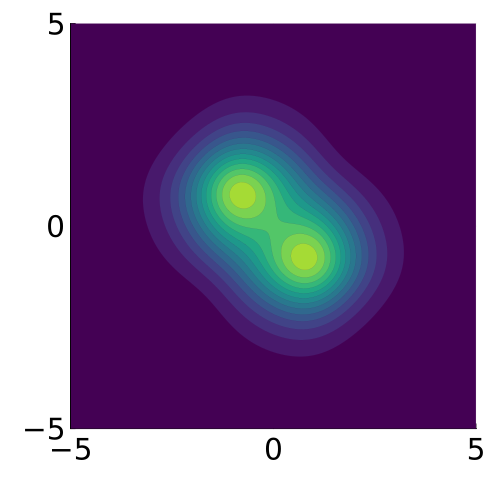} \\
         \includegraphics[width=0.13\textwidth]{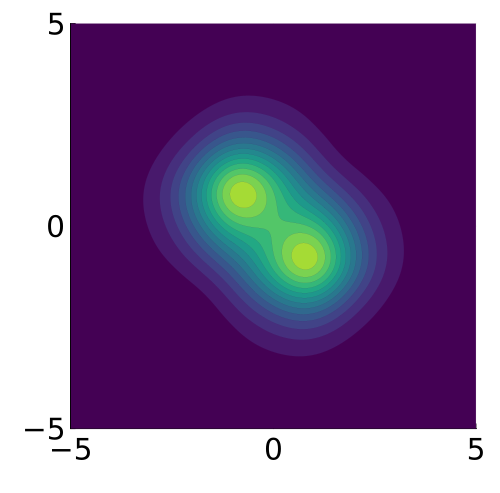} \\
         \includegraphics[width=0.13\textwidth]{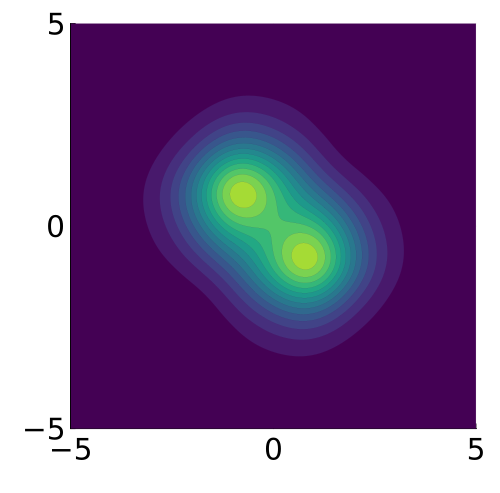} \\
         \includegraphics[width=0.13\textwidth]{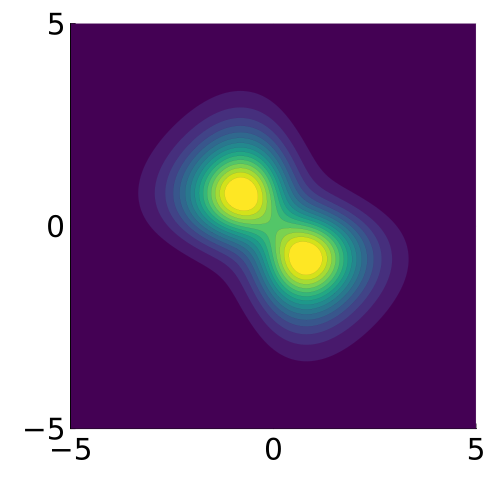} \\
         \includegraphics[width=0.13\textwidth]{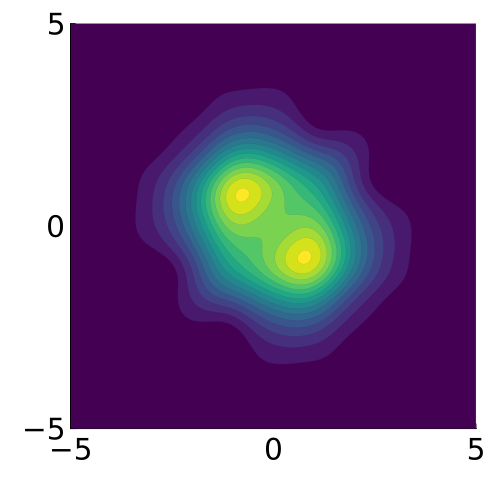}
    \end{tabular}
    }
    \subfigure[$a=1.5$]{
    \begin{tabular}{@{}c@{}}
         \includegraphics[width=0.13\textwidth]{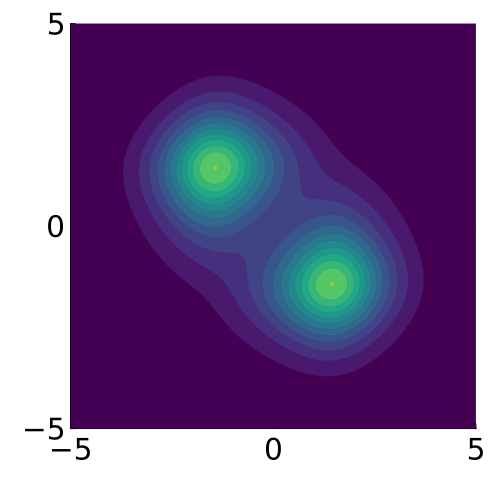} \\
        \includegraphics[width=0.13\textwidth]{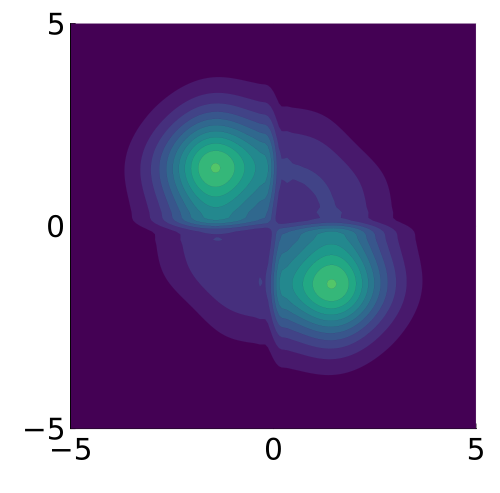} \\
         \includegraphics[width=0.13\textwidth]{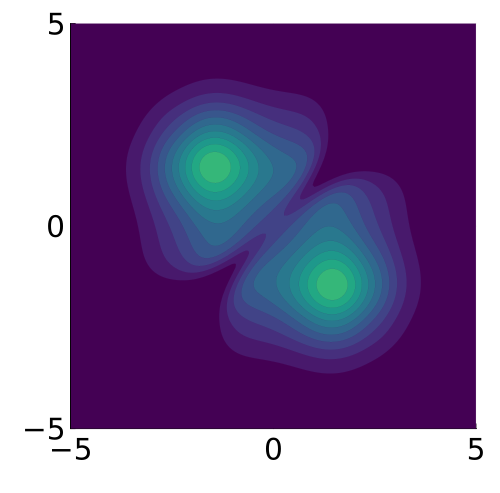} \\
         \includegraphics[width=0.13\textwidth]{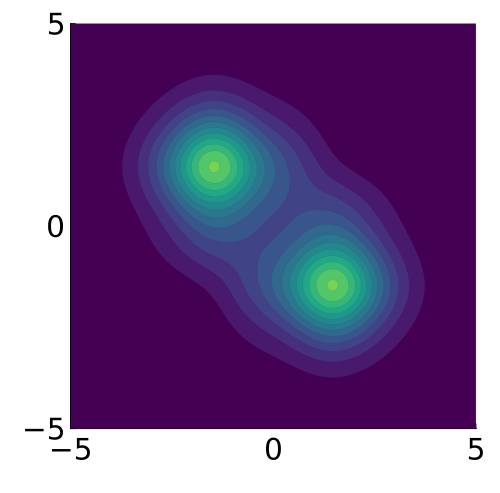} \\
         \includegraphics[width=0.13\textwidth]{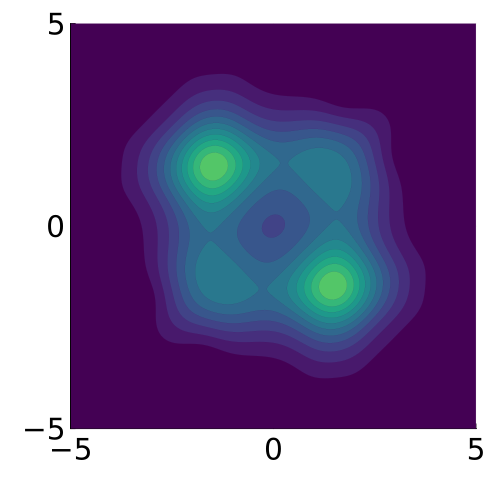}
    \end{tabular}
    }
    \subfigure[$a=2$]{
    \begin{tabular}{@{}c@{}}
         \includegraphics[width=0.13\textwidth]{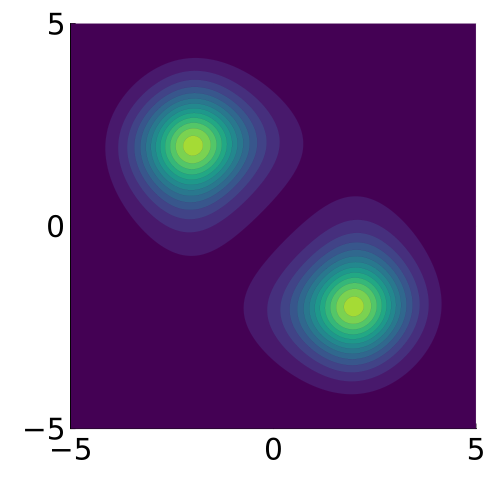} \\
        \includegraphics[width=0.13\textwidth]{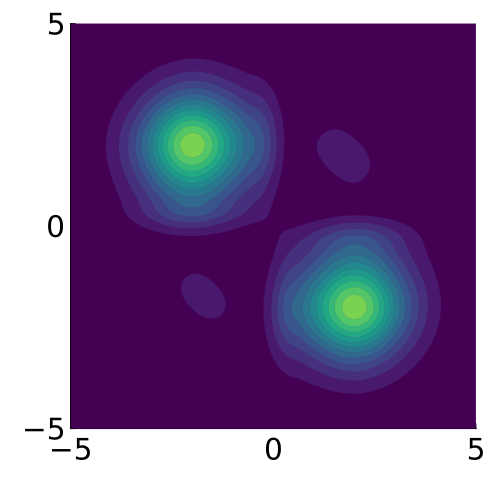} \\
         \includegraphics[width=0.13\textwidth]{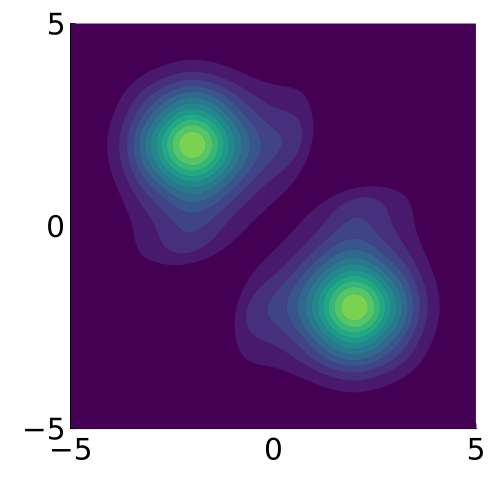} \\
         \includegraphics[width=0.13\textwidth]{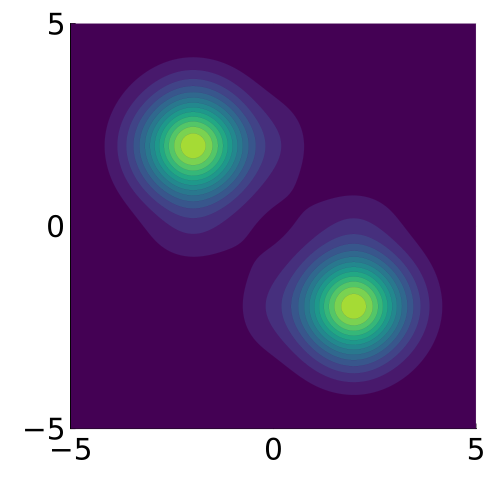}
         \\
         \includegraphics[width=0.13\textwidth]{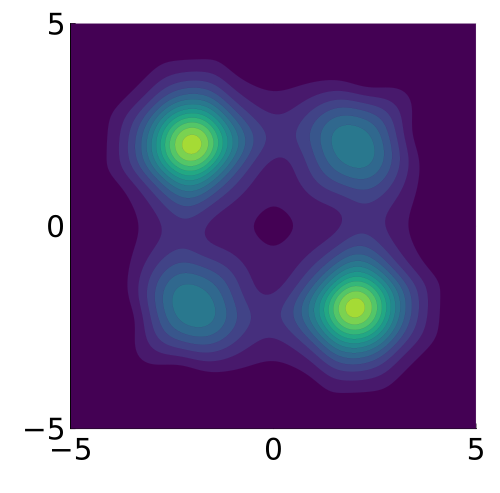}
    \end{tabular}
    }
    \subfigure[$a=2.5$]{
    \begin{tabular}{@{}c@{}}
         \includegraphics[width=0.13\textwidth]{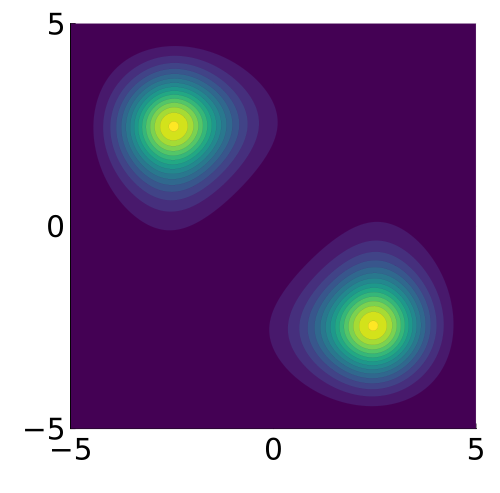} \\
         \includegraphics[width=0.13\textwidth]{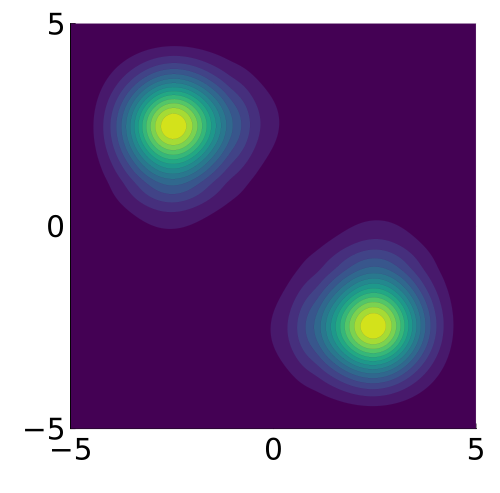} \\         \includegraphics[width=0.13\textwidth]{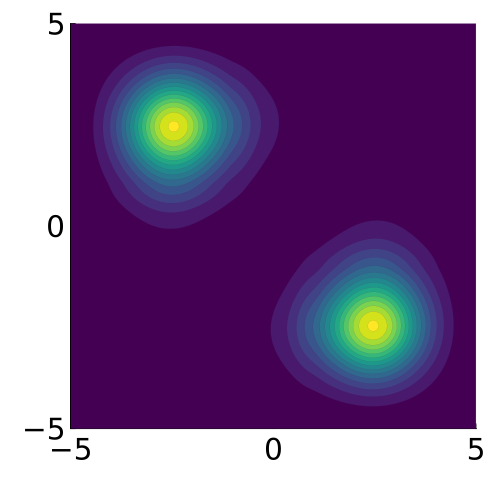} \\
         \includegraphics[width=0.13\textwidth]{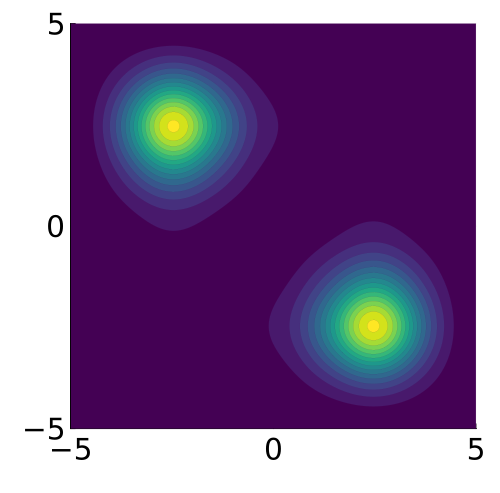}
         \\
         \includegraphics[width=0.13\textwidth]{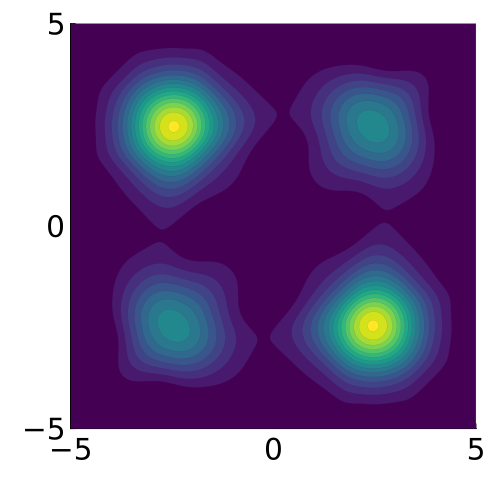}
    \end{tabular}
    }
    \subfigure[$a=3$]{
    \begin{tabular}{@{}c@{}}
         \includegraphics[width=0.13\textwidth]{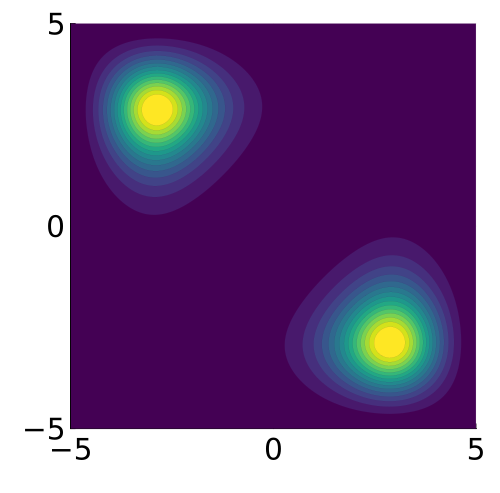} \\
         \includegraphics[width=0.13\textwidth]{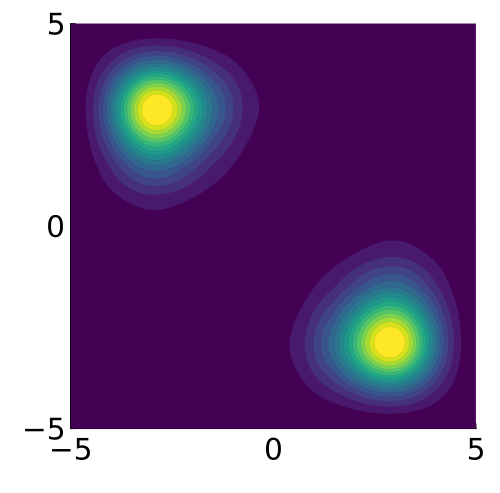} \\        
         \includegraphics[width=0.13\textwidth]{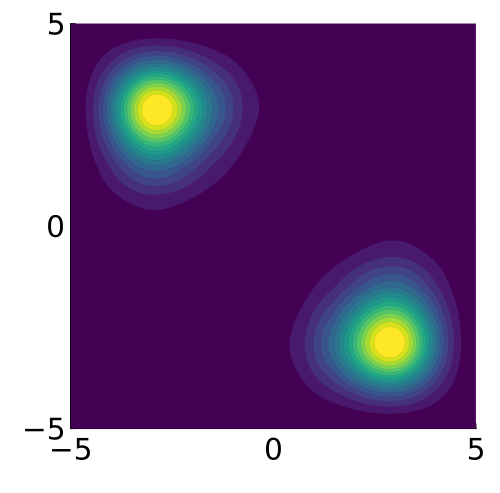} \\
         \includegraphics[width=0.13\textwidth]{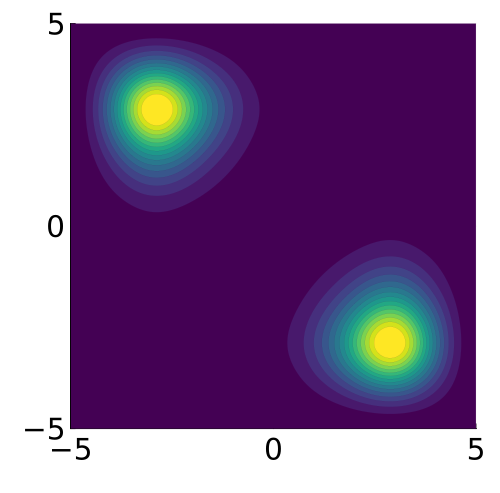}
         \\
         \includegraphics[width=0.13\textwidth]{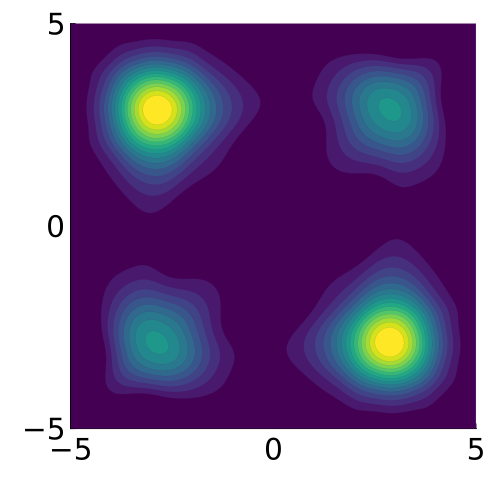}
    \end{tabular}
    }
    \caption{Pair densities for two-particle systems computed from different models. Exact (top row), 
    linear interpolation copula (second row),
    Wasserstein barycenter copula (third row), one-parameter neural net copula (fourth row), LDA-0 (bottom row). The color limits are 
    $(0, 0.1)$.
    The one-parameter neural net copula is seen to provide an excellent fit at all values of the bond length parameter $a$.}
    \label{fig:fit_rho2}
\end{figure*}

\begin{table}[]
    \centering
    \begin{tabular}{|c|c|c|c|}
    \hline
         & a = 1.5 & a=2 & a=2.5  \\
         \hline
        Linear barycenter copula & 1.02e-4 &
        1.85e-4 & 5.32e-5 \\
        \hline
        Wasserstein barycenter copula & 1.14e-4 & 1.10e-4 & 8.87e-4 \\
        \hline
        one-parameter neural net copula & 6.84e-5 & 1.16e-4 & 7.68e-5 \\
        \hline
    \end{tabular}
    \caption{Wasserstein-2 error on the pair density}
    \label{tab:W2}
\end{table}

\begin{table}[]
    \centering
    \begin{tabular}{|c|c|c|c|}
    \hline
         & a = 1.5 & a=2 & a=2.5  \\
         \hline
        Linear barycenter copula & 2.04e-2  & 1.74e-2
         &  5.26e-3 \\
        \hline
        Wasserstein barycenter copula & 2.70e-2 & 1.85e-2 & 6.21e-3 \\
        \hline
        Sigmoid copula & 1.01e-2 & 8.57e-3 & 4.79e-3 \\
        \hline
    \end{tabular}
    \caption{$L^2$ error on the pair density}
    \label{tab:L2}
\end{table}

The copula fits are ultimately used for computing approximations to the pair density, so we present the pair densities derived from the three types of copula approximations in Figure~\ref{fig:fit_rho2}. In the top row, we plot the exact pair densities corresponding to the setting above, i.e. $a=1,1.5,2,2.5,3$.

In the second row, we plot the pair density computed from the copula obtained using convex combinations as in Figure~\ref{fig:copula_fit}, second row. We observe that the pair density for $a=1.5$ exhibits a pretty large error, with a square shape, but is in good agreement for the other cases.

In the third row, we plot the pair density computed from the copula obtained using Wasserstein barycenters as in Figure~\ref{fig:copula_fit}, second row. Apart from the pair density for $a=1.5$, which exhibits a clearly visible error, the two other cases are in good agreement.
We remark that the barycenter of a pair density or copula does not necessarily have the same marginals as the barycenter of the underlying densities. In practice, in our numerical examples this difference was not too important. An alternative definition of Wasserstein barycenters  which preserves the marginals is proposed in~\cite{Dalery2024-gy}, and could be incorporated.

In the fourth row, we plot the pair density corresponding to the sigmoid fit, given by the fourth row in Figure~\ref{fig:copula_fit}. We observe that the pair density is very well approximated in all cases.

In Tables~\ref{tab:W2} and~\ref{tab:L2}, we quantify the errors of the pair density for these examples. It seems that the sigmoid fit gives the best results, however the differences are not so significant. 

\begin{table}[]
    \centering
    \begin{tabular}{|c|c|c|c|}
    \hline
         & a = 1.5 & a=2 & a=2.5  \\
         \hline
        Linear barycenter copula & 2.85e-2  & 4.03e-2
         & 1.07e-2  \\
        \hline
        Wasserstein barycenter copula &2.33e-2  & 9.49e-3 & 2.27e-2 \\
        \hline
        one-parameter neural net copula & 3.49e-3 & 1.82e-2 & 1.14e-2 \\
        \hline
        LDA-0 & 4.25e-1 & 7.54e-1 & 1.10 \\
        \hline
    \end{tabular}
    \caption{Relative Coulomb energy error}
    \label{tab:Coulomb}
    \vspace{-.5cm}
\end{table}

Finally, since the aim of constructing accurate pair densities is to be able to accurately compute quantities of interest such as the electron-electron interaction energy, we compute the relative error in the soft Coulomb energy, 
\begin{equation} \label{Vee}
    E[\rho_2] =  \int \rmv_{\rm ee}(x-y)\, \rho_2(x,y)\,  dx\, dy
\end{equation}
for the different approximations in Table~\ref{tab:Coulomb}. We also compare these errors to those of the exchange-only local density approximation (LDA-0) from Section \ref{sec:LDA}.

We observe in Table~\ref{tab:Coulomb} that the errors for the three approximations are one to two orders of magnitude better than the LDA and that the lowest errors are given by the Wasserstein barycenters, themselves very close to the sigmoid copula fit. The bad performance of the LDA near dissociation is well known\cite{Cohen2012-xk}, with or without an LDA correlation contribution; by contrast the copula models have the correct asymptotics at dissociation built in.

\begin{figure*}
    \centering
    \subfigure[$a=1.5$]{
    \begin{tabular}{@{}c@{}}
         \includegraphics[width=0.32\textwidth]{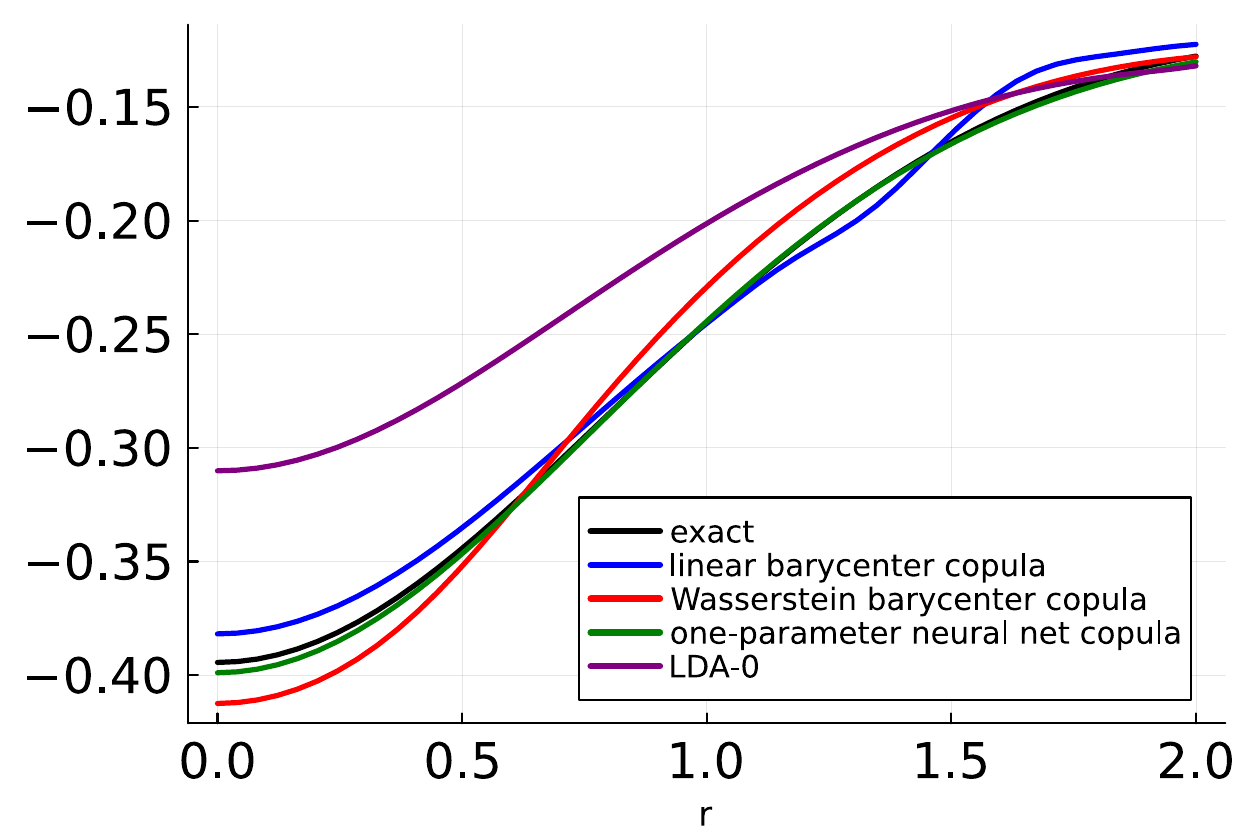} \\
    \end{tabular}
    }
    \subfigure[$a=2$]{
    \begin{tabular}{@{}c@{}}
         \includegraphics[width=0.32\textwidth]{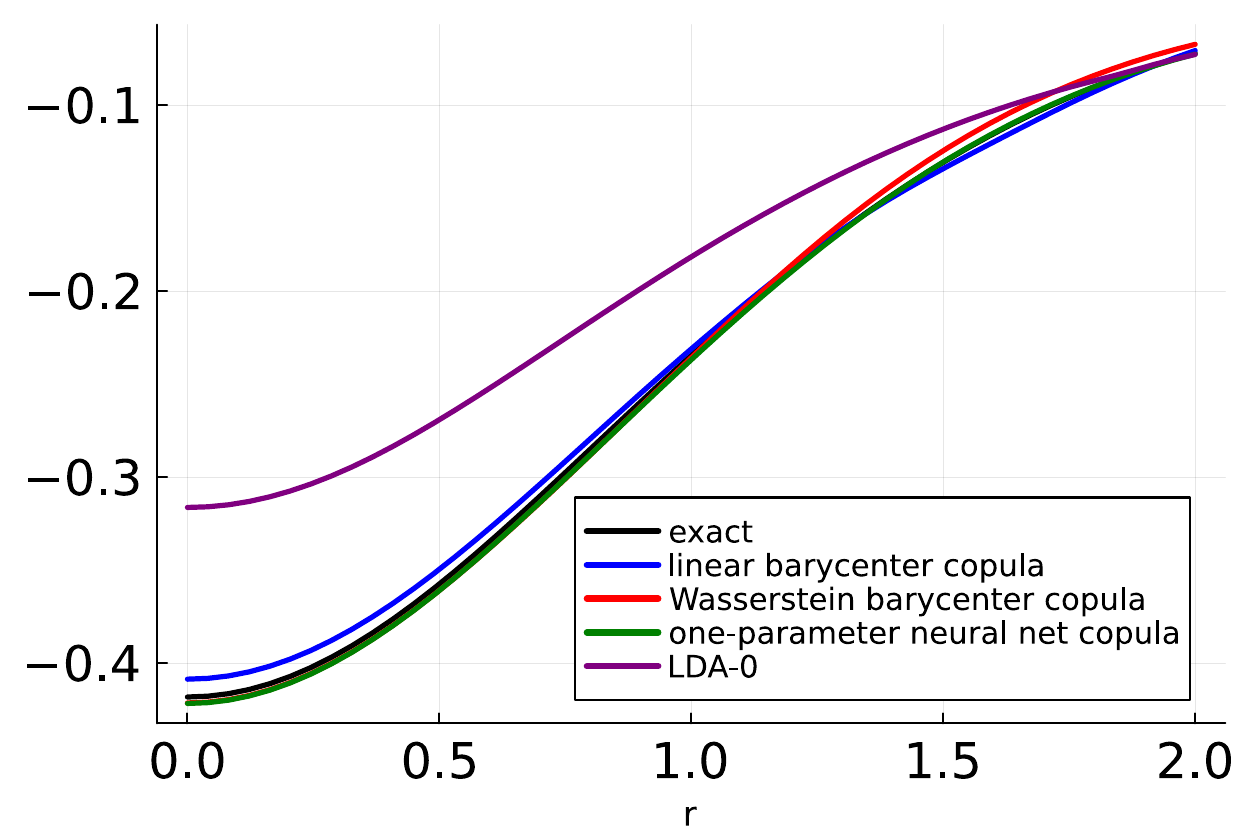} \\
    \end{tabular}
    }
    \subfigure[$a=2.5$]{
    \begin{tabular}{@{}c@{}}
         \includegraphics[width=0.32\textwidth]{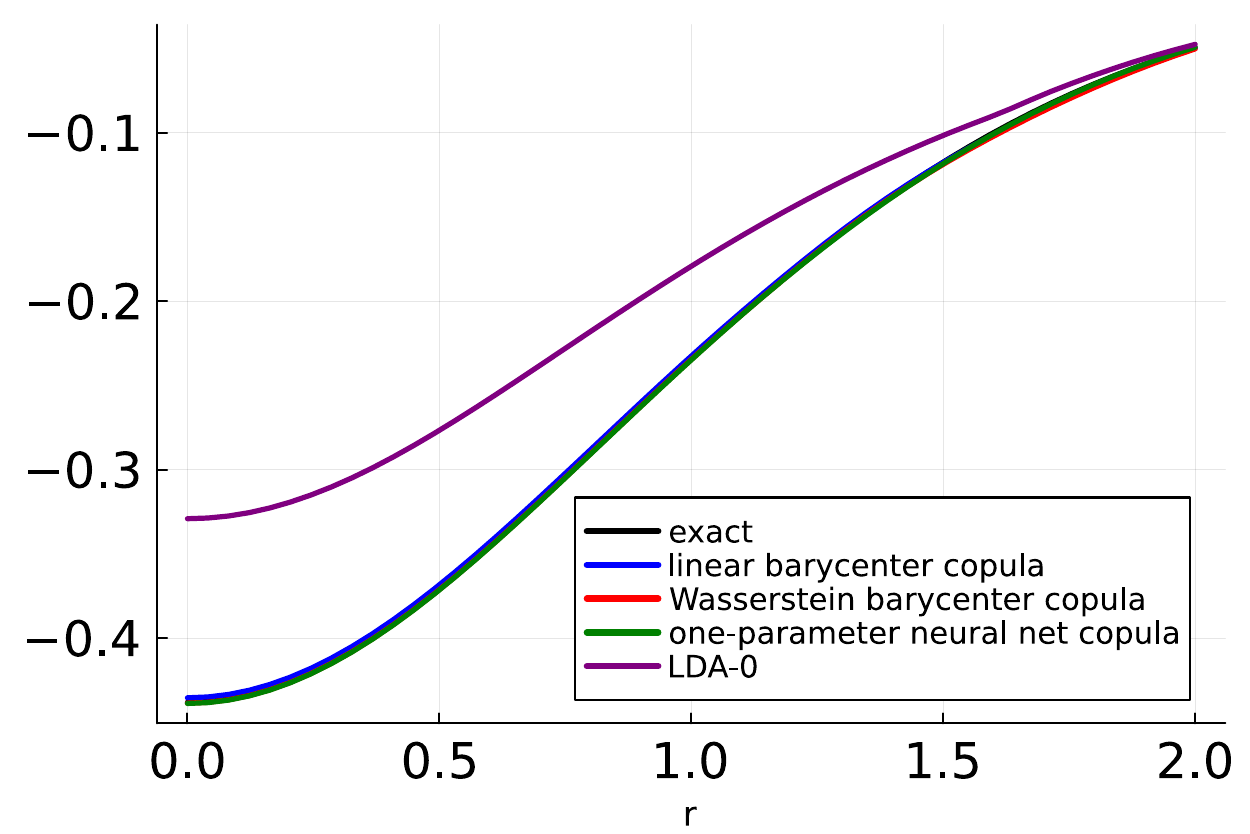} \\
    \end{tabular}
    }
    \caption{Exchange-correlation hole, various methods}
    \label{fig:fit_exhole}
\end{figure*}

In order to provide a different representation of the pair densities, we consider the (radially averaged) exchange-correlation hole, which in one dimension is defined for $x\in\R$ and $r>0$ as
{\small
\[
   h_{\rm xc}(x,r) = \frac12
   \left(\frac{\rho_2(x,x+r)}{\rho(x)} - \rho(x+r) +\frac{\rho_2(x,x-r)}{\rho(x)} - \rho(x-r)
   \right).
\]
}
In Figure~\ref{fig:fit_exhole}, we plot the exchange-correlation holes $h_{\rm xc}(x,r)$ for the different approximations defined above, with $x$ taken to be the position of one of the nuclei, and compare with the LDA. Figure \ref{fig:fit_exhole} shows that all our approximations are much better than the LDA, and the sigmoid approximation provides a near-perfect match of the exact hole, at all values of the bondlength parameter $a$.

\section{Acknowledgements}

The authors would like to thank Yvon Maday for fruitful discussions at the beginning of this project, and an anonymous referee for suggesting to include a discussion of $N$-representability through the angle of the copula.  GD
was supported by the French ‘Investissements d’Avenir’ program, project Agence Nationale de la Recherche (ISITE-BFC) (contract ANR-15-IDEX-0003), and by the Region Bourgogne Franche-Comté.

\appendix

\section{Proof of Theorem~\ref{thm:L:paird}}

We start by recalling the spin-dependent pair density $\rhotilde_2 \, : \, (\Omega\times\Z_2)^2\to\R$ of an $N$-electron wavefunction $\Psi\in L^2_a((\Omega\times\Z_2)^N)$, 
\vspace{-.3cm}
$$
  \rhotilde_2(\bfx_1,\bfx_2) = {N \choose 2} \int_{_{(\Omega\times\Z_2)^{N-2}}} 
  \!\!\!\!\!\!\!\!\!\!\!\!
  |\Psi(\bfx_1,\ldots,\bfx_N)|^2 d\bfx_3\ldots d\bfx_N,
$$
which reduces to the pair density by summing over spins,
$$
   \rho_2(\bfr_1,\bfr_2) = \sum_{\s_1,\s_2\in\Z_2} \rhotilde_2(\bfr_1,\s_1,\bfr_2,\s_2).
$$

It suffices to establish the analogous expression for the spin-dependent pair density $\rhotilde_2$ of $\Psi$ in terms of the spin-dependent subsystem pair densities and densities $\rhotilde_2^A$, $\rhotilde_2^B$ and $\rhotilde^A$, $\rhotilde^B$. 
Formula \eqref{L:paird} then follows by summing over spins, and formula \eqref{L:d} by integrating \eqref{L:paird} over $\bfr_2$. 

Substituting definitions leads to
the following double sum over permutations, 
\begin{equation} \label{doublesum}
 N! k! (N - k)! \; \rhotilde_2(\bfx_1,\bfx_2) = \frac{N(N\mi 1)}{2}\!\!\! \sum_{\sigma,\tau\in{\cal S}_N}\! \epsilon(\sigma)\epsilon(\tau)I_{\sigma,\tau}(\bfx_1,\bfx_2)
 \end{equation}
with 
{\small
\begin{eqnarray*}
   I_{\sigma,\tau} & (\bfx_1,\bfx_2)  =  \displaystyle \int
\Psi^A(\bfx_{\sigma(1)},\ldots,\bfx_{\sigma(k)})
\Psi^B(\bfx_{\sigma(k+1)},\ldots,\bfx_{\sigma(N)}) \\
   & \cdot \Psi^A(\bfx_{\tau(1)},\ldots,\bfx_{\tau(k)})^\ast
   \Psi^B(\bfx_{\tau(k+1)},\ldots,\bfx_{\tau(N)})^\ast d\bfx_3\ldots d\bfx_N. 
\end{eqnarray*}
}
Splitting the double sum into four double sums corresponding to the four possibilities of the indices $1$, $2$
belonging to $\sigma(\{1,\ldots,k\})$ respectively $\sigma(\{k+1,\ldots,N\})$ shows that the r.h.s. of \eqref{doublesum} equals 
{\small
    \begin{eqnarray} 
   & & \;\; \; \frac{N(N\mi 1)}{2}\Bigl[ \Bigr.
         \underbrace{\sum_{{\scriptstyle \sigma,\tau\in{\mathcal S}_{N} \atop \scriptstyle \{1,2\}\subseteq
         \sigma(\{1,\ldots,k\}) }} \! \epsilon(\sigma)\epsilon(\tau)I_{\sigma,\tau}(\bfx_1,\bfx_2)}_{=:\Sigma_1(\bfx_1,\bfx_2)} 
       \nonumber \\
   & & 
   + \underbrace{\sum_{{\scriptstyle \sigma,\tau\in{\mathcal S}_{N} \atop \scriptstyle 1\in\sigma(\{1,\ldots,k\}), \,
         2\in\sigma(\{k+1,\ldots,N\}) }} \!\!\!\! \epsilon(\sigma)\epsilon(\tau)I_{\sigma,\tau}(\bfx_1,\bfx_2)}_{=:\Sigma_2(\bfx_1,\bfx_2)} \nonumber \\
 \end{eqnarray}
}
{\small
    \begin{eqnarray} 
     & & 
   + \underbrace{\sum_{{\scriptstyle \sigma,\tau\in{\mathcal S}_{N} \atop \scriptstyle 2\in\sigma(\{1,\ldots,k\}), \,
         1\in\sigma(\{k+1,\ldots,N\}) }} \!\!\!\! \epsilon(\sigma)\epsilon(\tau)I_{\sigma,\tau}(\bfx_1,\bfx_2)}_{=:\Sigma_3(\bfx_1,\bfx_2)} 
         \nonumber \\
  & & 
   + \;\;\;\;\;\;\, \underbrace{\sum_{{\scriptstyle \sigma,\tau\in{\mathcal S}_{N} \atop \scriptstyle \{1,2\}\subseteq
         \sigma(\{k+1,\ldots,N\}) }} \;\;\;\, \epsilon(\sigma)\epsilon(\tau)I_{\sigma,\tau}(\bfx_1,\bfx_2)}_{=:\Sigma_4(\bfx_1,\bfx_2)}
   \Bigl.\Bigr] . \label{splitsum}
\end{eqnarray}
}

The four sums inside the square brackets will be dealt with separately. In the first sum,
$I_{\sigma,\tau}$
vanishes unless $\tau=(\tau_1\times\tau_2)\sigma$ for $\tau_1\in{\mathcal S}_k(\sigma(\{1,\ldots,k\}))$,
$\tau_2\in{\mathcal S}_{N-k}(\sigma(\{k+1,\ldots,N\}))$, because otherwise one coordinate which is integrated over appears both inside $\Psi^A$ and $\Psi^B$ in the definition of $I_{
\sigma,\tau}$. So the first sum simplifies as follows
    \begin{eqnarray*}
  \Sigma_1(\bfx_1,\bfx_2) && = \sum_{{\scriptstyle \sigma\in{\mathcal S}_{N} \atop \scriptstyle \{1,2\}\subseteq \sigma(\{1,\ldots,k\}) }} 
        \sum_{\tau_1\in{\mathcal S}_{k}}\;
       \sum_{\tau_2\in{\mathcal S}_{N-k}}
       \epsilon(\tau_1)\epsilon(\tau_2) \\
       && \quad \quad
       \int
       \Psi^A(\bfx_{\sigma(1)},\ldots,\bfx_{\sigma(k)})\Psi^B(\bfx_{\sigma(k+1)},\ldots,\bfx_{\sigma(N)}) \\
   & & \cdot \Psi^A(\bfx_{\tau_1\sigma(1)},\ldots,\bfx_{\tau_1\sigma(k)})^\ast
       \Psi^B(\bfx_{\tau_2\sigma(k+1)},\ldots,\bfx_{\tau_2\sigma(N)})^\ast  
       \\&& \qquad \qquad d\bfx_3\ldots d\bfx_N
       \\
   & & = \!\!\!\! \sum_{{\scriptstyle \sigma\in{\mathcal S}_{N} \atop \scriptstyle \{1,2\}\subseteq \sigma(\{1,\ldots,k\}) }}
         \!\!\!\!
       |{\mathcal S}_{k}| \; |{\mathcal S}_{N-k}| \; \\&& 
       \quad \int|\Psi^A(\bfx_1,\bfx_2,\bfy_3,\ldots,\bfy_k)|^2 d\bfy_3\ldots d\bfy_N 
       \\&&
   = \frac{k(k\mi 1)}{N(N\mi 1)}\; |{\mathcal S}_{N}| \; |{\mathcal S}_{k}| \; |{\mathcal S}_{N-k}| \;
       \frac{2}{k(k\mi 1)}\; {\rhotilde_2}^A(\bfx_1,\bfx_2).
\end{eqnarray*}

Here for the last equality we have used that the number of elements of $\sigma\in{\mathcal S}_{N}$ such that $\{1,2\}\subseteq\{\sigma(1),\ldots,\sigma(k)\}$ equals $|{\mathcal S}_{N}|$ times the number of pairs in $\{1,\ldots,k\}$ divided by the number of pairs
in $\{1,\ldots,N\}$, i.e. $\tfrac{k(k-1) }{ N(N-1) }|{\mathcal S}_{N}|$. Consequently
$$
  \frac{N(N-1)}{2} \; \Sigma_1 = N! k! (N-k)! \; \rhotilde_2^A. 
$$
Analogously, for the last sum we obtain 
$$
  \frac{N(N-1)}{2}\; \Sigma_4 = N! k! (N-k)! \; {\rhotilde_2}^B.
$$
Next, let us deal with the second sum, which we re-write as
$$
    \Sigma_2(\bfx_1,\bfx_2) = \sum_{i=1}^k \;\sum_{j=k+1}^N \sum_{{\scriptstyle \sigma\in{\mathcal S}_{N} \atop \scriptstyle \sigma(i)=1,\,
       \sigma(j)=2 }} \sum_{\tau\in{\mathcal S}_{N}}  \epsilon(\sigma)\epsilon(\tau)I_{\sigma,\tau}(\bfx_1,\bfx_2).
$$
For fixed $i$, $j$, by \eqref{TO} the integral $I_{\sigma,\tau}$ vanishes unless $\tau(\{1,\ldots,k\})=\sigma(\{1,\ldots,k\})$ and
 $\tau(\{k+1,\ldots,N\})=\sigma(\{k+1,\ldots,N\}$, so as before summation over $\tau\in{\mathcal S}_N$ can be replaced by summation over $\tau=(\tau_1\times \tau_2)\sigma$ where  
$\tau_1\in{\mathcal S}_{k}$, $\tau_2\in{\mathcal S}_{N-k}$, and we obtain 
\begin{widetext}
{\small
    \begin{eqnarray*}
 \Sigma_2(\bfx_1,\bfx_2) & = & \sum_{i=1}^k\; \sum_{j=k+1}^N \!\!\! \sum_{{\scriptstyle \sigma\in{\mathcal S}_{N} \atop \scriptstyle \sigma(i)=1,\,
      \sigma(j)=2 }}\!\!\! \sum_{\tau_1\in{\mathcal S}_{k}} \; \sum_{\tau_2\in{\mathcal S}_{N-k}} \epsilon(\tau_1)\epsilon(\tau_2) \int \Bigl\{ \Bigr.
      \Psi^A(\bfx_{\sigma(1)},\ldots,\bfx_{\sigma(k)})\Psi^A(\bfx_{\tau_1\sigma(1)},\ldots,\bfx_{\tau_1\sigma(k)})^\ast
      \\[-3mm]
  & & \hspace*{65mm} \cdot
     \Psi^B(\bfx_{\sigma(k+1)},\ldots,\bfx_{\sigma(N)})
     \Psi^B(\bfx_{\tau_2\sigma(k+1)},\ldots,\bfx_{\tau_2\sigma(N)})^\ast \Bigl.\Bigr\} d\bfx_3\ldots d\bfx_N \\
  & = & \sum_{i=1}^k \; \sum_{j=k+1}^N \!\!\!\sum_{{\scriptstyle \sigma\in{\mathcal S}_{N} \atop \scriptstyle \sigma(i)=1,\,
      \sigma(j)=2 }}\!\!\! |{\mathcal S}_{k}| \; |{\mathcal S}_{N-k}| 
   \int
      |\Psi^A(\bfx_{\sigma(1)},\ldots,\bfx_{\sigma(k)})|^2|\Psi^B(\bfx_{\sigma(k+1)},\ldots,\bfx_{\sigma(N)})|^2  \\
  & = & \sum_{i=1}^k\; \sum_{j=k+1}^N \!\!\!\sum_{{\scriptstyle \sigma\in{\mathcal S}_{N} \atop \scriptstyle \sigma(i)=1,\,
      \sigma(j)=2 }}\!\!\! 
      |{\mathcal S}_{k}| \; |{\mathcal S}_{N-k}| \; 
      \frac{1}{k}{\rhotilde}^A(\bfx_1)\frac{1}{N-k}{\rhotilde}^B(\bfx_2) 
  = \frac{k}{N} \, \frac{N-k}{N-1} \; |{\mathcal S}_{N}| \; |{\mathcal S}_{k}| \; |{\mathcal S}_{N-k}| \; \frac{1}{k}{\rhotilde}^A(\bfx_1) \frac{1}{N-k} {\rhotilde}^B(\bfx_2)
    .
\end{eqnarray*}
}
\end{widetext}
Here for the last equality we have used that the number of $\sigma\in{\mathcal S}_{N}$ such that
$1\in\{\sigma(1),\ldots,\sigma(k)\}$ and
$2\in\{\sigma(k+1),\ldots,\sigma(N)\}$ equals ${ \scriptstyle \frac{k}{N} \frac{N-k}{N-1} }|{\mathcal S}_{N}|$. Consequently 
$$
  \frac{N(N-1)}{2}\;\Sigma_2 = N!k!(N-k)! \; \cdot \; \frac12 \rhotilde^A\otimes\rhotilde^B.
$$
Finally, the third sum equals the second sum but with the roles of
$\Psi^A$, $\Psi^B$ and $k$, $N \mi k$ switched, whence
$$
  \frac{N(N-1)}{2}\;\Sigma_3 = N!k!(N-k)! \; \cdot \; \frac12 \rhotilde^B\otimes\rhotilde^A.
$$
Collecting terms and substituting into \eqref{splitsum} yields the desired expression for $\rhotilde_2$.

\bibliography{biblio}

\end{document}